\documentclass[hyper]{JHEP3}

\pdfoutput=1
\usepackage{epsfig}
\usepackage{latexsym}
\usepackage{amsfonts}
\usepackage{amsmath}
\usepackage{amsthm}
\usepackage{amssymb}
\usepackage{amsbsy}
\usepackage{multirow}

\newcommand{\cp}[1]{\ensuremath{\mathbb{CP}^{#1}}}
\newcommand{\wpr}[2]{\ensuremath{W\mathbb{P}^{#1}_{#2}}}
\def\be{\begin{equation}}
\def\ee{\end{equation}}
\def\bea{\begin{eqnarray}}
\def\eea{\end{eqnarray}}
\def\bi{\begin{itemize}}
\def\ei{\end{itemize}}
\def\ba{\begin{align}}
\def\ea{\end{align}}

\def\Dob{{\rm \overline{D0}}}
\def\Dsb{{\rm \overline{D6}}}
\def\Dtb{{\rm \overline{D2}}}

\title{\begin{center} Black hole meiosis \end{center}}
\author{\begin{center}{Walter Van Herck and Thomas Wyder}\end{center}\\
\begin{center}{Institute for Theoretical Physics}\\
{K.~U.~Leuven }\\
{Celestijnenlaan 200D}\\
{B-3001 Leuven, Belgium}\end{center}
\bigskip
\centerline{{\rm E-mail}: \email{waltervh@itf.fys.kuleuven.be, thomas@itf.fys.kuleuven.be}}}

\abstract{The enumeration of BPS bound states in string theory needs refinement. Studying partition functions of particles made from D--branes wrapped on algebraic Calabi--Yau 3--folds, and classifying states using split attractor flow trees, we extend the method for computing a refined BPS index, \cite{Collinucci:2008ht}. For certain D--particles, a finite number of microstates, namely polar states, exclusively realized as bound states, determine an entire partition function (elliptic genus). This underlines their crucial importance: one might call them the `chromosomes' of a D--particle or a black hole. As polar states also can be affected by our refinement, previous predictions on elliptic genera are modified. This can be metaphorically interpreted as `crossing--over in the meiosis of a D--particle'. Our results improve on \cite{Gaiotto:2007cd}, provide non--trivial evidence for a strong split attractor flow tree conjecture, and thus suggest that we indeed exhaust the BPS spectrum. In the D--brane description of a bound state, the necessity for refinement results from the fact that tachyonic strings split up constituent states into `generic' and `special' states. These are enumerated separately by topological invariants, which turn out to be partitions of Donaldson--Thomas invariants. As modular predictions provide a check on many of our results, we have compelling evidence that our computations are correct.}

\keywords{Elliptic genus, Calabi Yau, Topological String, Donaldson--Thomas invariant, Wall--crossing}

\preprint{KUL-TF-09/18}

\begin{document}
\section{Introduction}
\label{sect:intro}
By virtue of the correspondence principle, BPS black holes in 4d and 5d string theory compactifications can be studied as bound states of D--brane systems. These D--branes are wrapped such that they form a particle from a 4d (or 5d) point of view: a so--called \emph{D--particle}. If the D--branes carry a sufficient amount of charge, the system will form a black hole through the gravitational backreaction as soon as the string coupling is turned on. Also a D--particle carrying minimal charge is a very interesting toy model, which offers itself for studying the basic mathematical principles governing BPS black holes, although the direct interpretation as a black hole is lost.

In this paper, the partition functions of D--particles modeled as mixed (magnetic/electric charge) ensembles with branes wrapped on algebraic Calabi--Yau 3--folds are studied. The microstates of key interest are so--called polar states, which are given the interpretation of chromosomes of D--particles or black holes in subsection \ref{subsect:chromosomes}. One of the main virtues of polar states of black holes is that they are always multi--centered solutions. In the case of a D--particle, polar and some non--polar bound states are enumerated, and we propose a refined computational scheme to calculate indices for these bound states. At the same time, we split up Donaldson--Thomas invariants, used to enumerate constituents of these bound states, into \emph{Donaldson--Thomas partitions}. The index for which we propose a refinement was used in \cite{Denef:2007vg}. The authors of this paper also state that a moduli space of a bound state is in general a fibration. This was examined for an example in \cite{Collinucci:2008ht} and a refinement of the index enumerating a bound state was suggested there. This work can be seen as a further development of these latter ideas.\\
\\
\textbf{Interest in multi--centered BPS black hole solutions}\\
\\
In recent years, great interest has arisen in multi--centered black hole solutions within the BPS spectrum of type II string theory compactifications. This has several reasons, of which we shall list a few of special interest for the development of our work:
\bi
   \item According to the BPS black hole attractor mechanism, a spherically symmetric (single--centered) black hole solution can be found in $\mathcal{N}=2, d=4$ supergravity theories, independently of the chosen values of the vector multiplet scalar fields at infinite distance of the black hole $t_{\infty}:=t(r=\infty)$, where $t(r)$ denotes the value of the scalars in dependence of a radial spacetime coordinate $r$. The value of these moduli will be driven to the so--called `attractor value' at the event horizon, $t_{*}:=t(r=0)$. The values of these moduli form a line in moduli space (connecting $t_{\infty}$ and $t_{*}$), which will be referred to as a \emph{single flow}. G. Moore put forth a correspondence between spherically symmetric solutions to the BPS attractor equations and BPS states in string theory. Studies by G. Moore in 1998 \cite{Moore:1998pn,Moore:1998zu} showed that this correspondence does not hold. For a full correspondence, (at least) stationary multi--centered solutions must also be considered.
      \item The connection between the supergravity description of BPS black hole solutions and the D--brane description has received completely new impulses from the studies on multi--centered solutions and bound states over the last decade. In \cite{Denef:2002ru} it was shown how a D--brane system wrapped around a compact manifold, yielding a single point in space, can transform into a bound state of two (or more) constituents, in the gravitational description. Genuine bound states of black holes, subject to specific equilibrium distances for stability, were discovered by Denef and collaborators \cite{Denef:2000ar,Denef:2000nb}. A review how to construct these multi--centered solutions was given in \cite{Bates:2003vx}.  By now, it seems fair to say that it has been acknowledged that they form a very prominent part of the spectrum. A sample study of the BPS spectrum for a type II compactification on the quintic 3--fold provides a nice illustration \cite{Denef:2001xn}.
  \item The formulation of the OSV conjecture \cite{Ooguri:2004zv} released a period of very active research on the connection of black hole entropy and topological string theory; see \cite{Pioline:2006ni} for a review. Basically, the conjecture suggests a relation of a black hole partition function with the topological string partition function of the form, $\mathcal{Z}_{\textrm{BH}}\approx |\mathcal{Z}_{\textrm{top}}|^2$. Attempts to prove this conjecture have proven to be extremely difficult. One attempt by Denef and Moore, which is of special relevance for the present work, put multi--centered black hole solutions to the center of the stage \cite{Denef:2007vg}. The authors concretized the conjecture using a D4--D2--D0 partition function as a black hole partition function, built from a mixed ensemble, keeping the D4--charge fixed and varying over the D2/D0 charges. This partition function displays modular invariance and it turns out that it is completely determined by knowing the degeneracy of a finite number of microstates: \emph{the polar states}\footnote{These polar states are typically states with a low D2/D0--charge; upon increasing the D2/D0 charge, the states become non--polar at some point.}. Such a relation was derived by the authors of \cite{Denef:2007vg} and independently in \cite{deBoer:2006vg}, subject to certain amendments \cite{Manschot:2007ha}. These polar states are exclusively realized as multi--centered solutions, as will also be discussed in section \ref{sect:background}. A large part of the work presented in this paper consists of the enumeration of polar states for D--particles.
  \item Another finding of \cite{Denef:2007vg} is, that upon scaling up the charge of a brane system, the entropy of multi--centered microstates carrying that charge will always start to dominate over the entropy coming from single--centered states. This is puzzling, as the entropy of multi--centered solutions scales with the cube, $S\rightarrow \Lambda^3 S$ when scaling D--brane charges $Q\rightarrow\Lambda Q$, and not with the square, as is the case for single--centered solutions, $S\rightarrow \Lambda^2 S$. For a (large) black hole, that carries large charges, this implies that one would expect a far too large entropy and this puzzling discovery has been given the name \emph{entropy enigma}. This again stresses the importance of multi--centered solutions. A possible connection exists with another phenomenon discovered, namely the appearance of \emph{scaling solutions}. Scaling solutions appear as multi--centered BPS solutions, which however allow the continous variation of a parameter, the distance between constituents. This implies that one can vary this `scaling modulus' and choose two centers to lie so close together in spacetime that their throats virtually melt together, and the constituents become indistinguishable to an outside observer. The precise meaning of these solutions is at this moment still a point of ongoing discussion, and their implications for the split attractor flow conjecture (discussed in section \ref{subsect:splitflows}) are not completely clear. These issues have been recently addressed in \cite{deBoer:2008fk,deBoer:2009un,deBoer:2008zn}. We will discuss our results in the light of these issues in a separate section, \ref{scalingsolutions}.
 \ei

Just like single--centered BPS black holes, multi--centered BPS black hole solutions are also governed by an attractor mechanism. It was originally extended to multi--centered black holes in \cite{Denef:2000nb}. The image of the scalar moduli in moduli space forms a line, that splits (possibly several times) and runs from a background value $t_{\infty}$ to a split point (and maybe further split points), in order to end at two (or several) attractor points, $t_{1*}, t_{2*},...$, one for each center of the bound state. This image has been given the name \emph{split attractor flow tree} and has been conjectured to be an existence criterion for a multi--centered black hole solution. The meaning of split flow trees extends beyond supergravity, as has been confirmed by studies on the quiver description of D--brane systems in \cite{Denef:2002ru}. In fact, the \emph{split attractor flow tree conjecture} states that (single and) split flows completely classify the BPS spectrum of type II string theory. The results in this paper can be interpreted as strong evidence for this conjecture.\\
\\
\textbf{Main goal: index refinement for BPS bound states}\\
\\
In \cite{Collinucci:2008ht}, a study very similar to those in this paper was performed, and a refined prescription for calculating an index for  a BPS bound state was proposed. The reason for the necessity of this refinement lay in the fact that the tachyon fields connecting the two constituents of a bound state did not perceive all the individual microstates of a constituent generically. The investigations in this paper go in the same direction, however the reason why the tachyon fields do not perceive all constituent states of a bound states generically, will be different. This leads us to new techniques to calculate the refined index. Non--trivial checks on our method yield exact confirmation of the predictions from modularity on degeneracies of certain states, suggesting that our technique is indeed correct, and beyond that, that the strong split attractor flow conjecture might well be completely accurate. Whereas the `naive' index to enumerate a bound state is a simple product, we argue, as in \cite{Denef:2007vg,Collinucci:2008ht}, that the moduli space is rather a fibration. This means that the index falls into a sum of several pieces, which can be grouped logically. This leads us to distinguish between those states of a constituent, in our case a D6--brane system enumerated by a Donaldson--Thomas invariant, that are perceived generically, and those that are perceived as special states. This also allows the definition of \emph{Donaldson--Thomas partitions}, which enumerate those generic and special states separately. The main idea of our technique will be explained at the beginning of chapter \ref{sect:DTpartitions}, which also contains our results on the refined bound state index and on Donaldson--Thomas partitions.\\
\\
\textbf{Organization of this paper}
\bi
 \item Section \ref{sect:background} covers some basic material and may be skipped by the expert reader. Subsection \ref{subsect:dtinvariants} briefly recapitulates the index developed for the enumeration of BPS bound states in type II string theory compactifications in \cite{Denef:2007vg}, extensively used and eventially improved on in this paper. For the sake of stressing their importance, special bound states, namely \emph{polar states}, which determine elliptic genera through modularity, are given the interpretation of chromosomes of a D--particle / a black hole in subsection \ref{subsect:chromosomes}. The reader unfamiliar with techniques used in this paper will find short, but more or less self--contained appendices covering the relevant background material used in this paper. We mostly follow the setup of \cite{Denef:2007vg} and \cite{Collinucci:2008ht}. The reader will find more explanation on the setup of this paper, in particular D--branes wrapped on algebraic one--modulus Calabi--Yau varieties, in appendix \ref{sect:setup}, a short account of (modified) elliptic genera in appendix \ref{sect:ellGenera}, and a short introduction to split attractor flow trees as well as a formulation of the split attractor flow tree conjecture is presented in appendix \ref{subsect:splitflows}. Eventually, to concretely establish the existence of BPS states for our study models, we will use numerical techniques, involving the use of mirror symmetry, adapted from \cite{Denef:2001xn}. This is necessary, as we work with very small charges, and therefore instanton corrections to the central charges of brane systems become dominant. More details on these techniques can be found in appendix \ref{numericalmethods}. 
   \item Section \ref{sect:ellipticgenera} presents our results for polar states on two Calabi--Yau's, described as hypersurfaces in weighted projective spaces. These results allow a prediction on the partition function for a D--particle, and exactly match the results in \cite{Gaiotto:2007cd}.
  \item Section \ref{sect:DTpartitions} presents our most interesting findings. For some tractable non--polar states for two of our models, calculations based on the `naive computation' of a BPS index show a discrepancy from the result predicted by modularity. We demonstrate that our computations based on a refined index for BPS bound states however show an exact matching with the prediction. At the same time, we use our procedure to distinguish between constituent states that are perceived by the tachyon field generically, and those that are perceived differently: we use the terms `generic' and `special' states and call the invariants that enumerate these D6--brane systems, \emph{Donaldson--Thomas partitions}, $\mathcal{N}^{(g)}_{\textrm{DT}}(\beta ,n), \mathcal{N}^{(s)}_{\textrm{DT}}(\beta ,n)$, where the superscripts `g' and `s' stand for `generic' and `special', respectively, and where $N_{\textrm{DT}}(\beta ,n)=\mathcal{N}^{(g)}_{\textrm{DT}}(\beta ,n)+\mathcal{N}^{(s)}_{\textrm{DT}}(\beta ,n)$.
 \item In section \ref{refinedellipticgenera}, we show that the refined prescription to compute BPS indices for bound states also alters the enumeration of polar states, leading to a new prediction for the elliptic genus for a degree ten hypersurface, embedded in a weighted projective space. This prediction is slightly different from the results in \cite{Gaiotto:2007cd}. 
   \item Section \ref{sect:discussion} is a discussion of our results. We devote some time to the question whether anything can be learned about a `tentative' part of the classical BPS spectrum of a D--particle, namely scaling solutions, in a separate subsection, \ref{scalingsolutions}. In another subsection, \ref{subsect:meiosis}, we interpret our most interesting results, the necessity for the refinement of bound state indices, as an artefact of a `sort of meiosis for D--particles / black holes'. We conclude with some implications and possible directions for future research. 
   \ei

\section{Exact enumeration of BPS microstates of a D--particle}\label{sect:background}
We model D--particles using mixed ensembles of D4--D2--D0 branes wrapped on a Calabi--Yau 3--fold $X$, which we choose to be a hyperplane in a weighted projective space. A D4--brane with charge $p=1$ is wrapped on the hyperplane class divisor, and kept fixed, while varying over all possible $U(1)$ worldvolume fluxes $F\in H^2(X)$ and various numbers of bound $\Dob$--branes. We follow the setup of \cite{Denef:2007vg,Collinucci:2008ht}. The reader can find more details on our setup and conventions in appendix \ref{sect:setup}.

The existence of D4--D2--D0 BPS states will be inferred from the the existence of split attractor flow trees. The reader unfamiliar with this technique can find a short introduction in appendix \ref{subsect:splitflows}. As we inspect low--charged BPS states, higher order curvature corrections to the action become important, and central charges of brane systems will have to be calculated by exploiting mirror symmetry. Details on this procedure can be found in appendix \ref{numericalmethods}.

\subsection{An index for BPS bound states using Donaldson--Thomas (DT) invariants}\label{subsect:dtinvariants}

Our goal is to describe D4--D2--D0 configurations  as bound states of D6--D2--D0 and ${\rm \overline{D6}}$--D2--D0, possibly carrying U(1) fluxes, whereby the D4 charge is induced by the latter fluxes. This will allow us to factorize the indices of D4 systems as products of D6 and $\rm{\overline{D6}}$ indices. In order to compute the BPS indices of the D6 and $\rm{\overline{D6}}$ systems, we will make use of Donaldson--Thomas invariants.  An invariant $N_{\textrm{DT}}(\beta, n)$ computes the Witten index of a system with a D2 brane wrapping a curve of homology class $\beta$, and a collection of $\rm{\overline{D0}}$'s, such that the total D0 charge equals $n$. Although the U(1) flux on the D6 interacts with these lower branes, it does not alter the Witten index. In mathematical terms, the DT invariants compute the dimensions of the moduli spaces of the ideal sheaves corresponding to curves and points on the Calabi--Yau. They are indeed conjectured \cite{MNOP1, MNOP2} to contain equivalent information as the Gopakumar--Vafa invariants \cite{Gopakumar:1998ii, Gopakumar:1998jq}, which count the states of M2 branes with momentum, where the M2's are wrapped on holomorphic curves. By the conjectured identity between the generating functional for GV invariants and DT invariants \cite{Dijkgraaf:2006um}, one can easily obtain the DT invariants for the Calabi--Yau manifolds we use in this paper, from \cite{Huang:2006hq} and \cite{Klemm:2004km}. We will state the DT invariants of interest for the models we study, where they are directly applied, in chapters \ref{sect:ellipticgenera} and \ref{sect:DTpartitions}.

As was done in \cite{Collinucci:2008ht}, we will use the index for D4--D2--D0 BPS states of total charge $\Gamma$ from \cite{Denef:2007vg} to enumerate states:
\begin{equation}
  \Omega(\Gamma)=\sum_{\Gamma\rightarrow\Gamma_1+\Gamma_2}(-1)^{|\langle\Gamma_1,\Gamma_2\rangle |-1}|\langle\Gamma_1,\Gamma_2\rangle |\,\Omega (\Gamma_1)\,\Omega (\Gamma_2),
\end{equation}
with the sum running over all possible first splits $\Gamma\rightarrow\Gamma_1+\Gamma_2$, belonging to a full split flow tree, and $\langle \Gamma_1,\Gamma_2\rangle$ is the symplectic intersection of the two charges, as defined in appendix \ref{sect:setup}. The microscopic logic behind this formula is that all degrees of freedom in a D6/$\Dsb$ can be factorized as the degrees of freedom on the gauge theories of the D6 and $\Dsb$ plus the degrees of freedom of the tachyon field, which are counted by the intersection product. This formula is not accurate though, for all cases, as the moduli space of a bound state is in general a fibration: this is discussed in detail, in chapter \ref{sect:DTpartitions}.

In general, as discussed in appendix \ref{subsect:splitflows}, a specific charge can give rise to several split flow trees. A split flow tree will contribute a term to the index of the D4 system as follows:
\begin{equation}
  \Delta \Omega(\Gamma_{D4})=(-1)^{|\langle\Gamma_{D6},\Gamma_{\Dsb}\,\rangle |-1}\,|\langle\Gamma_{D6},\Gamma_{\Dsb}\rangle | \, N_{\textrm{DT}}(\beta_1, n_1) \, N_{\textrm{DT}}(\beta_2, n_2)\,,
\end{equation}
where 
\begin{equation}
n = \tfrac{1}{2}\,\chi(C_{\beta})+N
\end{equation}
with $n$ the total and $N$ the added $\Dob$--charge.

The research presented in this paper is closely connected to an `OSV--like statement' for D--particles. The topological string partition function is used to enumerate D6 (and lower dimensional brane charge) systems, or, in the mirror picture, to count D3--brane systems. Schematically, the relationships, of which we would like to gain a better understanding, can be expressed as
\be
    \mathcal{Z}_{D_{\textrm{particle}}}\sim|\mathcal{Z}_{\textrm{top}}|^2\approx |\mathcal{Z}_{\textrm{DT}}|^2\approx |\mathcal{Z}_{D6}|^2=|\mathcal{Z}_{D3}|^2,
\ee
where:
\bi
   \item $\mathcal{Z}_{D_{\textrm{particle}}}\sim|\mathcal{Z}_{\textrm{top}}|^2$ is an OSV--like statement for a D--particle.
   \item $|\mathcal{Z}_{\textrm{top}}|^2\approx |\mathcal{Z}_{\textrm{DT}}|^2$ expressed that one uses the asymptotic expansion of the topological string partition function (which is not known exactly) known as the Donaldson--Thomas (DT) partition function.
   \item $|\mathcal{Z}_{\textrm{DT}}|^2\approx |\mathcal{Z}_{D6}|^2$ expressed that one uses DT invariants to enumerate D6--D4--D2--D0 BPS states\footnote{The closer relationship between D6--D2--D0 states and DT invariants has been clarified in \cite{Denef:2007vg}.}
   \item $|\mathcal{Z}_{D6}|^2=|\mathcal{Z}_{D3}|^2$ expressed that mirror symmetry comes into play, and that we examine the mirror D3--brane systems, as these allow computation of exact central charges.
\ei

The partition function of a D--particle $\mathcal{Z}_{D_{\textrm{particle}}}$ is identical with the modified elliptic genus, associated to the M--theory picture of the worldvolume description of these ensembles of BPS D--brane states. Elliptic genera are roughly sketched in appendix \ref{sect:ellGenera}. The rough scheme discussed above is followed up in this paper, resulting in various elliptic genera for specific CY 3--fold study models, obtained from a new perspective.

\subsection{Polar states: the chromosomes of D--particles / black holes}\label{subsect:chromosomes}
\begin{figure}[h]
\setlength{\unitlength}{1cm}
\centering 
\begin{picture}(5,4)
\put(-4.5,0) {\includegraphics[width=0.415\textwidth,angle=0]{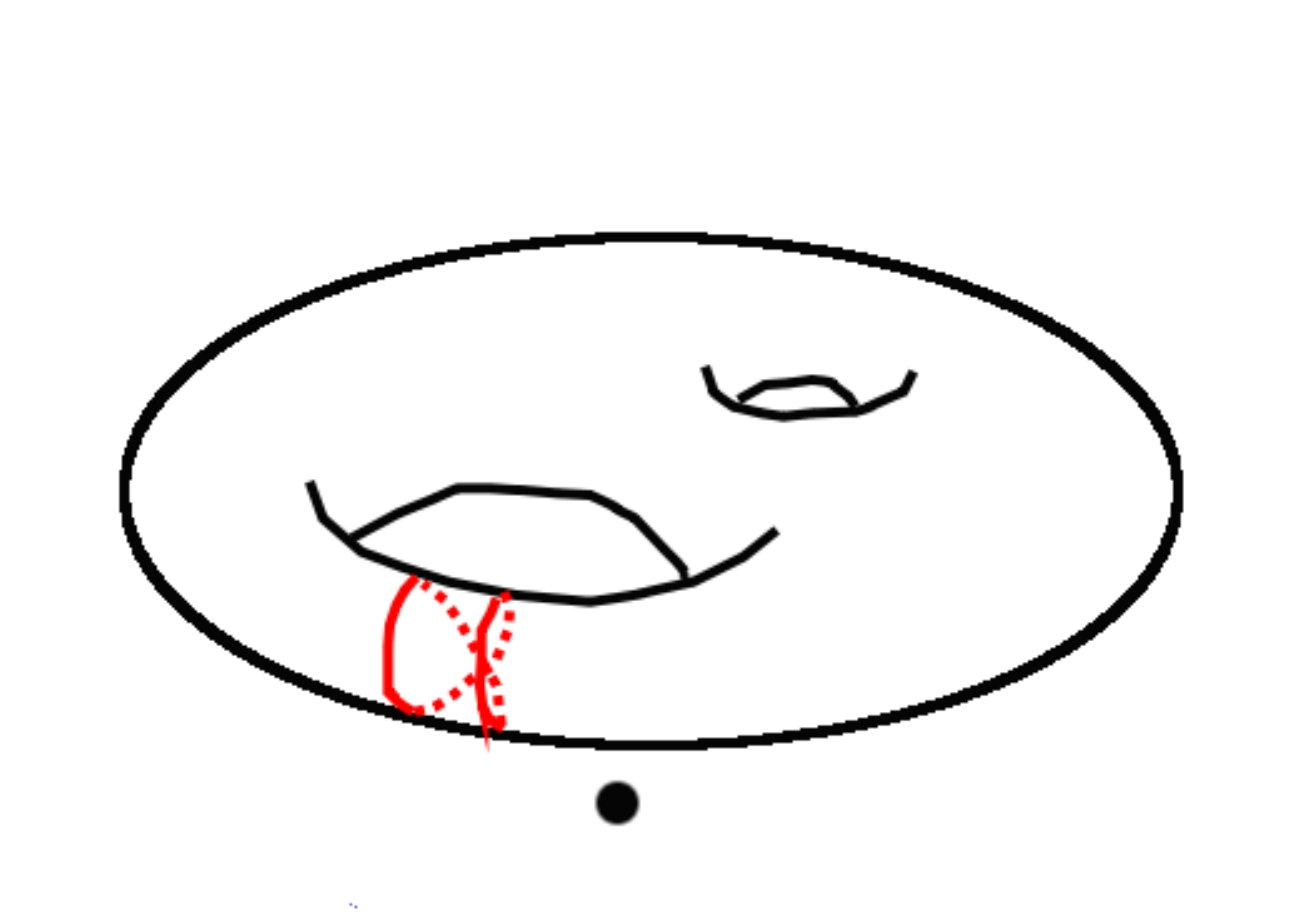}} 
\put(2.5,0) {\includegraphics[width=0.415\textwidth,angle=0]{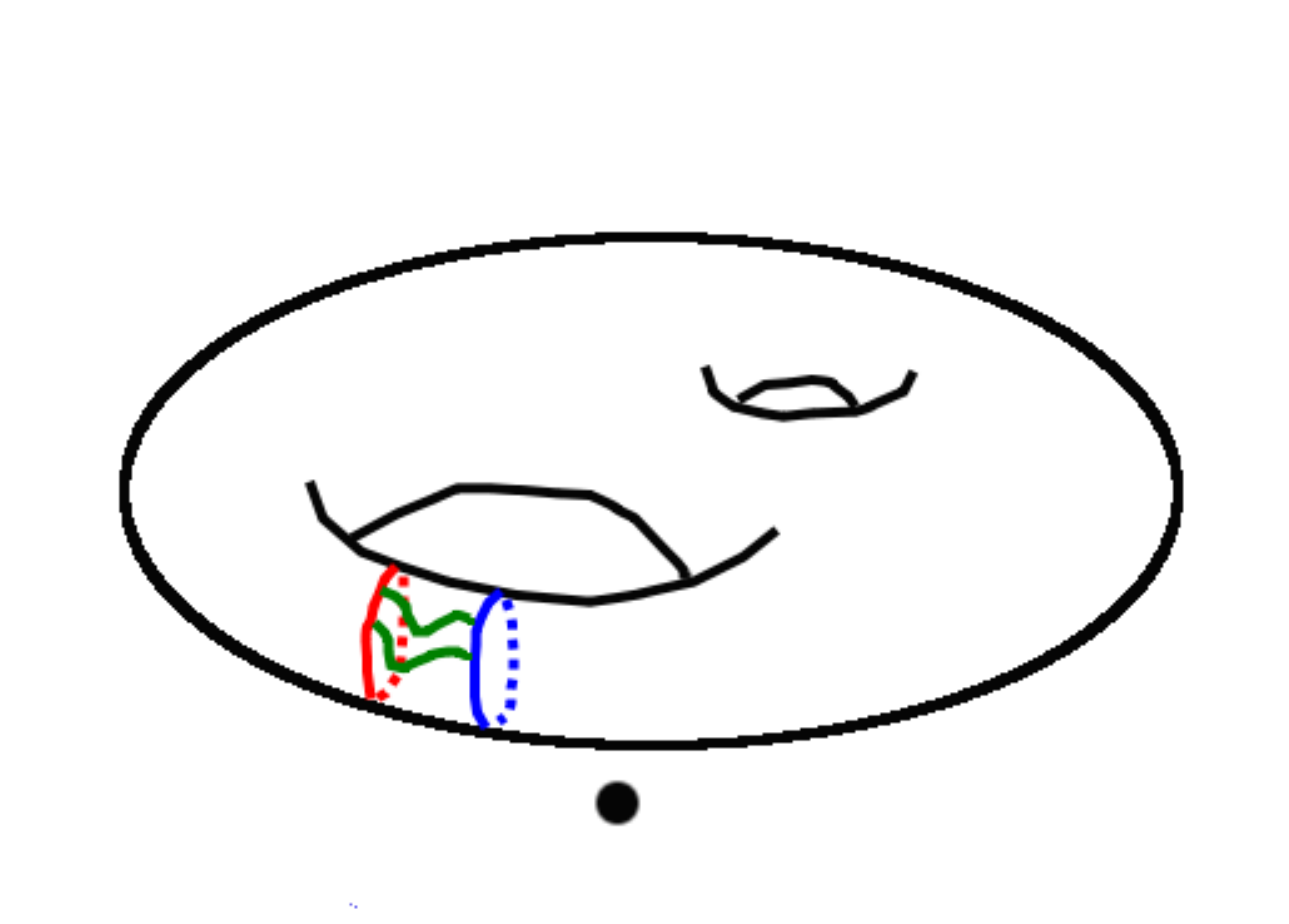}}
\put(-3.6,0.05) {\small{random BPS microstate}}
\put(4.5,0.05) {\small{polar state}}
\end{picture}
\caption{A D--particle's entropy is explained by the many possible ways to model BPS microstates by wrapping branes around cycles in compact dimensions. Among the many BPS microstates of a D--particle modeled as a mixed ensemble, a finite number of polar states (pictured on the right) determine the entire partition function. Polar states come as bound states: the two branes (red and blue) are glued together by tachyonic string modes (green).}
\label{dparticledegeneracies}
\end{figure}

For partition functions of D--particles (whether they have enough mass to backreact a black hole or not) modeled using a mixed ensemble of branes\footnote{In the original OSV paper, \cite{Ooguri:2004zv}, the authors argue, that the choice of a mixed ensemble is natural from the viewpoint of topological string theory (TST), as electric vs. magnetic for topological strings is like a choice of position vs. momentum. TST computes topological invariants enumerating constituent BPS states, thus there is also some motivation for the connection of TST with BPS bound states.}, the degeneracy of a finite number of BPS microstates, \emph{the polar states}, determines all other degeneracies through modularity. This is illustrated in figure \ref{dparticledegeneracies}.

Although there are of course many shortcomings in the analogy we are about to propose, there is a similarity between how the degeneracy of polar states dictates `the whole rest of possible microstates of a D--particle', and how knowledge of the genome on chromosomes in microbiology dictates --- we allow ourselves to simplify things considerably --- `all possible molecular states of an organism'. We thus like to think of polar states as a sort of `chromosomes for D--particles (and black holes)'. Additionally, just like chromosomes come in pairs for (nearly all) mammals (`diploid' organisms), polar states are realized as bound states, thus they come in pairs of `parent chromosomes'. This is illustrated in the following figure:
\begin{figure}[h]
\setlength{\unitlength}{1cm}
\centering 
\begin{picture}(5,4)
\put(0,0) {\includegraphics[width=0.4\textwidth,angle=0]{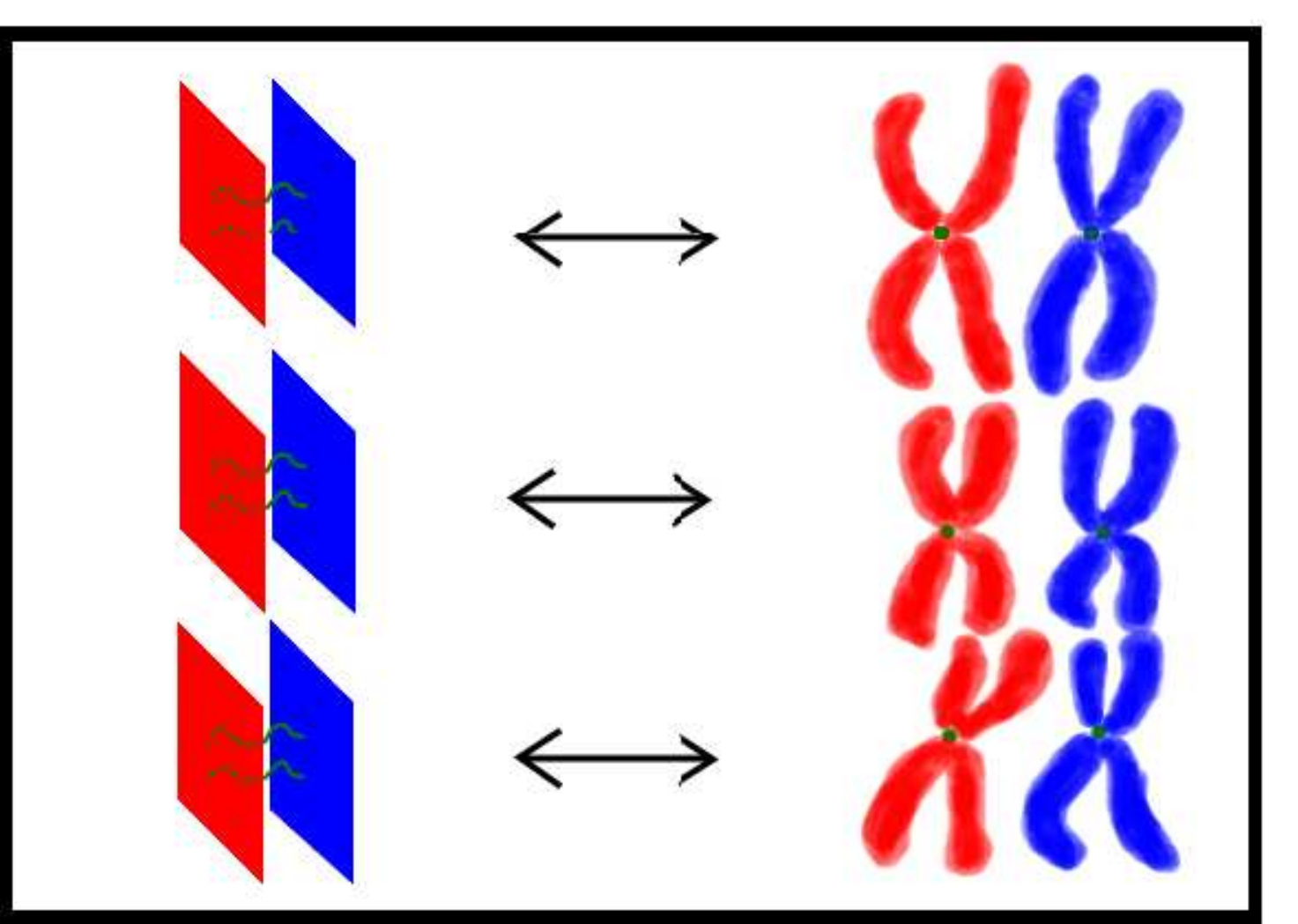}} 
\end{picture}
\caption{\textbf{Polar states as chromosomes}: The finite number of polar states determine the entire partition function of a D--particle/black hole, and they appear as pairs (held together by tachyonic string modes). Also chromosomes for diploid organisms come in pairs, and they encode the entire information of the `microstructure' of the organism.}
\end{figure}

We will come back to this metaphor, when we discuss our results in section \ref{sect:discussion}. Of course, these thoughts are nothing else than some intuitive interpretations of the modular properties and the information encoded by the poles of a weak Jacobi form. For now, this may serve as a motive to focus on a D--particle's polar states. We will however also investigate some of the `almost polar' states\footnote{By almost polar, we mean non--polar states, very close to the polarity--bound, thus carrying charges which are very close to charges of polar states.}, which will provide the most interesting checks on our methods.

\section{Elliptic genera from split flow trees and DT invariants}\label{sect:ellipticgenera}
This section is a short account of our results on all polar states for two Calabi--Yau (CY) varieties, which allows us to predict the corresponding elliptic genera. We start with a CY given as a sextic hypersurface.

\subsection{Polar states on the sextic hypersurface in $\wpr{4}{11112}$}
In $\wpr{4}{11112}$, the adjunction formula shows that one can obtain a CY hypersurface by choosing a degree six polynomial. The total Chern class reads $c(X)=\frac{(1+H)^4(1+2H)}{1+6H}=1+14H^2-68H^3$, and using $\int_X H^3=\int_{\wpr{4}{11112}}6H^4=3$, one gets $\chi (X)=-204$. As $\int_X H^3=3$, the weak Jacobi form is in this case three--dimensional:
\begin{equation}
Z(q, \bar q, z) = \sum_{k=0}^{2} Z_k(q)\, \Theta_k (\bar q, z)\,,
\end{equation}
which means that we only have to determine $Z_0$ and $Z_1$. By choosing a basis for $L_X^{\perp}$ one can easily see that there are two gluing vectors, but by the symmetry $\gamma =-\gamma$ one knows that one has just one `fundamental' gluing vector $\gamma_1$. According to our previous intuition, this means that in order to determine the complete elliptic genus, one will have to enumerate states in the classes $[0,\hat{q}_0]$ and $[\gamma_1,\hat{q}_0]$. For convenience, we list the DT invariants for the sextic of interest:
\begin{center}
\begin{tabular}{|c|c|c|c|c|}
    \hline
    \multicolumn{5}{|c|}{\textbf{Donaldson--Thomas invariants: sextic}}\\
    \hline
    & $\mathbf{n=0}$ & $\mathbf{n=1}$ & $\mathbf{n=2}$ & $\mathbf{n=3}$\\
    \hline
    $\mathbf{\beta=0}$ & 1 & 204 & 20'298 & 1'311'584\\
    \hline
    $\mathbf{\beta=1}$ & 0 & 7884 & 1'592'568 & 156'836'412\\
    \hline
    $\mathbf{\beta=2}$ & 7884 & 7'636'788 & 1'408'851'522 & 136'479'465'324\\
    \hline
    $\mathbf{\beta=3}$ & 169'502'712 & 443'151'185'260 & 5'487'789'706'776 & 440'554'251'409'968\\
    \hline
\end{tabular}
\end{center}
\begin{enumerate}
\item $\mathbf{\Delta q=0, \Delta q_0=0, \qquad [0,\frac{45}{24}]}$:\\
The pure D4--brane carries half a unit of flux to ensure anomaly cancellation, and has total charge $\Gamma = H+\frac{H^2}{2}+(\frac{\chi (P)}{24}+\frac{1}{2}F^2)\,\omega =H+\frac{1}{2}H^2+\frac{3}{4}H^3 =(0,1,\frac{3}{2},\frac{9}{4})$, where we have introduced the notation
\be
    \Gamma= (p^0,p,q,q_0),
\ee
which we will equally use from now on. As in \cite{Collinucci:2008ht}, we will also label charge systems by their deviation in D2--brane charge $\Delta q$ and D0--brane charge $\Delta q_0$ as measured from the most polar state. In the `charge shift' notation it is denoted as $\Delta q=0, \Delta q_0=0$. As explained in \cite{Denef:2007vg}, various charges are related by flux shifts. Charge equivalence classes contain the same entropy, and they can be labeled by the (flux) gluing vector (see appendix \ref{sect:ellGenera}), as well as the reduced D0--brane charge. A charge equivalence class can thus be labeled by $[\gamma ,\hat{q}_0]$, and the most polar state for the sextic lies in the class $[0,\frac{45}{24}]$. One finds a split flow tree with centers
\begin{eqnarray*}   
\Gamma_1&=&(1,1,\frac{13}{4},\frac{9}{4}),\\   \Gamma_2&=&(-1,0,-\frac{7}{4},0).
\end{eqnarray*}
It looks and is enumerated as follows:
\begin{figure}[h]
\setlength{\unitlength}{1cm}
\centering 
\begin{picture}(3,2.3)
\put(0,0) {\includegraphics[width=0.25\textwidth,angle=0]{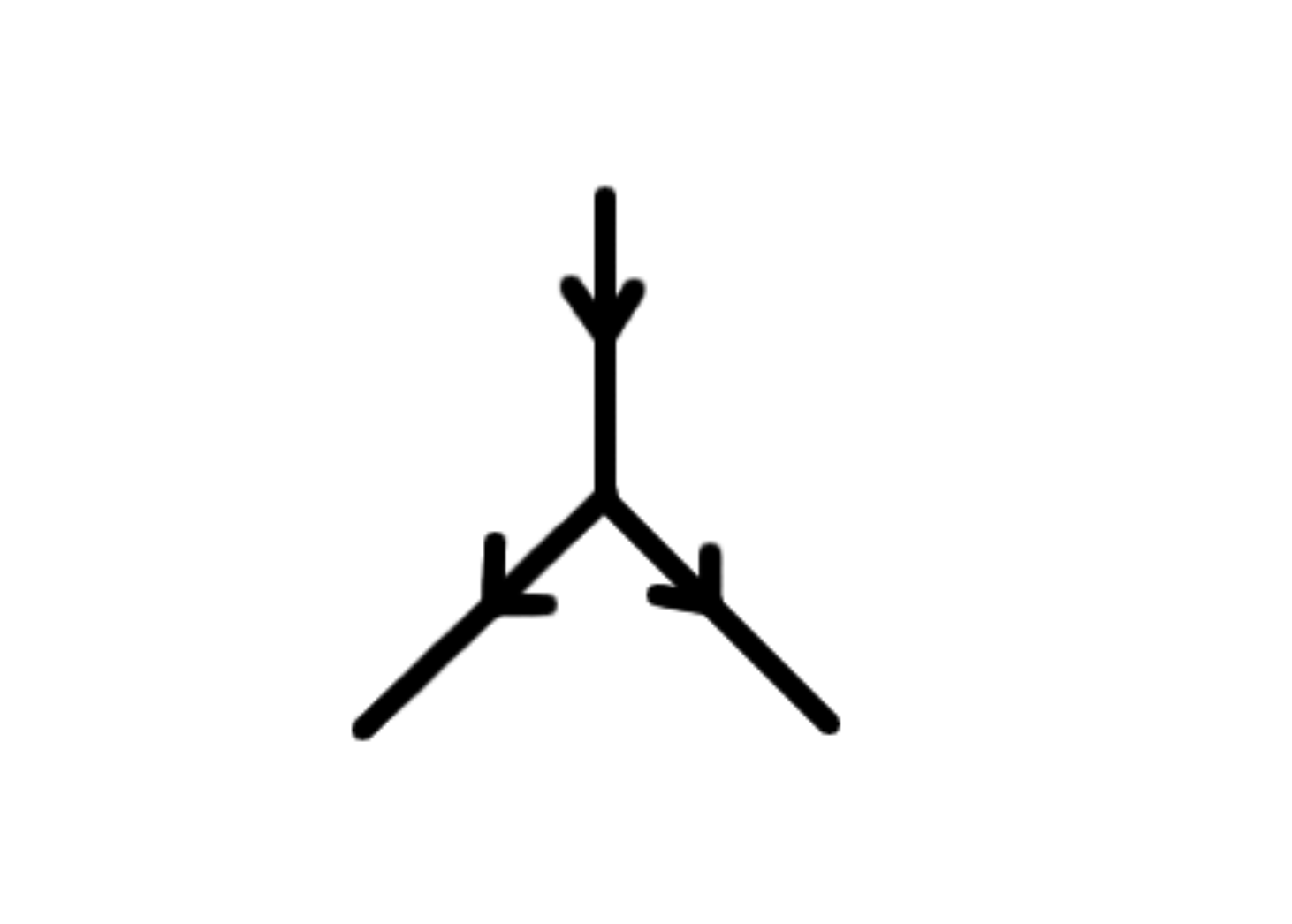}}
\put(1.5,2.25) {$D4_{\frac{H}{2}}$}
\put(0.6,0.15) {$D6_H$}
\put(2.25,0.15) {$\overline{D6}$}
\put(4,1.3) {$=$}
\put(5,1.3) {$-4$.}
\end{picture}
\end{figure}

The BPS index reads
\begin{equation}
  \Omega=(-1)^{|\langle\Gamma_1,\Gamma_2\rangle|-1}|\langle\Gamma_1,\Gamma_2\rangle|N_{\textrm{DT}}(0,0)\cdot N_{\textrm{DT}}(0,0)=(-1)^3\cdot 4\cdot 1\cdot 1=-4.
\end{equation}
Note that the fact that the intersection number between $\Gamma_1$ and $\Gamma_2$ nicely corresponds with the fact that the moduli space of the hyperplane $H\subset X$ is a $\cp 3$, hence $\chi (\cp 3)=4$, because the coordinate with weight $2$ can of course not be used to define a hyperplane.

\item $\mathbf{\Delta q=0, \Delta q_0=-1,\qquad [0,\frac{21}{24}]}$:\\
Adding one $\Dob$, one gets the total charge $(0,1,\frac{3}{2},\frac{5}{4})$, with reduced D0--brane charge $\hat{q}_0=\frac{21}{24}$. The flow tree is analogous to what was found for the quintic \cite{Collinucci:2008ht}(the side of the $\Dob$ after the first split can be chosen, according to where one is with respect to the appropriate threshold wall). The charges of the centers after the first split read
\bea
    \Gamma_1&=&(1,1,\frac{13}{4},\frac{9}{4}),\nonumber\\
    \Gamma_2&=&(-1,0,-\frac{7}{4},-1),\nonumber
\eea
and the flow tree looks like

\begin{figure}[h]
\setlength{\unitlength}{1cm}
\centering 
\begin{picture}(3,2.3)
\put(0,0) {\includegraphics[width=0.25\textwidth,angle=0]{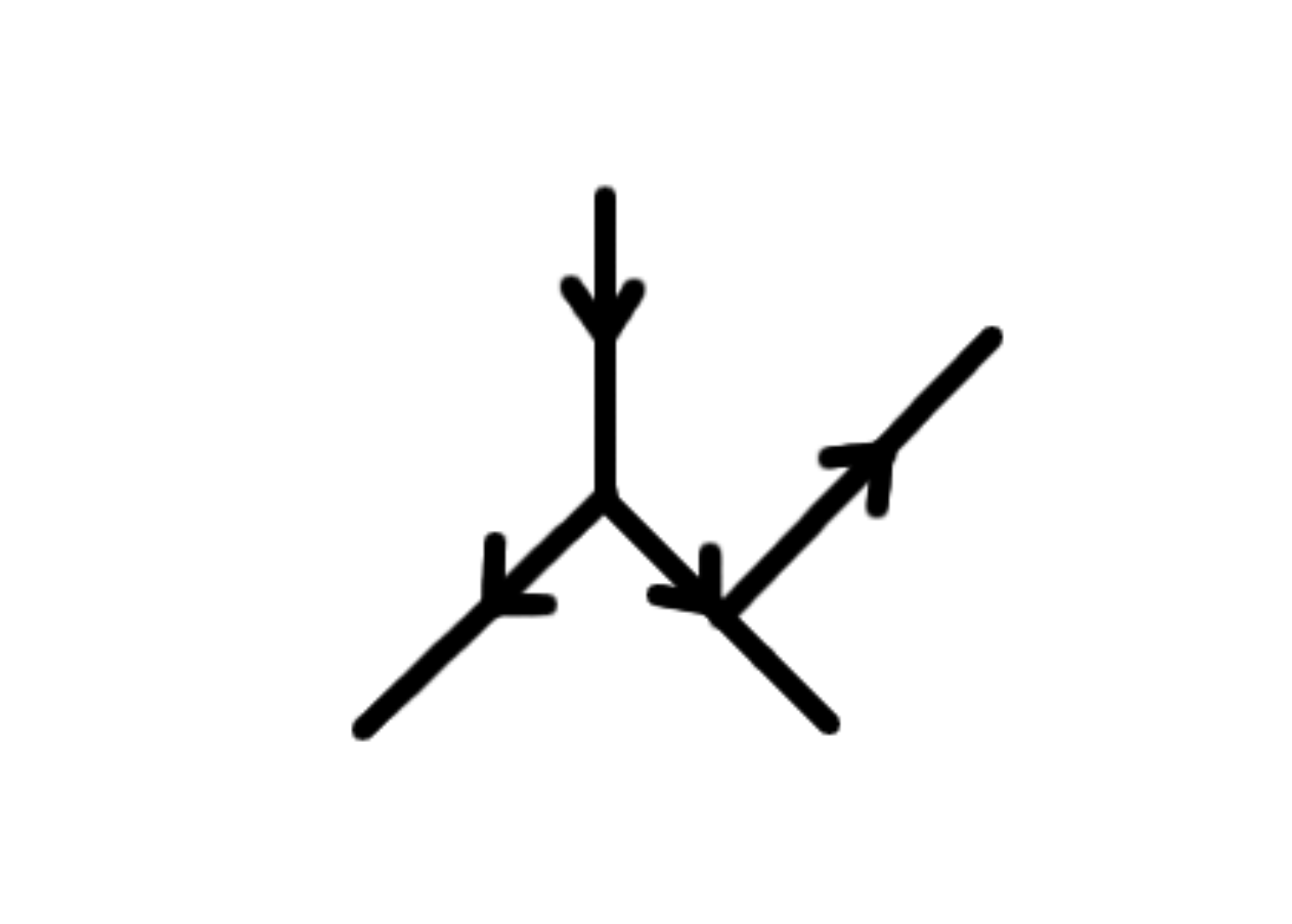}}
\put(1,2.35) {$D4_{\frac{H}{2},\overline{D0}}$}
\put(0.6,0.15) {$D6_H$}
\put(2.25,0.15) {$\overline{D6}$}
\put(2.65,1.8) {$\overline{D0}$}
\put(4,1.3) {$=$}
\put(5,1.3) {$612$,}
\end{picture}
\end{figure}

where of course
\begin{equation}
  \Omega=(-1)^{|\langle\Gamma_1,\Gamma_2\rangle|-1}|\langle\Gamma_1,\Gamma_2\rangle|N_{\textrm{DT}}(0,0)\cdot N_{\textrm{DT}}(0,1)=(-1)^2\cdot 3\cdot 1\cdot 204=612.
\end{equation}
\item $\mathbf{\Delta q=1, \Delta q_0=-1,\qquad [\gamma_1,\frac{5}{24}]}$:\\
One can now consider a flux, which will involve what we called the relevant gluing vector $\gamma_1$. According to our intuition, which receives further support at this point, this means turning on an extra flux dual to a degree one rational curve. This leads to the total charge $(0,1,\frac{5}{2},\frac{5}{4})$, and to the reduced D0--brane charge $\hat{q}_0=\frac{5}{24}$: thus, there is only one polar state in this $\gamma_1$--class. One finds the split flow tree with a pure D6 one one side, and a $\Dsb$ with a D2 on a degree one rational curve, as expected. The charges read
\bea
    \Gamma_1&=&(1,1,\frac{13}{4},\frac{9}{4}),\nonumber\\
    \Gamma_2&=&(-1,0,-\frac{3}{4},-1),\nonumber
\eea
with the split flow tree
\begin{figure}[h]
\setlength{\unitlength}{1cm}
\centering 
\begin{picture}(3,2.3)
\put(0,0) {\includegraphics[width=0.25\textwidth,angle=0]{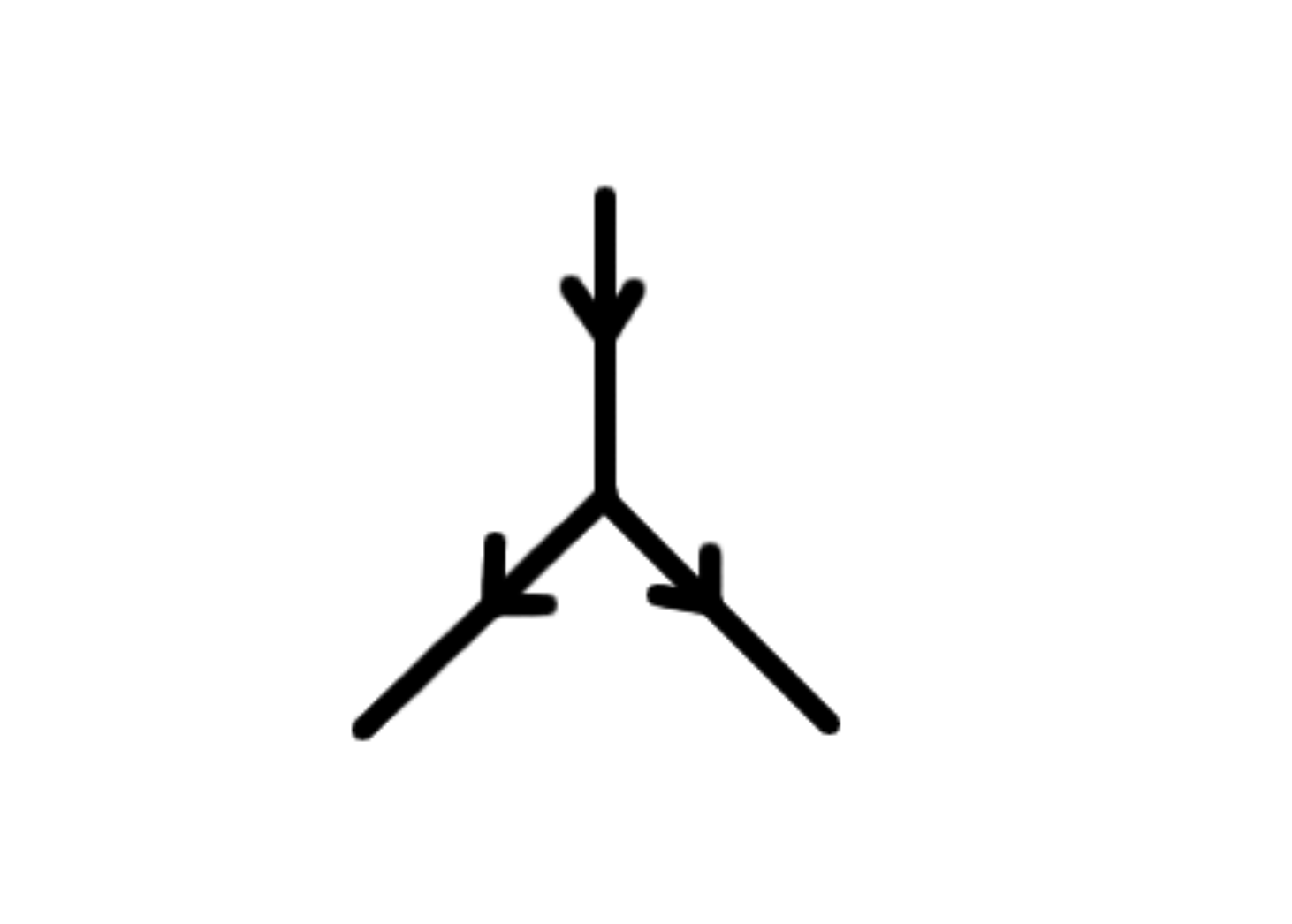}}
\put(1.5,2.25) {$D4_{\frac{H}{2}+F(C_1{}^{g=0})}$}
\put(0.6,0.15) {$D6_H$}
\put(2.25,0.15) {$\overline{D6}_{D2(C_1{}^{g=0})}$}
\put(4,1.3) {$=$}
\put(5,1.3) {$=-15'768$.}
\end{picture}
\end{figure}

The BPS index is calculated according to
\begin{equation}
  \Omega=(-1)^{|\langle\Gamma_1,\Gamma_2\rangle|-1}|\langle\Gamma_1,\Gamma_2\rangle|N_{\textrm{DT}}(0,0)\cdot N_{\textrm{DT}}(1,1)=(-1)^1\cdot 2\cdot 1\cdot 7'884=-15'768.
\end{equation}

\end{enumerate}
Using a basis for modular forms of the right weight, one can use these numbers to determine the modular form to be given by
\begin{align}
  Z_0(q)&=&q^{-\frac{45}{24}}(-4+612q-40'392q^2+146'464'860q^3...) \label{eq:z0sextic} \\
  Z_1(q)=Z_2(q)&=&q^{-\frac{29}{24}}(-15'768q+7'621'020q^2+...)), \label{eq:z1sextic}
\end{align}
whose uniqueness follows from \cite{Denef:2007vg, deBoer:2006vg, Manschot:2007ha}.
This agrees with the findings of \cite{Gaiotto:2007cd} (up to an overall sign), which is of course not a surprise, given that the small number of polar states supporting split flow tree realizations apparently do not involve subtleties.

\subsection{Polar states on the octic hypersurface in $\wpr{4}{11114}$}
For $\wpr{4}{11114}$, the adjunction formula shows that one can obtain a CY hypersurface by choosing a degree eight polynomial. The total Chern class reads $c(X)=\frac{(1+H)^4(1+4H)}{1+8H}=1+22H^2-148H^3$, and using $\int_X H^3=\int_{\wpr{4}{11114}}8H^4=2$, one gets $\chi (X)=-296$. As $\int_X H^3=2$, the weak Jacobi form is in this case two--dimensional:
\begin{equation}
Z(q, \bar q, z) = \sum_{k=0}^{1} Z_k(q)\, \Theta_k (\bar q, z)\,,
\end{equation}
which means that we have to determine $Z_0$ and $Z_1$. By choosing a basis for $L_X^{\perp}$ one can easily see that there is only one gluing vector, $\gamma_1$. According to our previous intutition, this means that in order to determine the complete elliptic genus, one will have to enumerate states in the classes $[0,\hat{q}_0]$ and $[\gamma_1,\hat{q}_0]$. For convenience, we also list the DT invariants of interest, for the octic:
\begin{center}
\begin{tabular}{|c|c|c|c|c|}
    \hline
    \multicolumn{5}{|c|}{\textbf{Donaldson--Thomas invariants: octic}}\\
    \hline
    & $\mathbf{n=0}$ & $\mathbf{n=1}$ & $\mathbf{n=2}$ & $\mathbf{n=3}$\\
    \hline
    $\mathbf{\beta=0}$ & 1 & 296 & 43'068 & 4'104'336\\
    \hline
    $\mathbf{\beta=1}$ & 0 & 29'504 & 8'674'176 & 1'253'300'416\\
    \hline
    $\mathbf{\beta=2}$ & 564'332 & 204'456'696 & 45'540'821'914 & 6'127'608'486'208\\
    \hline
    $\mathbf{\beta=3}$ & 8'775'447'296 & 6'313'618'655'104 & 1'225'699'503'521'536 & 141'978'726'005'461'504\\
    \hline
\end{tabular}
\end{center}

We will be quite brief on the summary of results on polar states, given the analogy to the previously discussed cases.
\begin{enumerate}
\item $\mathbf{\Delta q=0, \Delta q_0=0, \qquad [0,\frac{23}{12}]}$:\\
The most polar state is the D4--brane carrying flux $\frac{H}{2}$ for anomaly cancellation, with total charge $(0,1,1,\frac{13}{6})$. The reduced D0--brane charge can be calulcated to be $\hat{q}_0=\frac{23}{12}$, and the state is of course in the class $[0,\frac{23}{12}]$. One finds a split flow tree with centers
\bea   
\Gamma_1&=&(1,1,\frac{17}{6},\frac{13}{6}),\nonumber \\   \Gamma_2&=&(-1,0,-\frac{11}{6},0),\nonumber
\eea
with split flow tree

\begin{figure}[h]
\setlength{\unitlength}{1cm}
\centering 
\begin{picture}(3,2.3)
\put(0,0) {\includegraphics[width=0.25\textwidth,angle=0]{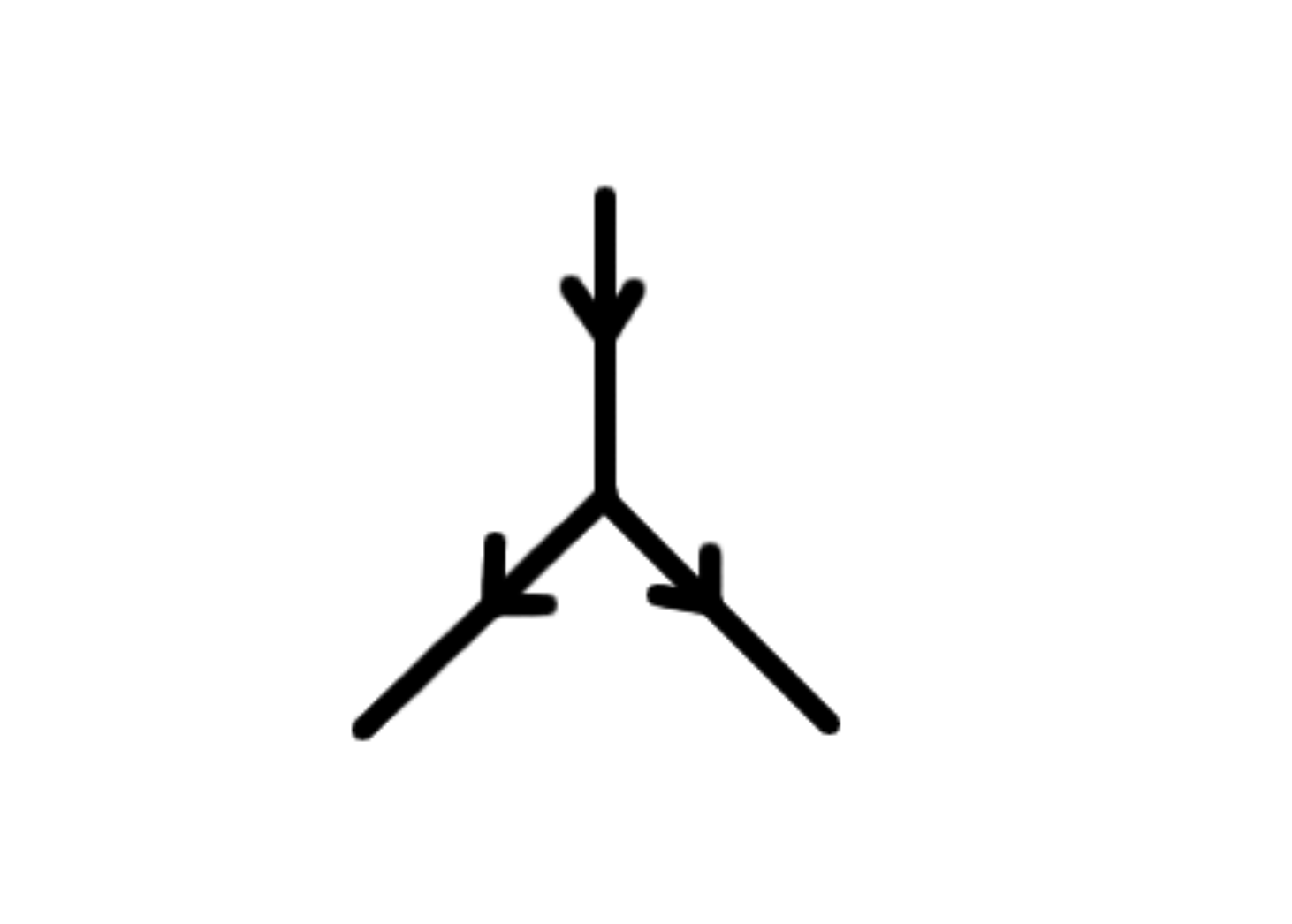}}
\put(1.5,2.25) {$D4_{\frac{H}{2}}$}
\put(0.6,0.15) {$D6_H$}
\put(2.25,0.15) {$\overline{D6}$}
\put(4,1.3) {$=$}
\put(5,1.3) {$-4$.}
\end{picture}
\end{figure}

where the BPS index is calculated as
\begin{equation}
  \Omega=(-1)^{|\langle\Gamma_1,\Gamma_2\rangle|-1}|\langle\Gamma_1,\Gamma_2\rangle|N_{\textrm{DT}}(0,0)\cdot N_{\textrm{DT}}(0,0)=(-1)^3\cdot 4\cdot 1\cdot 1=-4.
\end{equation}
Note again the fact that the intersection number between $\Gamma_1$ and $\Gamma_2$ equals $-4$ corresponds with the fact that one cannot use one of the five coordinates to define a hyperplane, in the D4--picture.
\item $\mathbf{\Delta q=0, \Delta q_0=-1,\qquad [0,\frac{11}{12}]}$:\\
Adding one $\Dob$, one gets the total charge $(0,1,1,\frac{7}{6})$, with reduced D0--brane charge $\hat{q}_0=\frac{11}{12}$. The flow tree is again analogous to what we found for the sextic (the side of the $\Dob$ after the first split can be chosen, according to where one is with respect to the appropriate threshold wall). The charges of the centers after the first split read
\bea
    \Gamma_1&=&(1,1,\frac{17}{6},\frac{13}{6}),\nonumber\\
    \Gamma_2&=&(-1,0,-\frac{11}{6},-1),\nonumber
\eea
and the flow tree looks like

\begin{figure}[h]
\setlength{\unitlength}{1cm}
\centering 
\begin{picture}(3,2.3)
\put(0,0) {\includegraphics[width=0.25\textwidth,angle=0]{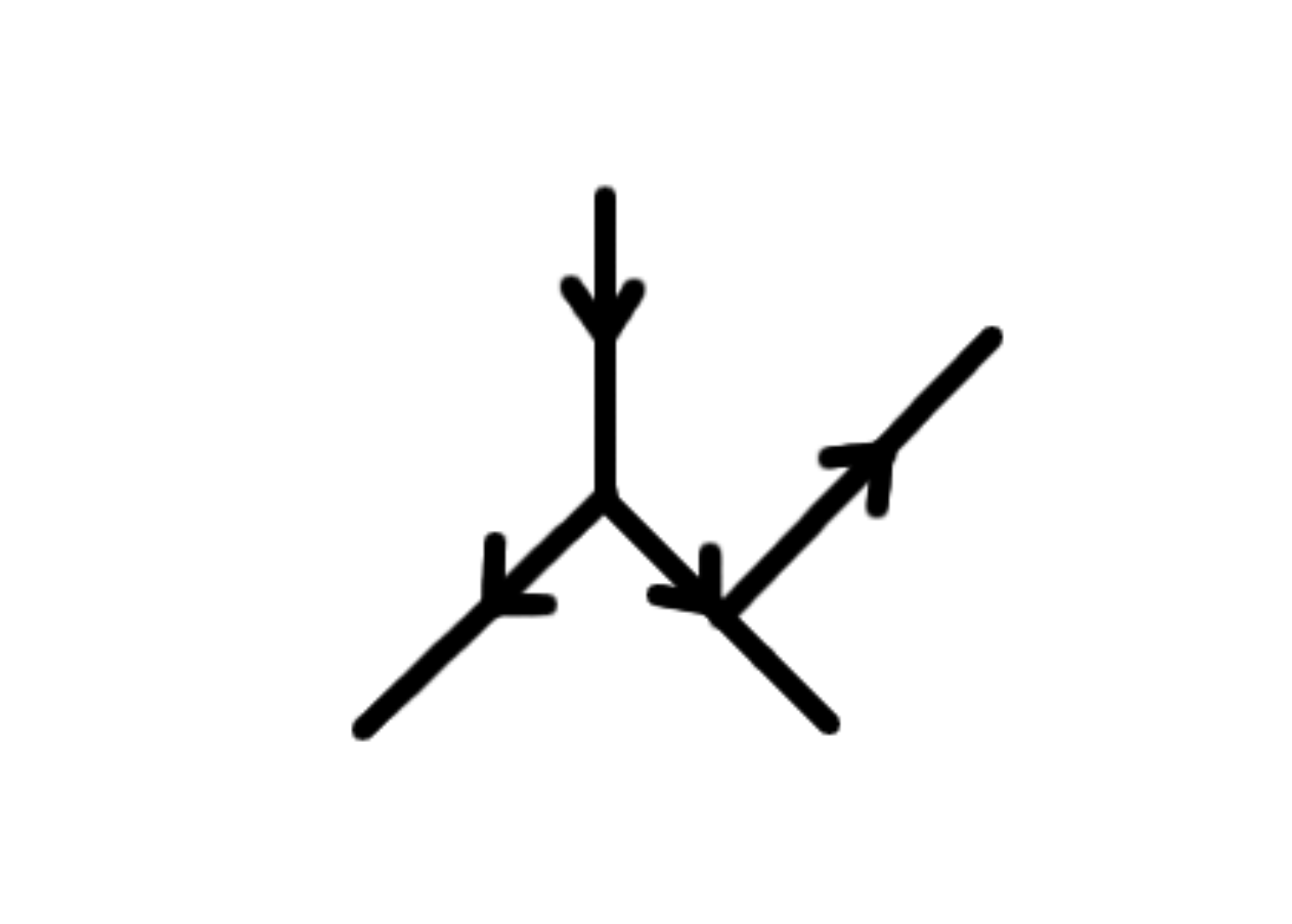}}
\put(1,2.35) {$D4_{\frac{H}{2},\overline{D0}}$}
\put(0.6,0.15) {$D6_H$}
\put(2.25,0.15) {$\overline{D6}$}
\put(2.65,1.8) {$\overline{D0}$}
\put(4,1.3) {$=$}
\put(5,1.3) {$888$,}
\end{picture}
\end{figure}

where of course
\begin{equation}
  \Omega=(-1)^{|\langle\Gamma_1,\Gamma_2\rangle|-1}|\langle\Gamma_1,\Gamma_2\rangle|N_{\textrm{DT}}(0,0)\cdot N_{\textrm{DT}}(0,1)=(-1)^2\cdot 3\cdot 1\cdot 296=888.
\end{equation}
\item $\mathbf{\Delta q=1, \Delta q_0=-1,\qquad [\gamma_1,\frac{1}{6}]}$:\\
One can now consider a flux, involving the relevant gluing vector $\gamma_1$. As previously seen, this means turning on an extra flux dual to a degree one rational curve. This leads to the total charge $(0,1,2,\frac{7}{6})$, and to the reduced D0--brane charge $\hat{q}_0=\frac{1}{6}$: thus, there is again only one polar state in this $\gamma_1$--class. One finds the split flow tree with a pure D6 one one side, and a $\Dsb$ with a D2 on a degree one rational curve, as expected. The charges read
\bea
    \Gamma_1&=&(1,1,\frac{17}{6},\frac{13}{6}),\nonumber\\
    \Gamma_2&=&(-1,0,-\frac{5}{6},-1),\nonumber
\eea
with the split flow tree
\begin{figure}[h]
\setlength{\unitlength}{1cm}
\centering 
\begin{picture}(3,2.3)
\put(0,0) {\includegraphics[width=0.25\textwidth,angle=0]{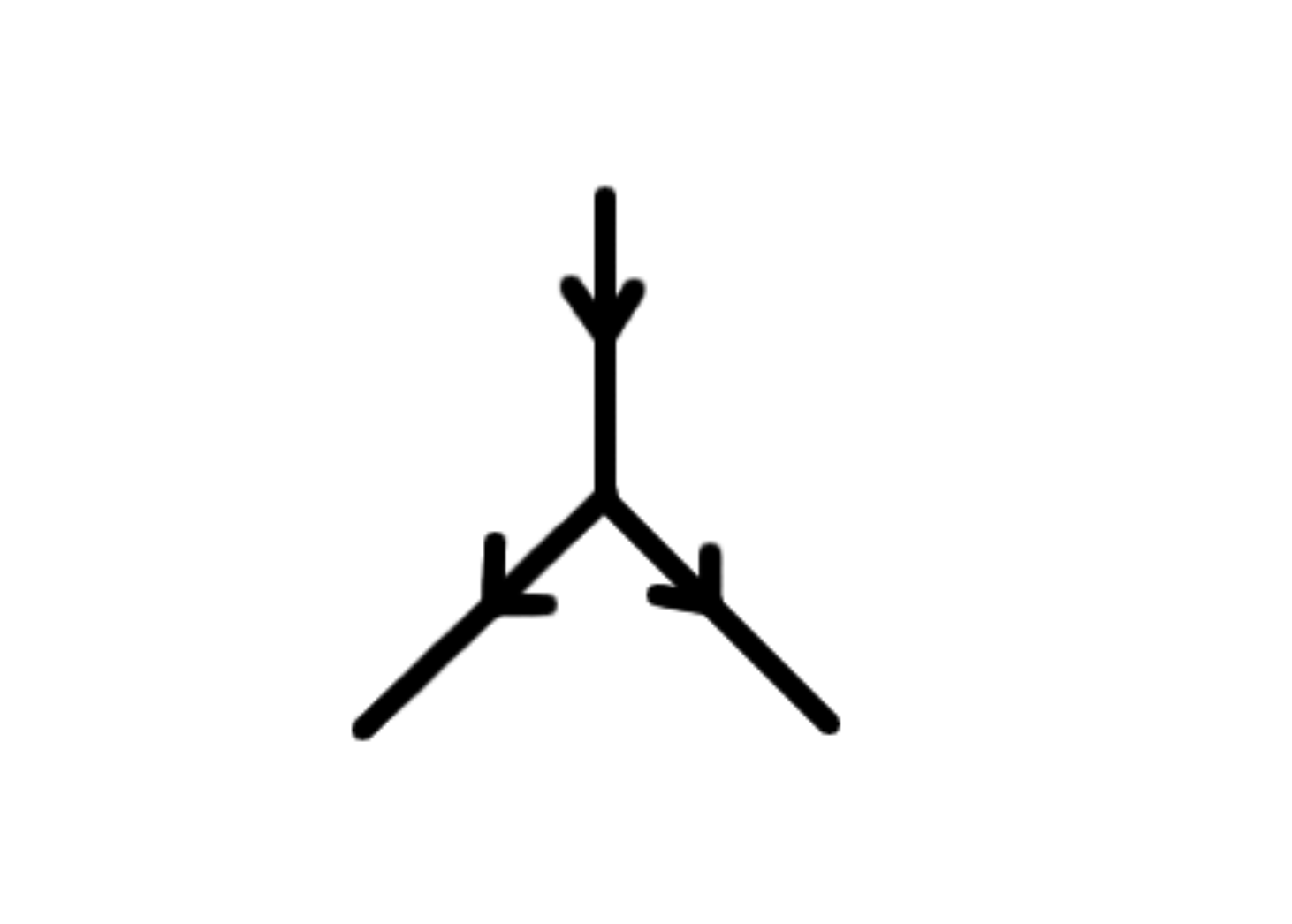}}
\put(1.5,2.25) {$D4_{\frac{H}{2}+F(C_1{}^{g=0})}$}
\put(0.6,0.15) {$D6_H$}
\put(2.25,0.15) {$\overline{D6}_{D2(C_1{}^{g=0})}$}
\put(4,1.3) {$=$}
\put(5,1.3) {$-59'008$.}
\end{picture}
\end{figure}

The BPS index is calculated according to
\begin{equation}
  \Omega=(-1)^{|\langle\Gamma_1,\Gamma_2\rangle|-1}|\langle\Gamma_1,\Gamma_2\rangle|N_{\textrm{DT}}(0,0)\cdot N_{\textrm{DT}}(1,1)=(-1)^1\cdot 2\cdot 1\cdot 29'504=-59'008.
\end{equation}
\end{enumerate}
Using a basis for modular forms of the right weight, one can use these numbers to determine the modular form to be given by
\begin{align}\label{octicellipticgenus}
  Z_0(\tau )&=&q^{-\frac{23}{12}}(-4+888q-86'140q^2+131'940'136q^3...) \\
  Z_1(\tau )=Z_2(\tau)&=&q^{-\frac{7}{6}}(-59'008q+8'615'168q^2+...)
\end{align}
This again agrees with the findings of \cite{Gaiotto:2007cd} (up to an overall sign).

\section{Refined enumeration of bound states and elliptic genera revisited}\label{sect:DTpartitions}
Using the BPS index for bound states explained in part \ref{subsect:dtinvariants}, we were able to predict the elliptic genera for two study models, and we found exact agreement with the authors of \cite{Gaiotto:2007cd}. In this section, we will discuss the non--polar states that demand the use of a refined prescription for the computation of the BPS index for bound states. In \cite{Collinucci:2008ht}, this was demonstrated for a non--polar BPS state on the quintic 3--fold: the refined computation yielded an exact match with the result predicted from modularity. In our case, the reason for the necessity for refined calculations is different, but we again find exact agreement with the result predicted by modularity, for all tractable examples. We will show that refinements can also alter the enumeration of polar states, in the next section.

We start with a subsection explaining the main reason why refinement is needed, followed by a subsection explaining the more specific techniques used in our computations. We then move on to the various explicit examples, and we will also show how our refined computations allow us to introduce partitions of the Donaldson--Thomas invariants used to enumerate the BPS states belonging to a D6--D2--D0 constituent of a bound state.

\subsection{The main idea: non--trivially fibered moduli spaces of bound states}
In the microscopic D--brane picture, a bound state of a D6--brane system carrying a flux $F_1$ and bound to $N$ $\Dob$'s, and a $\Dsb$ carrying a flux $F_2$ is held together by tachyonic string modes. For the pure D6--$\Dsb$ bound state, the tachyon field $T$ (\cite{Collinucci:2008pf}) can be understood as a map $T:F_1\rightarrow F_2$, where $F_{1,2}$ are the line bundles corresponding the fluxes. Alternatively, the tachyon can be understood as a section of
\be
   T\in \Gamma (F_2^{*}\otimes F_1),
\ee
and the Riemann--Roch theorem allows us to compute an index $\mathcal{I}_{\textrm{T}}$, counting the number of basis elements of this space of sections, $ \Gamma (F_2^{*}\otimes F_1)$:
\be\label{tachyonindex}
 \mathcal{I}_{\textrm{T}}=\int_X\textrm{ch}(F_2^{*})\textrm{ch}(F_1)Td(X),
\ee
where $\textrm{ch}(F)$ denotes the Chern class of $F$, and $\textrm{Td}(X)$ denotes the Todd class for the base space (in this case the CY). If one adds a $\Dob$--brane, the tachyon field needs to vanish at an additional point. The map is then a section of $ \Gamma (F_2^{*}\otimes F_1\times\mathcal{I}_{\textrm{p}})$, where $\mathcal{I}_{\textrm{p}}$ denotes the ideal sheaf on the point where the $\Dob$ resides. When $N$ $\Dob$'s are bound to the D6--brane, the tachyon field $T$ needs to vanish also at these points, which means that one has to impose $N$ constraints on $T$: the number of independent sections of the tachyon fields is reduced by $N$ (as opposed to the case where one studies a D6--$\Dsb$--system without any extra bound $\Dob$'s). In this way an index is computed, accounting for the degrees of freedom associated with the tachyon moduli space $\mathcal{M}_{\textrm{T}}$.

The index presented in subsection \ref{subsect:dtinvariants} is based on the assumption, that the moduli space of a bound state factorizes into the tachyon moduli space, the moduli space of the D6--brane, and the moduli space of the $\Dsb$--brane:
\be
   \mathcal{M}=\mathcal{M}_{\textrm{T}}\times\mathcal{M}_{D6}\times\mathcal{M}_{\Dsb}.
\ee
To compute an index one naively uses, \cite{Denef:2007vg} (compare also with subsection \ref{subsect:dtinvariants}),
\be \begin{array}{ccc}
   \underbrace{(-1)^{|\langle\Gamma_1,\Gamma_2\rangle |}|\langle\Gamma_1,\Gamma_2\rangle |}\quad\cdot & \quad\underbrace{N_{\textrm{DT}}(\beta_1,n_1)}\quad\cdot & \quad\underbrace{N_{\textrm{DT}}(\beta_2,n_2)},\\
   \mathcal{M}_\textrm{T}& \mathcal{M}_{D6} & \mathcal{M}_{\Dsb}
\end{array}\nonumber\ee
where $(-1)^{|\langle\Gamma_1,\Gamma_2\rangle |}|\langle\Gamma_1,\Gamma_2\rangle |$ and $\mathcal{I}_{\textrm{T}}$ agree up to a sign.

In general, this moduli space is non--trivially fibered \cite{Collinucci:2008ht}, and thus the dimensionality of the fiber can perform jumps. As a consequence, the index \ref{tachyonindex} is not always accurate: it computes a virtual dimension \cite{Vonk:2005yv}, which does not always equal the real dimension of the moduli space. Basically the reason is that the tachyon fields do not perceive all the constituent states generically. For simplicity, we will restrict the discussion in the following to the case that only one of the two constituents is not generically perceived, namely the $\Dsb$--consistuent. We will also focus on the case where $\Dob$--branes are not perceived generically by the tachyon fields, although we will show for an example, that an analogous phenomenon holds for special D2--states (and the curves on which these are wrapped). Generalizations of the presented scheme will become clear from our examples.

We show that the moduli space of a BPS bound state splits up for our examples, and can be grouped into two pieces:
\be
   \mathcal{M}=\mathcal{M}_{\textrm{T}_g}\times\mathcal{M}_{D6}\times\mathcal{M}_{\Dsb_g}\qquad\oplus \qquad \mathcal{M}_{\textrm{T}_s}\times\mathcal{M}_{D6}\times\mathcal{M}_{\Dsb_s}.
\ee
The first part, with the superscripts `g', stands for the part where the tachyon fields perceive the constituent states `generically', and hence the virtual dimension of $\mathcal{M}_{\textrm{T}}$ is actually the real dimension, and the part with the superscripts `s' stands for the part where the tachyon fields perceive the constituent states as `special states' (in the present case, always for the $\Dsb$--system). This usually happens because the tachyon fields do not perceive all $\Dob$'s that are bound to the $\Dsb$, which results in a \emph{constraint loss} on the tachyon field $T$. For example, generically three $\Dob$'s mean that the independent sections have to vanish at three points. If the tachyon is blind to one of the three, this number is reduced to two, resulting in a jump in the dimension of the fiber of the moduli space of the bound state. Typically, the dimension of $\mathcal{M}_{\textrm{T}_s}$ will thus be greater by one, as opposed to $\mathcal{M}_{\textrm{T}_g}$.

In \cite{Collinucci:2008ht}, it was proposed to define Donaldson--Thomas densities (in analogy to a top Chern class), that integrate over moduli space to a Donaldson--Thomas invariant, in order to be able to define a product formula for the bound state index on the level of index densities. One can also decide to integrate these densities on the various partitions of moduli space that have a constant dimension of the fiber of the tachyon moduli space. We call these integrated densities \emph{Donaldson--Thomas partitions} (DT partitions). They separately enumerate generic and special D6--D2--D0 states, and we will calculate a series of examples of such DT partitions. Using these, one can write down how our refined indices come about:
\newpage
\be \begin{array}{ccc}
   \underbrace{(-1)^{|\langle\Gamma_1,\Gamma_2\rangle |}|\langle\Gamma_1,\Gamma_2\rangle |}\quad\cdot & \quad\underbrace{N_{\textrm{DT}}(\beta_1,n_1)}\quad\cdot & \quad\underbrace{\mathcal{N}^{(g)}_{\textrm{DT}}(\beta_2,n_2)}\\
   \mathcal{M}_{\textrm{T}_g} & \mathcal{M}_{D6} & \mathcal{M}_{\Dsb_g}
\end{array}\nonumber\ee
\be\begin{array}{ccc}
+\qquad \underbrace{(-1)^{|\langle\Gamma_1,\Gamma_2\rangle |}\left( |\langle\Gamma_1,\Gamma_2\rangle |+1\right) } \quad\cdot & \quad\underbrace{N_{\textrm{DT}}(\beta_1,n_1)}\quad\cdot & \quad\underbrace{\mathcal{N}^{(s)}_{\textrm{DT}}(\beta_2,n_2)}.\\  \mathcal{M}_{\textrm{T}_s} & \mathcal{M}_{D6} & \mathcal{M}_{\Dsb_s}
\end{array}\nonumber
\ee
Before giving our explicit examples, we will now give a more detailed instruction on how to separate the `generic' from the `special' states, resulting in a calculation of Donaldson--Thomas partitions.

\subsection{Algebraic techniques to deal with special constituent states}
In \cite{Collinucci:2008ht},  $D4-3\Dob$ states had to be treated more carefully, because the tachyon fields of a D6--($\Dsb$--3$\Dob$)--bound states do not perceive three \emph{collinear} $\Dob$'s generically. Rather, they appear only as two particles. In terms of algebraic geometry, this is simple to express: the three constraints on the tachyon field are not independent (only two of them are independent). For our examples, this \emph{constraint loss} occurs for two reasons, which we shall refer to as `special loci', and as `special tangent directions' (which are important when blowups are performed, required for dealing with coincident loci for two $\Dob$'s). The special loci occur because the Calabi--Yau varieties are embedded in weighted projective spaces, which means that there is not a complete democracy amongst coordinates: If a $\Dob$ sits at a position with non--zero coordinates of a higher weight only, it will not impose constraints on the tachyon, as the higher weight coordinates cannot be included in the definition of the tachyon map. Let us elaborate on this in more detail.
\begin{enumerate}
\item
\textbf{Constraint loss because of special loci}\\
For simplicity, consider a Calabi--Yau embedded in $\wpr{4}{1111n}$ with $n>1$, coordinates $(x_1,...,x_5)$ and a bound state of a D6--brane with flux $F=H$ (one unit of flux) and a $\Dsb$ without flux. In this case, $H$ is a line bundle of which the coordinates $x_i$ form sections. However, $x_5$ is forbidden as a section, as it has a higher weight. This means that there are only four instead of five independent sections for the tachyon $T\in\Gamma (H)$ ($F_2^{*}$ is the trivial bundle in this case). The most general tachyon field reads $T=a_1x_1+...+a_4x_4$. In general, placing a $\Dob$ on the $\Dsb$ means imposing one constraint on the tachyon field. A good way to think about this, is by treating the $\Dob$ on the $\Dsb$ with point particle quantum mechanics, (check \cite{Witten:1982im}). If one puts the particle at $x_5=1$ (and all other coordinates zero), it will not impose a constraint on the tachyon field. This might be of relevance and it might not: one still has to check whether this point actually lies on the Calabi--Yau. We will encounter cases where situations analogous to this fictitious example arise.
\item \textbf{Constraint loss because of special tangent directions}\\
This was not the whole story, though. An additional complication arises as soon as one considers two (or more) $\Dob$--particles: orbifold singularities arise when particles meet. This is dealt with by performing a blowup, which imposes a distinction between the particles, that inhabit the same spot on the brane (and hence on the CY). One can intuitively picture this as considering an infinitesimally short time period before they meet, and distinguishing the particles upon all the different tangent directions (which of course have to be tangent directions to the Calabi--Yau variety under consideration), from which the two particles can approach each other. Two particles lying on the same spot in the Calabi--Yau would originally only impose one constraint on the tachyon field, according to the previous discussion. After performing a blowup, an additional constraint arises, from the splitting through the tangent direction. This means, that the tachyon moduli space fiber does not jump (at least generically) for states, where particles lie at the same locus. The point is: it \emph{can} jump. Namely, one (or several) tangent directions might be built from coordinates that do not impose constraints. Thus, also after performing blowups, one still has to distinguish between generic and special states of particles, that lie at coincident loci.
\end{enumerate}
Let us summarize the two situations, when special states occur (in this case for $\Dob's)$, once more:
\bi
  \item For non--coincident $\Dob$'s, one needs to check whether the particles lie at `special loci', where they do not impose a constraint on the tachyon fields. One could refer to these cases as the `special non--blowup loci'.
    \item For a bound state, there are cases, when various $\Dob$'s are coincident: in this case one needs to perform blowups. These blown--up states have to be separated into generic and special states, according to whether the tangent direction (arising from the blowup) imposes a constraint on the tachyon fields or not. One might refer to these latter cases as the `special blowup loci'.

\ei
As will become clear from an example later on, analogous implications arise for bound D2--branes, depending on whether the curves (on which those branes are placed) impose `enough' constraints on the tachyon fields (to make the bound state `generic').

\subsection{Special points and DT partitions $\mathcal{N}^{(g,s)}_{\textrm{DT}}(0,2)$}
In the following, a few interesting non--polar states on the sextic and the octic CY's for which the elliptic genera were predicted in the previous section, are examined. The refined caculations match the predictions from modularity.\\
\\
\textbf{The state $\mathbf{\Delta q=0, \Delta q_0=-2}$ on the sextic}\\
\\
For this charge system, one finds a split flow tree with centers
\bea
    \Gamma_1&=&(1,1,\frac{13}{4},\frac{9}{4}),\nonumber\\
    \Gamma_2&=&(-1,0,-\frac{7}{4},-2),\nonumber
\eea
\newpage
of the form

\begin{figure}[h]
\setlength{\unitlength}{1cm}
\centering 
\begin{picture}(3,2.3)
\put(0,0) {\includegraphics[width=0.25\textwidth,angle=0]{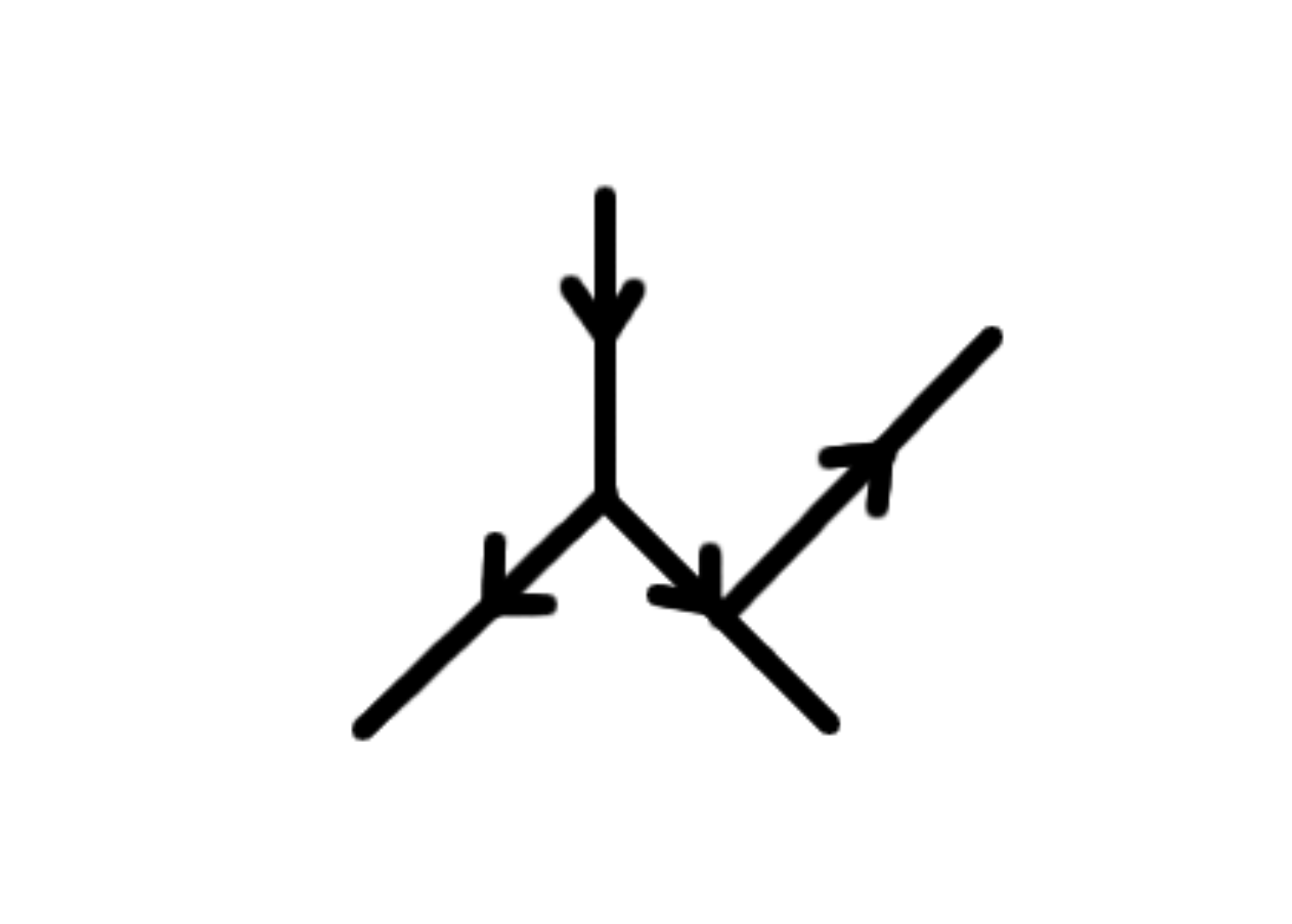}}
\put(1,2.35) {$D4_{\frac{H}{2},\overline{2D0}}$}
\put(0.6,0.15) {$D6_H$}
\put(2.25,0.15) {$\overline{D6}$}
\put(2.65,1.8) {$\overline{2D0}$}
\put(5,1.3) {$=-2\cdot 20'298=-40'596$,}
\end{picture}
\end{figure}

and obviously, the index obtained naively $\Omega_{\textrm{naive}}$ differs from the exact index, which is given by $\Omega_{\textrm{exact}}=-40'392$ (see equation \ref{eq:z0sextic}). This can be put right using the refined index, as the non--trivially fibered moduli space of this bound state dictates.

One can either argue from the D6--picture, or from a purely algebraic geometrical D4--perspective. Essentially the two arguments are identical. In the first picture, one chooses two point--like sheaves to lie on the vanishing locus of the polynomial describing the tachyon; in the second case, one places two $\Dob$--branes on the D4.

As dicussed before, for the most polar state, the most general tachyon map is of the form
\be\label{sextictachyon}
   T=a_1x_1+...a_4x_4,
\ee
choosing coordinates $(x_1,...,x_4,x_5)$ on $\wpr{4}{11112}$ transforming with weights $(1,1,1,1,2)$. Setting two $\Dob$--branes on the $\Dsb$ means that the map has to `pass' through two points, so in general, the moduli space of the tachyon will be reduced to from $\cp{3}$ to $\cp{1}$, yielding $\chi(\cp{1})=2$. This is where the intersection number $2$ comes from. It will now be shown that the dimension of this fiber jumps (as was observed for a state on the quintic 3--fold, though for a different reason, in \cite{Collinucci:2008ht}). One might say, that two $\Dob$--particles behave as one on a specific `special' locus. The scheme is to analyze the remaining moduli describing this tachyon, demanding that two points as well as the tangent directions at these points lie on the CY as well as on the zero locus of the tachyon.\\
\\
\textbf{Finding special tangent directions arising from blowups}\\
\\
We discussed in the previous section that the fiber of the tachyon moduli space jumps in two cases. Firstly, if a particle is placed at a `special locus', directly. Secondly, if two particles are coincident and the tangent direction arising from the blowup procedure is a `special tangent direction'. Before putting all the bits into place for the state under current investigation, we will look for the \emph{special tangent directions} for the blowups which we have to perform. Before continuing, let us stress that this is a mechanism special to Calabi--Yau varieties embedded in weighted projective spaces, but will probably play an even more prominent role when working with more general Calabi--Yau varieties, embedded in toric ambient spaces.

Generically, demanding that two points lie in the zero of (\ref{sextictachyon}), imposes two independent constraints, reducing the number of moduli by two. Denote the coordinates of the two points by $x_i^{P_1}$ and $x_i^{P_2}$. This is not that clear if the two points lie on top of each other $x^{P_1}_i=x^{P_2}_i$. However, this is generically resolved by the blowup procedure, where the tangent direction of the blowup gives an independent constraint, again leading to two constraints. However, as one is working with a weighted projective space, the coordinate $x_5$ does not appear in (\ref{sextictachyon}). Different $x_5$--coordinates are not `seen' by the tachyon. The points do however have to lie on the sextic, for which one can choose a representative given by a transverse polynomial, \cite{Candelas:1989hd},
\be\label{sexticpolynomial}
    p_{\textrm{sextic}}=x_5^3+p^{(6)}(x_1,x_2,x_3,x_4)=0.
\ee 
where $p^{(6)}$ is a degree 6 homogeneous polynomial in the indicated coordinates.
This means that the case $x_5^{P_1}=e^{\frac{2\pi i k}{3}}x_5^{P_2}$ for $k=0,1,2$ with all other coordinates equal ($x_i^{P_1}=x^{P_2}_i$ for $i=1,2,3,4$), is possibly of interest.

If the first four coordinates of the two points are identical, this amounts to imposing only one instead of two constraints on the tachyon field. After the usual blowup procedure (if the fifth coordinate is also equal), the particles are however distinguished by a tangent direction, which can be interpreted as the direction from which they `approach' each other.

Consider thus a tangent vector $X^i\partial_i$, and demand that it is a direction tangent to the sextic CY;
\be\label{tangentsextic}
  \nabla_X p_{\textrm{sextic}}=0,
\ee

But it also acts on (\ref{sextictachyon})
\be\label{tangentvectorconstraint}
   \nabla_XT=X^1a_1+...+X^4a_4.
\ee
Setting $\nabla_XT=0$, is like a second constraint. The particles are distinguished by the tachyon field after a blowup. This is also the reason why a coincident locus does not in general lead to a jump in the tachyon fiber above that locus. Namely, (\ref{tangentvectorconstraint}) does not lead to an extra constraint, iff
\be\label{rankofmatrix}
  \textrm{rank}\left(\begin{array}{cccc} x_1 & x_2 & x_3 & x_4 \\ X^1 & X^2 & X^3 & X^4 \end{array}\right) <2,
\ee
which can happen only if either $X^1=...=X^4=0$ or if $X^i=\lambda x_i$, with $\lambda \in \mathbb C^*$. We will show where this can happen in the present case. In the following distinction of cases, as well as in similar computations later on in this paper, we will always refer to coordinates of the coincident particles. $x_5\neq 0$ for example means, that the two (a priori) coincident particles lie on a locus with $x_5\neq 0$. In the present case, one can distinguish between
\bi
  \item $x_5\neq 0$: This means one can choose affine coordinates with $x_5=1$. Thus, in these coordinates, one knows that $X^5=0$ for the tangent vector (and hence the case $X^1=...=X^4=0$ is ruled out and only the case $X^i=\lambda x_i$ remains). The tangent vectors should remain in the CY (\ref{tangentsextic}), which leads to $6\lambda p^{(6)}=0$. Since $\lambda\neq 0$, this means $p^{(6)}=0$, so $x_5=0$ upon plugging this into (\ref{sexticpolynomial}), which contradicts our assumption. If $x_5\neq 0$, the two particles are distinguished after the blowup.
  \item $x_5=0$: In this case, fix (w.l.o.g.) $x_1=1$. This means $X^1=0$ for the tangent vector. To have only one instead of two `constraint' equations for the tachyon, one thus needs $X^1=...=X^4=0$. This occurs when the tangent vector equals $X^5\partial_5$, which is possible at the locus $x_5=0$. Note that there is \emph{one} tangent direction and \emph{one} locus ($x_5=0$) for which this happens. This is where the fiber of the tachyon moduli space jumps and needs to be taken into account.
\ei
\textbf{Calculating the refined index}\\
\\
Starting from the naive form for the index of this bound state,
\be
   \chi (\textrm{two particles})\cdot \chi (\textrm{tachyon})=\chi(X)^2\cdot\chi(\cp 1)+\textrm{corrections},
\ee
one can nicely see how the refinement comes in. The refined index receives the following contributions (where $\chi(X)=-204$ denotes the Euler character of the sextic).
\bi
   \item $\frac{1}{2}(\chi(X)^2-3\chi(X)+2\chi_0)\cdot\chi(\cp 1)$: this is the generic locus, where the two particles are separated and the locus where $x_i^{P_1}=x_i^{P_2}$ for $i=1,2,3,4$ has been substracted. Note that one has to be careful not to substract the locus where the first four coordinates are identical and $x_5=0$ more than once. This has been taken into account with the $+2\chi_0$ term. $\chi_0$ is the Euler character of the locus $x_5=0$ in the CY and can be calculated from the adjunction formula to give $\chi_0=108$. The factor $\frac 1 2$ accounts for the fact that interchanging the two particles gives the same configuration.
  \item $(\chi -\chi_0)\cdot \chi(\cp 2)\cdot\chi(\cp 1)$: this accounts for the locus where the two $\Dob$s coincide, without the locus $x_5=0$. Note that the $\chi (\cp 2)$ results from the blowup of a codimension $3$ locus.
  \item $2\cdot \frac{1}{2}(\chi -\chi_0)\cdot\chi(\cp 2)$: this takes into account the loci where $x_5^{P_1}=e^{\frac{2\pi i k}{3}}x_5^{P_2}$ for $k=1$ and $k=2$ (hence the overall factor of two, as these loci both contribute equally). Note that the fiber of the tachyon has jumped, these two particles are seen as one.
 \item $\chi_0\cdot(\chi(\cp 2)-1)\cdot\chi (\cp 1)$: here, the locus $x_i^{P_1}=x_i^{P_2}$ for $i=1,2,3,4$ and $x_5^{P_i}=0$ is dealt with. In principle, one just has to do a blowup of a codimension $3$ locus (hence a factor of $\chi(\cp 2)$). After the blowup, the tachyon `sees' two $\Dob$'s. However, one needs to substract the one tangent direction we found in the above analysis, because one loses one of the two `constraints' on the tachyon. This one tangent direction is taken into account on the next line.
 \item $\chi_0 \cdot 1\cdot\chi(\cp 2)$: for this blowup direction (for which the $1$ stands for the index), the tachyon again sees only one particle.
\ei
Note that had one ignored the subtlety with the locus $x_5$, one would have retrieved the calculation where one finds a product of the tachyon index $2$, and the Donaldson--Thomas invariant $N_{\textrm{DT}}(0,2)$. Collecting all the pieces linked to the value $2$ or $3$ for the tachyon index (up to a sign), one can state the correct index in the form
\be
   \Omega_{\textrm{exact}} =-2\cdot (20'298+204)-3\cdot (-204)=-40'392.
\ee
This allows us to state the \emph{Donaldson--Thomas partitions} $\mathcal{N}^{(g,s)}_{\textrm{DT}}(0,2)$ for the sextic:
\bea
   \mathcal{N}^{(g)}_{\textrm{DT}}(0,2)&=&20'502,\\
   \mathcal{N}^{(s)}_{\textrm{DT}}(0,2)&=&-204.
\eea
$\mathcal{N}^{(g)}_{\textrm{DT}}(0,2)$ counts the generic D6--2$\Dob$ BPS states, for which the tachyon perceives two $\Dob$'s, and $\mathcal{N}^{(s)}_{\textrm{DT}}(0,2)$ counts the special D6--2$\Dob$ BPS states, for which the tachyon perceives only one $\Dob$--brane. Note that there is a sign difference between these indices.
\be
   N_{\textrm{DT}}(0,2)= \mathcal{N}^{(g)}_{\textrm{DT}}(0,2)+ \mathcal{N}^{(s)}_{\textrm{DT}}(0,2).
\ee
Let us consider a second example.\\
\\
\textbf{The state $\mathbf{\Delta q=0, \Delta q_0=-2}$ on the octic}\\
\\
For this charge system, one finds a split flow tree with centers
\bea
    \Gamma_1&=&(1,1,\frac{17}{6},\frac{13}{6}),\nonumber\\
    \Gamma_2&=&(-1,0,-\frac{11}{6},-2),\nonumber
\eea
of the form
\begin{figure}[h]
\setlength{\unitlength}{1cm}
\centering 
\begin{picture}(3,2.3)
\put(0,0) {\includegraphics[width=0.25\textwidth,angle=0]{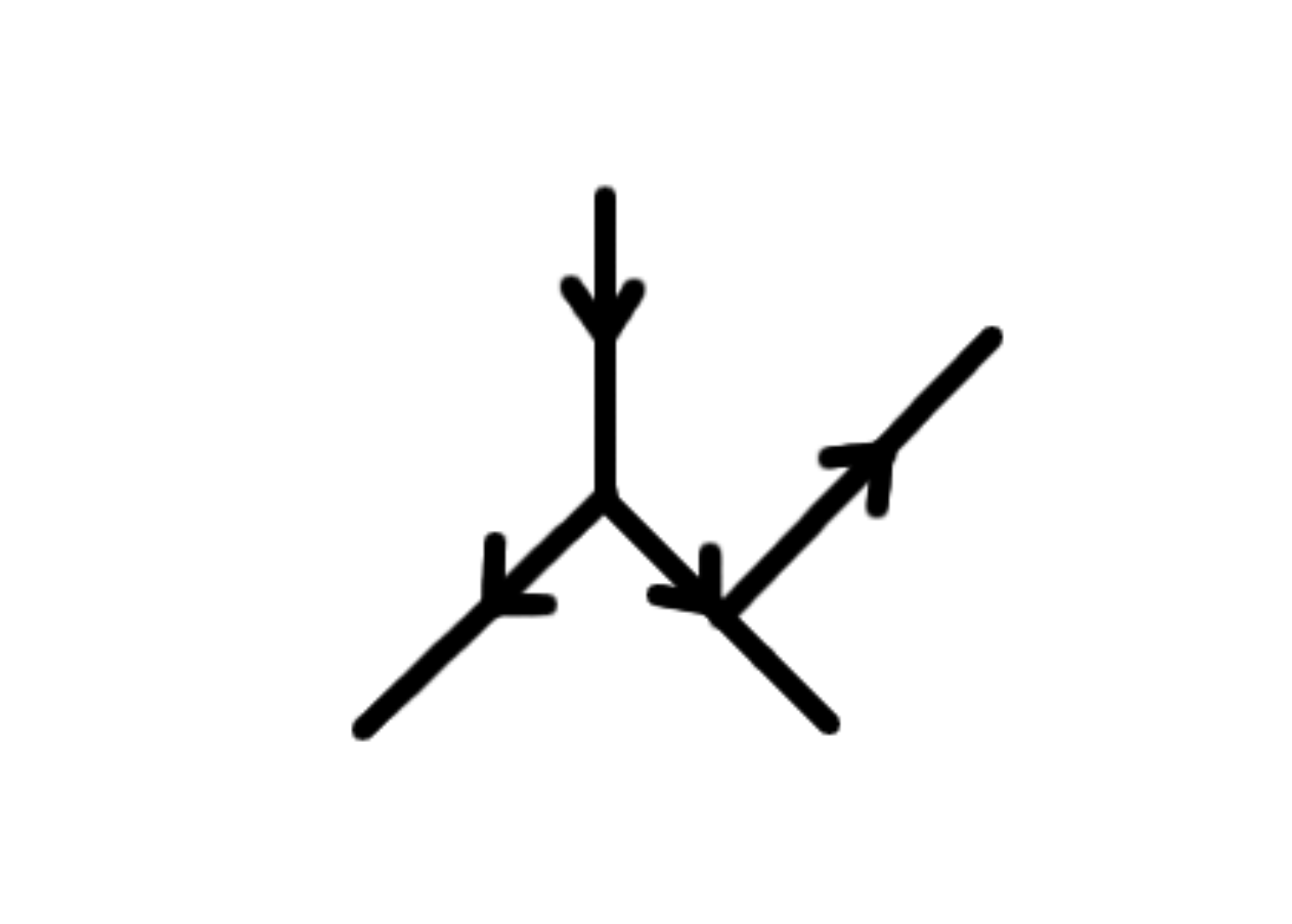}}
\put(1,2.35) {$D4_{\frac{H}{2},\overline{2D0}}$}
\put(0.6,0.15) {$D6_H$}
\put(2.25,0.15) {$\overline{D6}$}
\put(2.65,1.8) {$\overline{2D0}$}
\put(4,1.3) {$=$}
\put(5,1.3) {$=-2\cdot 43'068=-86'136$.}
\end{picture}
\end{figure}

For the octic, one can choose 
\be\label{octicpolynomial}
    p_{\textrm{octic}}=x_5^2+p^{(8)}(x_1,x_2,x_3,x_4)=0
\ee 
as a transverse polynomial. Again, the cases where the particles have four equal coordinates, and where $x_5^{P_1}=e^{\frac{2 \pi i k}{2}}x_5^{P_2}$, are possibly of interest. The analysis of the locus where the tachyon field perceives the BPS state with two $\Dob$'s differently and where the fiber changes dimension is analogous to the sextic case, and the discussion will therefore be brief. We again start by searching for the `special tangent directions' by considering (\ref{rankofmatrix}):
\bi
  \item $x_5\neq 0$: This means one can choose affine coordinates with $x_5=1$. Again, one knows that $X^5=0$ for the tangent vector (and hence the case $X^1=...=X^4=0$ is ruled out and only the case $X^i=\lambda x_i$ remains). The tangent vectors should remain in the CY, which leads to $8\lambda p^{(8)}=0$. $p^{(8)}=0$ would imply $x_5=0$, contradicting the first assumption.
  \item $x_5=0$: In this case, fix (w.l.o.g.) $x_1=1$. This means $X^1=0$ for the tangent vector. To have only one instead of two `constraint' equations for the tachyon, one thus needs $X^1=...=X^4=0$. This occurs when the tangent vector equals $X^5\partial_5$, which is possible at the locus $x_5=0$. Note that there is again \emph{one} tangent direction for which this occurs. This is where the fiber of the tachyon jumps and needs to be taken into account.
\ei

\textbf{Calculating the exact index}\\

The index receives similar contributions as in the case of the sextic, but the calculation is slightly simpler.
\bi
\item $\frac{1}{2}(\chi(X)^2-2\chi(X)+\chi_0)\cdot\chi(\cp 1)$: this again is the generic locus, but as in this case $x_5^{P_1}=e^{\frac{2 \pi i k}{2}}x_5^{P_2}$ for $k=0,1$ one substracts two instead of three loci with index $\chi(X)$. Instead of substracting the locus $x_5=0$ three times, one does this twice, and needs to compensate once. In this case, $\chi_0=304$.
  \item $(\chi -\chi_0)\cdot \chi(\cp 2)\cdot\chi (\cp 1)$: this accounts for the locus where the two $\Dob$s coincide, without the locus $x_5=0$. Note that the $\chi (\cp 2)$ results from the blowup of a codimension $3$ locus.
  \item $\frac{1}{2}(\chi -\chi_0)\cdot\chi(\cp 2)$: this takes into account the locus where $x_5^{P_1}=-x_5^{P_2}$ and other coordinates equal. This is again a locus where the tachyon index has jumped.
 \item $\chi_0\cdot (\chi(\cp 2)-1) \cdot\chi(\cp 1)$: here, the locus $x_5=0$ and other coordinates equal is dealt with. In principle one just has to do a blowup (hence a factor of $\chi(\cp 2)$). After the blowup, the tachyon `sees' two $\Dob$'s. Again, one needs to substract the one tangent direction we found above, because one loses one of the two `constraints' on the tachyon. This tangent direction is taken into account on the next line.
 \item $\chi_0 \cdot 1\cdot\chi (\cp 2)$: for this one blowup direction, the tachyon again sees only one particle.
\ei
Collecting all the pieces linked to the value $2$ or $3$ for the tachyon index (up to a sign), one can state the correct index in the form (to be compared with equation \ref{octicellipticgenus})
\be
   \Omega_{\textrm{exact}} =-2\cdot (43'068-4)-3\cdot (+4)=-86'140.
\ee
This means that the \emph{Donaldson--Thomas partitions} $\mathcal{N}^{(g,s)}_{\textrm{DT}}(0,2)$ for the octic read:
\bea
   \mathcal{N}^{(g)}_{\textrm{DT}}(0,2)&=&43'064,\\
   \mathcal{N}^{(s)}_{\textrm{DT}}(0,2)&=&4.
\eea
Note that the sum of the partitions yields $N_{\textrm{DT}}(0,2)$.

\subsection{Special curves and DT partitions $\mathcal{N}^{(g,s)}_{\textrm{DT}}(1,2)$}
Up until now, refinements were presented, which were necesary due to the fact that the tachyon fields did not perceive $\Dob$'s generically. An example will be given now, where an analogous refinement is necessary because of the fact that a D2/D0--state (with the D2 wrapped on a curve) is not perceived generically. A different way of expressing this is to state that there are \emph{special $\Dsb$--D2/D0 bound states}.\\
\newpage
\textbf{The state $\mathbf{\Delta q=1, \Delta q_0=-2}$ on the octic}\\
\\
For this charge system, one finds a split flow tree with centers
\bea
    \Gamma_1&=&(1,1,\frac{17}{6},\frac{13}{6}),\nonumber\\
    \Gamma_2&=&(-1,0,-\frac{5}{6},-2),\nonumber
\eea
of the form

\begin{figure}[h]
\setlength{\unitlength}{1cm}
\centering 
\begin{picture}(3,2.3)
\put(0,0) {\includegraphics[width=0.25\textwidth,angle=0]{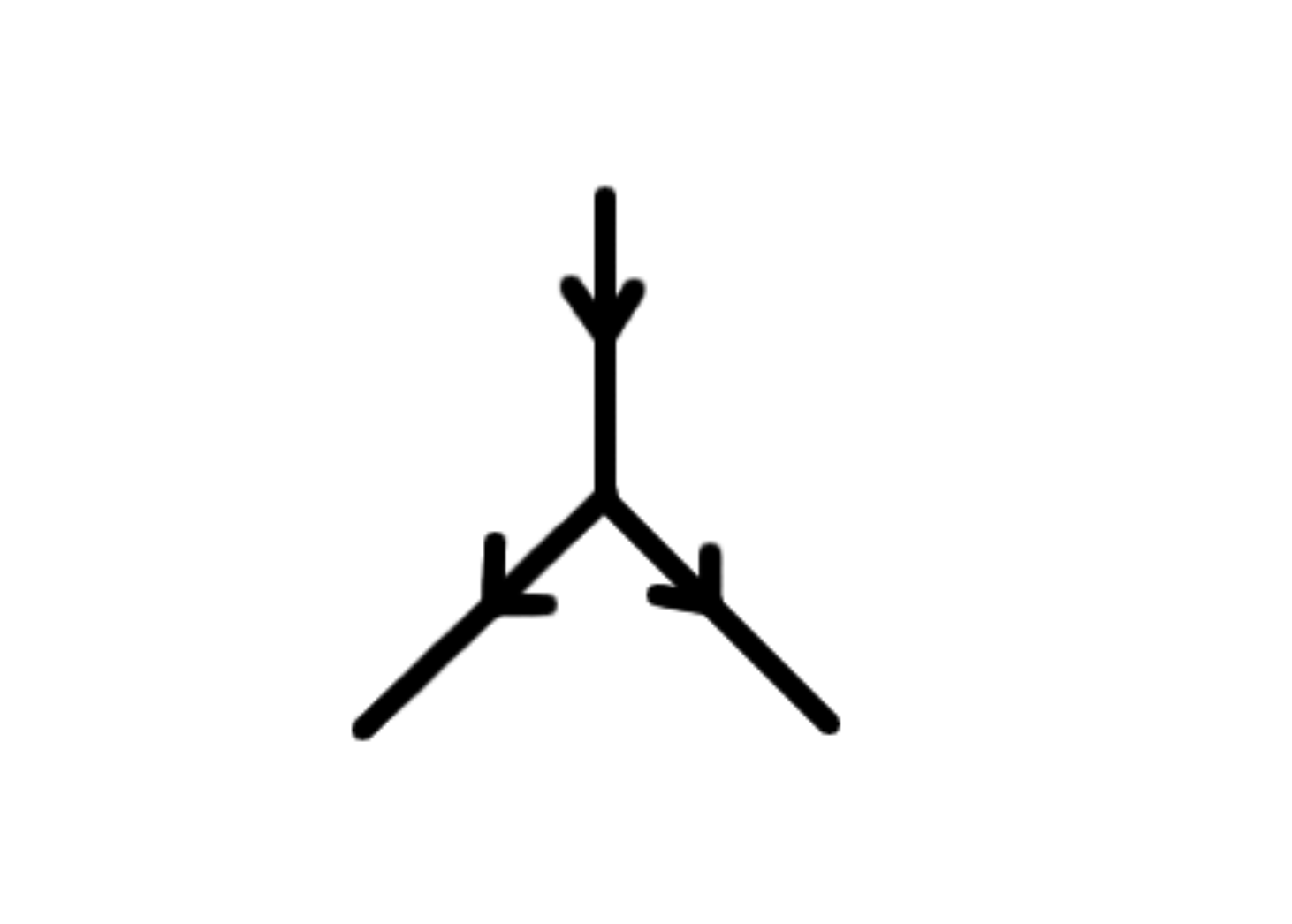}}
\put(1.5,2.25) {$D4_{\frac{H}{2}+F(C_1{}^{g=0})}$}
\put(0.6,0.15) {$D6_H$}
\put(2.25,0.15) {$\overline{D6}-D2(C_1{}^{g=0}),\overline{D0}$}
\put(4,1.3) {$=$}
\put(5,1.3) {$8'674'176$.}
\end{picture}
\end{figure}

with an index naively calculated as
\begin{equation}
  \Omega=(-1)^{|\langle\Gamma_1,\Gamma_2\rangle|-1}|\langle\Gamma_1,\Gamma_2\rangle|N_{\textrm{DT}}(0,0)\cdot N_{\textrm{DT}}(1,2)=(-1)^0\cdot 1\cdot 1\cdot 8'674'176=8'674'176.
\end{equation}

This naive index agains needs refinement. In this case, for simplicity, choose the Fermat--polynomial for the octic:
\be\label{octicfermat}
    p_{\textrm{octic}}=x_5^2+x_1^8+x_2^8+x_3^8+x_4^8=0
\ee 
The following discussion is nonetheless also valid for more general transverse polynomials. A degree one rational curve on the octic can be represented as a degree one map from $\cp 1$ to the Calabi--Yau. Consider for example the map
\be\label{octicratcurve}
    (s,t)\rightarrow (s,e^{\frac{\pi i}{8}} s,t,e^{\frac{\pi i}{8}} t,0).
\ee

This generically imposes two constraints on the tachyon field, reducing its moduli space to $\cp 1$. Adding an extra D--particle will then reduce this moduli space to $\cp 0$, unless something special happens:
\bi
  \item The particle ($\Dob$) does not sit on the curve, but nevertheless produces no extra constraint. It is easy to verify that this cannot possibly happen for this example.
  \item The particle lies on the curve, which means that a blowup needs to be performed in the directions normal to the curve. Again, one might encounter special tangent directions, which do not impose an extra constraint on the tachyon. Following a similar procedure as in the previous examples, one can indeed verify that this is the case for the direction $X^5\partial_5$. As $x_5=0$ lies on the curve (\ref{octicratcurve}), this direction is automatically also tangent to the octic.
\ei
\textbf{Calculating the exact index}\\

The various contributions to the exact index according to the refined prescription read:
\bi
\item $N_{DT}(1,1)(\chi(X)-\chi_C)\chi(\cp 0)$, where $\chi_C=2$ is the Euler characteristic of the curve. This term deals with the case, when the $\Dob$ is placed at a locus different from the curve, thereby reducing the tachyon moduli space to $\cp 0$.
\item $N_{DT}(1,1)\chi_C[\chi(\cp 1)-1]\chi(\cp 0)$, dealing with the case that the $\Dob$ is located on the curve, but the blowup tangent direction leads to an extra constraint on the tachyon.
\item $N_{DT}(1,1)\chi_C\cdot 1\cdot\chi(\cp 1)$, which deals with the case, when the $\Dob$ lies on the curve, and a blowup is performed leading to a special tangent direction. This is an example of what was referred to as a special D2/D0 bound state, or alternatively just as a special curve. In this case, the tachyon field moduli space remains a $\cp 1$.
\ei

In total, this leads to the index
\be
   \Omega_{\textrm{exact}} = 1\cdot (8'674'176+59'008)+2\cdot (-59'008)=8'615'168.
\ee
Spectacularly, by comparing this number to the prediction from modularity (\ref{octicellipticgenus}), one also finds exact agreement for this case! One can thus state the \emph{Donaldson--Thomas partitions} $\mathcal{N}_{\textrm{DT}}(1,2)$ for the octic:
\bea
   \mathcal{N}^{(g)}_{\textrm{DT}}(1,2)&=&8'733'184,\\
   \mathcal{N}^{(s)}_{\textrm{DT}}(1,2)&=&-59'008.
\eea
Again, note that the sum of the partitions yields $N_{\textrm{DT}}(1,2)$.

To summarize, we conclude that all our results obtained in this section provide exact agreement with the predictions from modularity. This clearly is strong evidence that our procedure is correct. These results also provide a non--trivial and successful test for the split attractor flow conjecture. We will now discuss states on another CY manifold, for which also a polar state requires calculation of a refined bound state index.

\section{Refined predictions for elliptic genera}\label{refinedellipticgenera}
In this section, we show how our refined index computation alters the prediction for an elliptic genus of a CY, realized as a degree ten hypersurface in $\wpr{4}{11125}$. We choose coordinates $(x_1,...,x_4,x_5)$ with the weights $(1,1,1,2,5)$. We will refer to this CY as the \emph{decantic}, in the following. The total Chern class of this space reads $c(X)=\frac{(1+H)^4(1+2H)(1+5H)}{1+10H}=1+34H^2-288H^3$, and using $\int_X H^3=\int_{\wpr{4}{11125}}10H^4=1$, one obtains $\chi (X)=-288$. We again wrap a D4--brane on the hyperplane class divisor $P=H$. For the lattice of fluxes, one finds in this case that the pullback $L_X=i^{*}_P(H^2(X,\mathbb{Z})$ and it's orthongonal complement, $L_X\oplus L_X^{\perp}$, are already unimodular, thus no gluing vectors exist and the elliptic genus is a `one--dimensional vector',
\begin{equation}
Z(q, \bar q, z) = Z_0(q)\, \Theta_0 (\bar q, z)\, .
\end{equation}
Again, the list of DT invariants of interest is stated below.
\begin{center}
\begin{tabular}{|c|c|c|c|c|}
    \hline
    \multicolumn{5}{|c|}{\textbf{Donaldson--Thomas invariants: decantic}}\\
    \hline
    & $\mathbf{n=0}$ & $\mathbf{n=1}$ & $\mathbf{n=2}$ & $\mathbf{n=3}$\\
    \hline
    \hline
    $\mathbf{\beta=0}$ & 1 & 288 & 40'752 & 3'774'912\\
    \hline
    $\mathbf{\beta=1}$ & 1150 & 435'827 & 89'103'872 & 11'141'118'264\\
    \hline
    $\mathbf{\beta=2}$ & -64'916'198 & 40'225'290'446 & 9'325'643'249'563 & 1'119'938'319'168'004\\
    \hline
\end{tabular}
\end{center}

\begin{enumerate}
\item $\mathbf{\Delta q=0, \Delta q_0=0, \qquad [0,\frac{35}{24}]}$:\\
As usual, the most polar state is the D4--brane carrying flux $\frac{H}{2}$ for anomaly cancellation, with total charge $(0,1,\frac{1}{2},\frac{19}{12})$. The reduced D0--brane charge can be calulcated to be $\hat{q}_0=\frac{35}{24}$, thus the state lies in the class $[0,\frac{35}{24}]$. One finds a split flow tree with centers
\bea   
\Gamma_1&=&(1,1,\frac{23}{12},\frac{19}{12}),\nonumber \\   \Gamma_2&=&(-1,0,-\frac{17}{12},0),\nonumber
\eea
with split flow tree
\begin{figure}[h]
\setlength{\unitlength}{1cm}
\centering 
\begin{picture}(3,2.3)
\put(0,0) {\includegraphics[width=0.25\textwidth,angle=0]{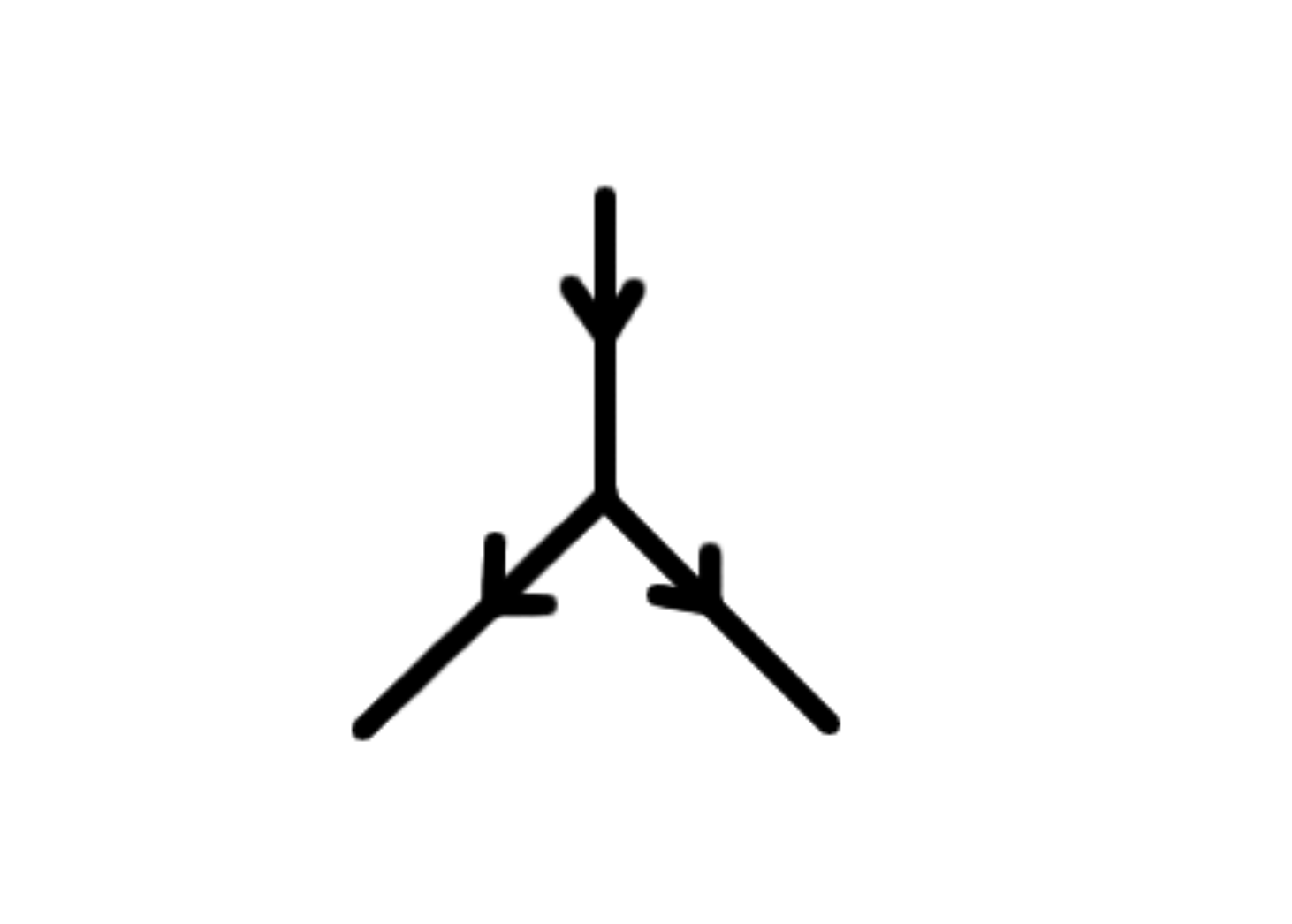}}
\put(1.5,2.25) {$D4_{\frac{H}{2}}$}
\put(0.6,0.15) {$D6_H$}
\put(2.25,0.15) {$\overline{D6}$}
\put(4,1.3) {$=$}
\put(5,1.3) {$3$,}
\end{picture}
\end{figure}

where the BPS index is calculated according to
\begin{equation}
  \Omega=(-1)^{|\langle\Gamma_1,\Gamma_2\rangle|-1}|\langle\Gamma_1,\Gamma_2\rangle|N_{\textrm{DT}}(0,0)\cdot N_{\textrm{DT}}(0,0)=(-1)^2\cdot 3\cdot 1\cdot 1=3.
\end{equation}
Again, the number $3$ can nicely be understood along the previous lines. In $\wpr{4}{11125}$ one can only use three coordinates to define a hyperplane, and $\chi (\cp 2)=3$, so the correspondence between divisor moduli in the D4--picture and tachyonic degrees between the D6 and the $\Dsb$ again works out.
\item $\mathbf{\Delta q=0, \Delta q_0=-1,\qquad [0,\frac{11}{24}]}$:\\
Adding one $\Dob$, one obtains the total charge $(0,1,\frac{1}{2},\frac{7}{12})$, with reduced D0--brane charge $\hat{q}_0=\frac{11}{24}$. The flow tree is again analogous to the previous findings, the side of the $\Dob$ after the first split being governed by the appropriate threshold wall. The charges of the centers after the first split read
\bea
    \Gamma_1&=&(1,1,\frac{23}{12},\frac{19}{12}),\nonumber\\
    \Gamma_2&=&(-1,0,-\frac{17}{12},-1).\nonumber
\eea
\newpage
The flow tree looks like
\begin{figure}[h]
\setlength{\unitlength}{1cm}
\centering 
\begin{picture}(3,2.3)
\put(0,0) {\includegraphics[width=0.25\textwidth,angle=0]{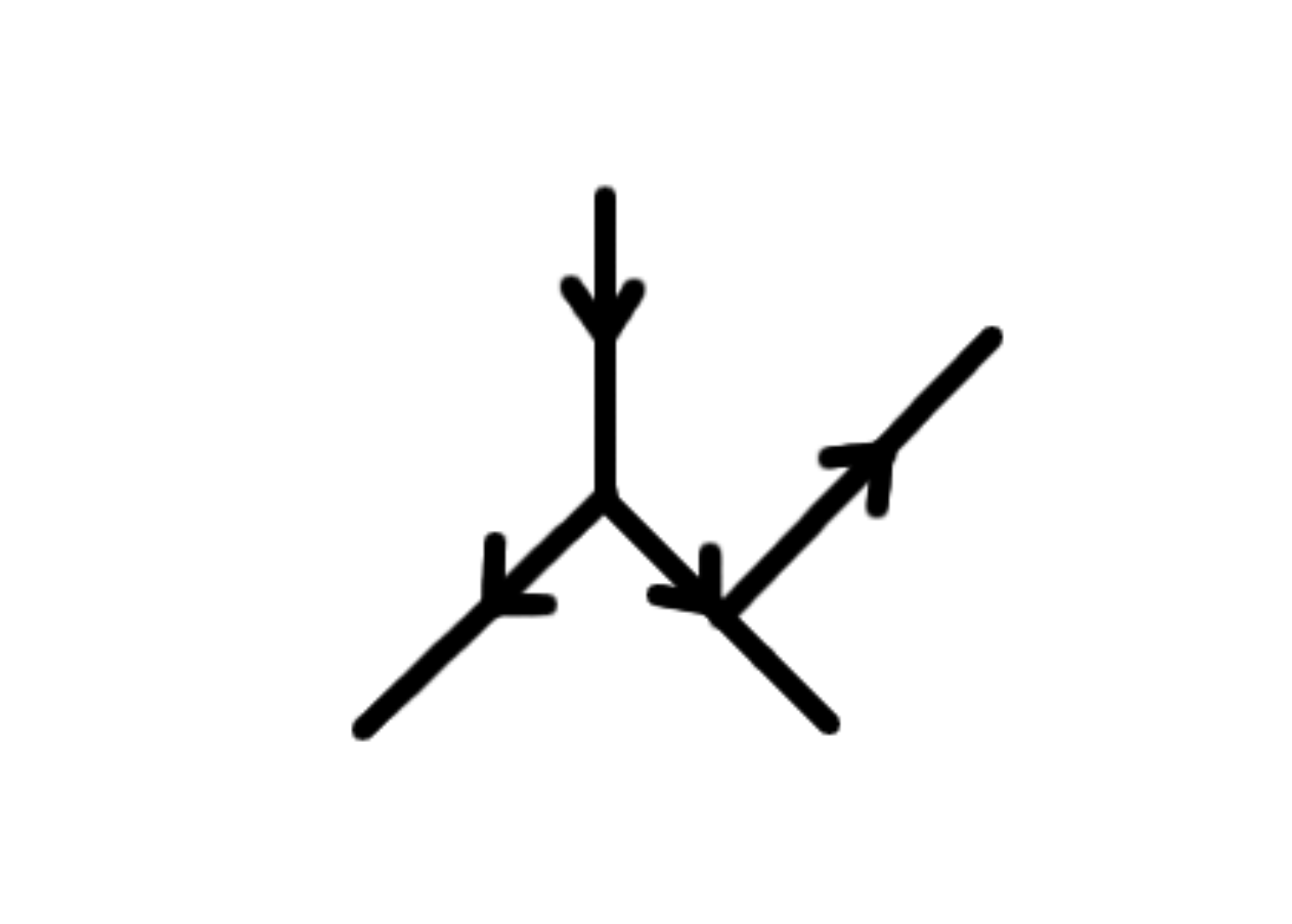}}
\put(1,2.35) {$D4_{\frac{H}{2},\overline{D0}}$}
\put(0.6,0.15) {$D6_H$}
\put(2.25,0.15) {$\overline{D6}$}
\put(2.65,1.8) {$\overline{D0}$}
\put(4,1.3) {$=$}
\put(5,1.3) {$-576$,}
\end{picture}
\end{figure}

from which one is tempted to conclude
\begin{equation}\label{naivedecantic01}
  \Omega=(-1)^{|\langle\Gamma_1,\Gamma_2\rangle|-1}|\langle\Gamma_1,\Gamma_2\rangle|N_{\textrm{DT}}(0,0)\cdot N_{\textrm{DT}}(0,1)=(-1)^1\cdot 2\cdot 1\cdot 288=-576.
\end{equation}
\end{enumerate}
Using these two polar degeneracies, the elliptic genus would be determined to be
\begin{equation}\label{wrongellipticgenusdecantic}
  Z_0(q)=q^{-\frac{35}{24}}(3-576q+271'704q^2+206'401'533q^3+21'593'767'647q^4...),
\end{equation}
which agrees with the findings in \cite{Gaiotto:2007cd}. We will now argue that this is not quite correct. We will predict the elliptic genus to be 
\begin{equation}
  Z_0(q)=q^{-\frac{35}{24}}(3-575q+271'955q^2+206'406'410q^3+21'593'817'025q^4...).
\end{equation}
The reason for this lies in the fact that the index (\ref{naivedecantic01}) is not correct, because the state $\Delta q=0,\Delta q_0=1$ has a non--trivially fibered moduli space. This will allow us to calculate partitions of $N_{\textrm{DT}}(0,1)$, for the decantic.\\
\\
\textbf{DT partitions $\mathcal{N}^{(g,s)}_{\textrm{DT}}(0,1)$: generic and special $D6-\Dob$ states}\\
\\
Note that the tachyon map for the most polar state is a section of the bundle $H$, and is of the general form
\be\label{decantictachyon}
   T=a_1x_1+...a_3x_3,
\ee
as the coordinates $x_4$ and $x_5$ are `forbidden' (weight too high). This yields a moduli space with Euler character $\chi (\cp 2)=3$, accounting for the degeneracy of the most polar state. If one adds a $\Dob$--brane to the system (this is also discussed in \cite{Collinucci:2008ht} for examples on the quintic 3--fold), the tachyon map also has to vanish on an additional point. Inserting this in \eqref{decantictachyon} generically eliminates one of the moduli, reducing to a tachyon moduli space with Euler character $\chi (\cp 1)=2$. The problem is that one can place the $\Dob$--brane at the locus  $x_1=x_2=x_3=0$, which is indeed a point lying on the decantic $X$. Placing the $\Dob$ on this point means that this `particle' will not imply a constraint on \eqref{decantictachyon}. The Euler character of this locus ($x_1=x_2=x_3$) $\chi_0$ can easily be determined: $\chi_0=1$ (this is trivial, as the locus is just a point).

Thus, the correct index for the $\Delta q=0, \Delta q_0=-1$ system reads as follows:
\be
     \Omega = -2\cdot (\chi_0-\chi (X))-3\cdot(-\chi_0 )=-575.
\ee
In other words, $N_{\textrm{DT}}(0,1)$ can be partitioned, into the DT--partitions
\bea
   \mathcal{N}^g_{\textrm{DT}}(0,1)&=&\chi_0-\chi (X) = 289,\\
   \mathcal{N}^s_{\textrm{DT}}(0,1)&=&-\chi_0 = -1.
\eea
Again, note that $N_{DT}(0,1)=\mathcal{N}^g_{\textrm{DT}}(0,1)+\mathcal{N}^{s}_{\textrm{DT}}(0,1)$. The superscript $g$ stands for generic, and $\mathcal{N}^g_{\textrm{DT}}(0,1)$ counts the number of D6--$\Dob$ states, which are perceived by the tachyon generically. The superscript $s$ stands for special. Accordingly, $\textrm{N}^s_{\textrm{DT}}(0,1)$ counts the number of D6--$\Dob$ states, where the $\Dob$ sits at a special locus, where the tachyon matrix does not perceive the particle.\\ 

Of course, this also means, that the index for the split flow tree must be stated correctly:
\begin{figure}[h]
\setlength{\unitlength}{1cm}
\centering 
\begin{picture}(3,2.3)
\put(0,0) {\includegraphics[width=0.25\textwidth,angle=0]{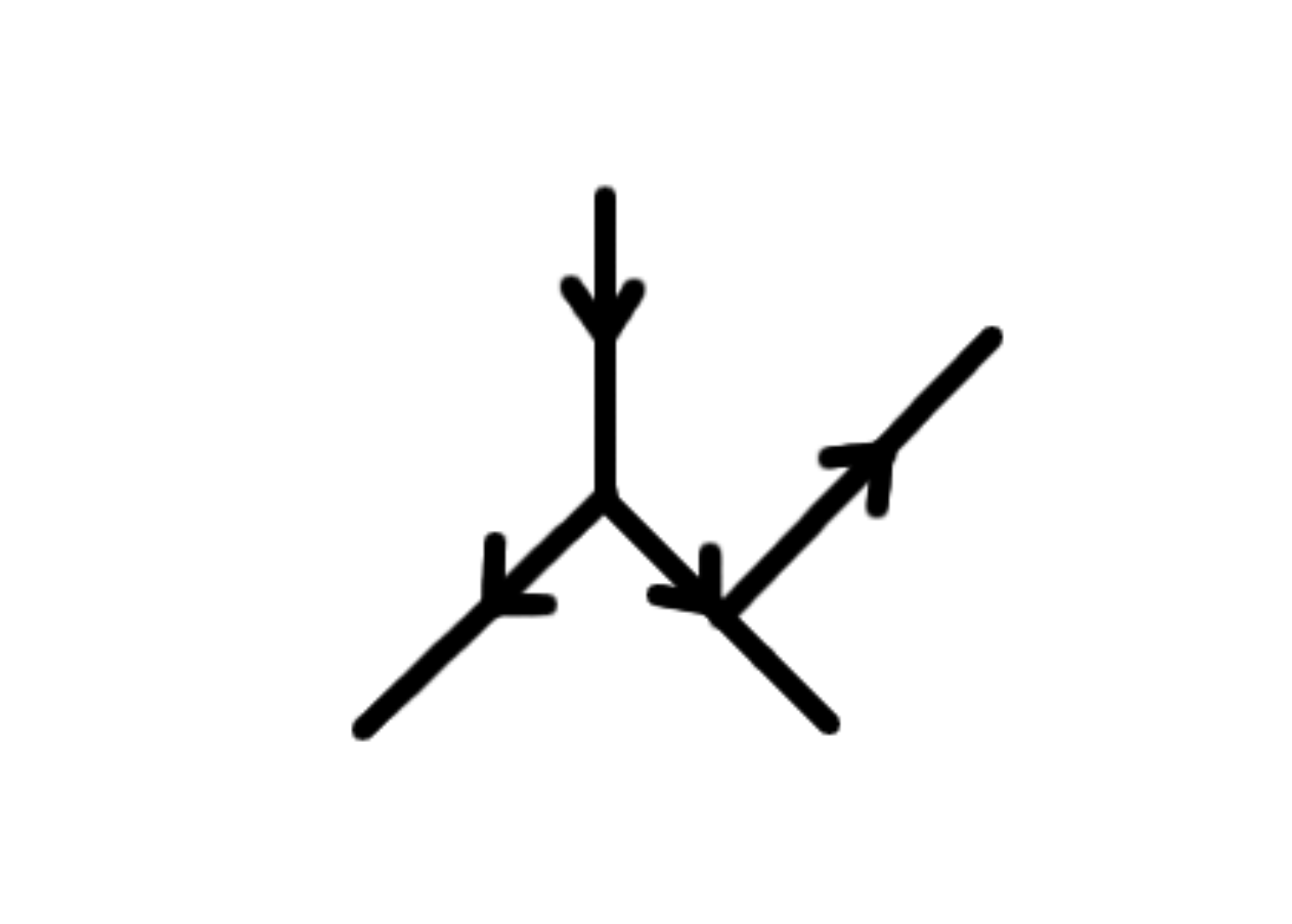}}
\put(1,2.35) {$D4_{\frac{H}{2},\overline{D0}}$}
\put(0.6,0.15) {$D6_H$}
\put(2.25,0.15) {$\overline{D6}$}
\put(2.65,1.8) {$\overline{D0}$}
\put(4,1.3) {$=$}
\put(5,1.3) {$-575$.}
\end{picture}
\end{figure}

This leads to the elliptic genus
\begin{equation}\label{ellipticgenusdecantic}
  Z_0(q)=q^{-\frac{35}{24}}(3-575q+271'955q^2+206'406'410q^3+21'593'817'025q^4...).
\end{equation}
It is interesting to note that the authors of \cite{Gaiotto:2007cd}, after having predicted a slightly deviating elliptic genus (as explained above), find $271'952$ as a prediction for the number of BPS states of the system, which we denote as $\Delta q=0,\Delta q_0=-2$. This is only off by $3$ of the modular result we predict, as opposed to the $248$ from the result predicted by the `naive elliptic genus' (\ref{wrongellipticgenusdecantic}). This might be seen as an indication that the new prediction is indeed correct. Unfortunately, our technique does not allow to check this prediction, as a single flow exists for this state and it is thus not possible for us to confirm our elliptic genus with absolute certainty, although we do believe that we have collected strong evidence for our computational scheme. It is nevertheless interesting to predict Donaldson--Thomas partitions $\mathcal{N}^{(g,s)}(0,2)$ for the decantic.\\
\\
\textbf{The state $\mathbf{\Delta q=0, \Delta q_0=-2}$ on the decantic}\\
\\
The total charge for this system reads $\Gamma =(0,1,\frac{1}{2},-\frac{5}{12})$, which implies $\hat{q}_0=-\frac{13}{24}$: this is thus a non--polar state. One finds a split flow tree with the centers
\bea
    \Gamma_1&=&(1,1,\frac{23}{12},\frac{19}{12}) \nonumber\\
    \Gamma_2&=&(-1,0,-\frac{17}{12},-2),\nonumber
\eea
\newpage
and a flow tree of the form

\begin{figure}[h]
\setlength{\unitlength}{1cm}
\centering 
\begin{picture}(3,2.3)
\put(0,0) {\includegraphics[width=0.225\textwidth,angle=0]{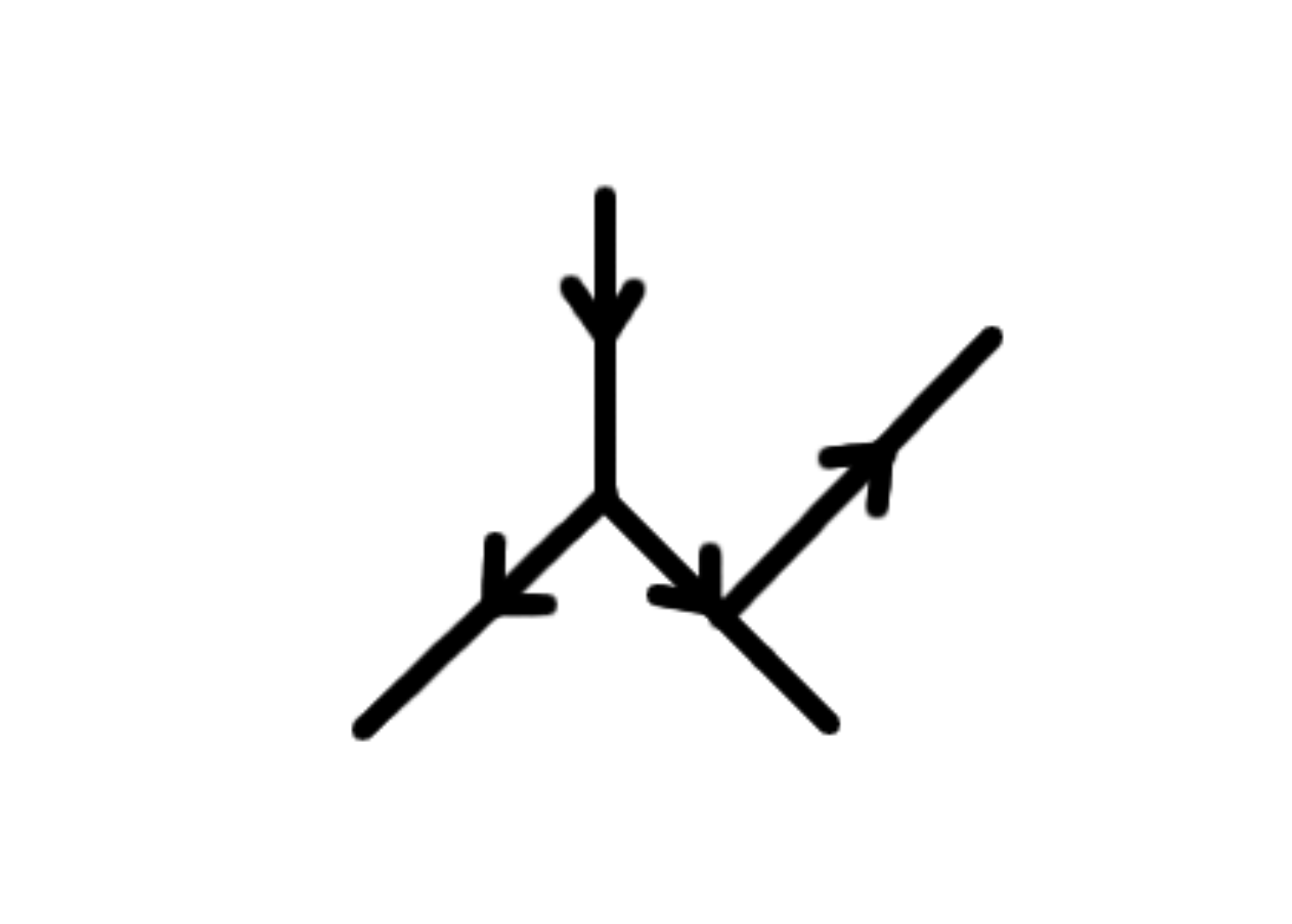}}
\put(1,2.35) {$D4_{\frac{H}{2},\overline{2D0}}$}
\put(0.6,0.15) {$D6_H$}
\put(2.25,0.15) {$\overline{D6}$}
\put(2.65,1.8) {$\overline{2D0}$}
\put(4,1.3) {$=$}
\put(5,1.3) {$=40'752$,}
\end{picture}
\end{figure}

which would naively yield an index
\begin{equation}
  \Omega=(-1)^{|\langle\Gamma_1,\Gamma_2\rangle|-1}|\langle\Gamma_1,\Gamma_2\rangle |N_{\textrm{DT}}(0,0)\cdot N_{\textrm{DT}}(0,2)=(-1)^0\cdot 1\cdot 1\cdot 40'752=40'752.
\end{equation}
For the decantic, one can choose 
\be\label{decanticpolynomial}
    p_{\textrm{decantic}}=x_5^2+x_4^5+p^{(10)}(x_1,x_2,x_3)=0
\ee 
as a transverse polynomial. Note that the moduli space for the tachyon was $\cp 2$ for the most polar state. Generically, this is reduced to $\cp 0$ when placing two $\Dob$s, but there are a lot of subtleties involved. Namely, the cases when the $\Dob$s have three equal coordinates $x_1,x_2,x_3$ and $x_4^{P_1}=e^{\frac{2\pi i j}{5}}x_4^{P_2}, x_5^{P_1}=e^{\frac{2\pi i k}{2}}x_5^{P_2}$ (with $j=0,1,2,3,4$ and $k=0,1$) are of special interest. Additionally, the locus $x_1=x_2=x_3=0$ is special. Again, constraint loss will happen on some loci directly, but will also result from blowups, when placing the two $\Dob$s on the same locus. Thus, we will start by analyzing which tangent directions are special, and thus require special treatment when performing a blowup.

The condition for a contraint loss to occur after a blowup reads
\be
  \textrm{rank}\left(\begin{array}{cccc} x_1 & x_2 & x_3 \\ X^1 & X^2 & X^3 \end{array}\right) <2,
\ee
which can happen either if $X^1=...=X^3=0$ or if $X^i=\lambda x_i$.
\bi
  \item $x_4\neq 0$ or $x_5\neq 0$: Assuming that not all coordinates $x_1, x_2, x_3$ vanish at the same time (this case will be dealt with separately), an analysis shows that for the cases that either (or both) of the coordinates $x_4, x_5$ do not vanish, one finds one tangent direction for which a constraint loss occurs. These subcases shall be discussed briefly:
  \bi
  \item $x_4\neq 0, x_5\neq 0$: In this case set $x_4=1$, thus $X^4=0$. One can easily check that $X^1=X^2=X^3=0$ is not possible as it would imply $X^5=0$. Thus, set $X^i=\lambda x_i$ ($i=1,2,3$), which leads to $\nabla_X p_{\textrm{decantic}}=10\lambda p^{(10)}+2X^5x_5=0$. Combining this with \eqref{decanticpolynomial} yields $x_5^2+1=\frac{1}{5\lambda}X^5x_5$. This completely fixes the tangent vector, thus there is one direction for which constraint loss occurs.
  \item $x_4\neq 0, x_5= 0$: Set $x_4=1$, thus $X^4=0$.  Choosing $X^1=X^2=X^3=0$ yields $\nabla_X p_{\textrm{decantic}}=2X^5x_5=0$, thus one tangent direction.
  \item $x_4=0, x_5\neq 0$: Set $x_5=1$, thus $X^5=0$.  Choosing $X^1=X^2=X^3=0$ yields $\nabla_X p_{\textrm{decantic}}=5X^4x_4^5=0$, thus one tangent direction.
  \ei

  \item $x_4=x_5=0$:  In this case, one can choose $x_1=1$, and thus $X^1=0$. So, a general tangent vector reads $X_2\partial_2+...+X^5\partial_5$. Setting $X^1=X^2=X^3=0$, and plugging this into $\nabla_X p_{\textrm{decantic}}=0$ yields $5X^4x_4+2X^5x_5=0$, which is always satisfied. Thus, the tangent directions for which there is a constraint loss, form a $\cp 1$.
 \item $x_1=x_2=x_3=0$: In this case, set $x_5=1$. In that case \eqref{decanticpolynomial} reads $x_4^5+1=0$. One might think that one has found five points on the CY, but taking the equivalence relation into account under the group action, one realizes that this is only one point. In this case $X^4=X^5=0$, and thus $X^1=X^2=X^3=0$ is not possible. Therefore, there is no (extra) constraint loss when considering a blowup, when the two $\Dob$s coincide at this point on the decantic.

\ei

\textbf{Calculating the exact index}\\

Using the adjunction formula, one can calculate the Euler character associated to a number of loci of interest for the following calculation:
\begin{enumerate}
   \item $x_4=0: \chi_4=76$.
   \item $x_5=0: \chi_5=295$.
   \item $x_4=x_5=0: \chi_{45}=-70$.
   \item $x_1=x_2=x_3=0:$ Recall that this is only one point. The Euler character is thus $\chi_0=1$.
\end{enumerate}
A careful calculation reveals the following contributions:
\bi
\item $\frac{1}{2}( (\chi(X)-\chi_0)^2-10(\chi(X)-\chi_0-\chi_4-\chi_5+\chi_{45})-2(\chi_4-\chi_{45})-5(\chi_5-\chi_{45})$\\$-\chi_{45} )\cdot\chi(\cp 0)$: this is the generic locus. The case $x_1=x_2=x_3=0$ has been substracted from the beginning on, and additionally, also the cases when  the coordinates $x_1,x_2,x_3$ of the two $\Dob$s are identical and $x_4^{P_1}=e^{\frac{2\pi i j}{5}}x_4^{P_2}, x_5^{P_1}=e^{\frac{2\pi i k}{2}}x_5^{P_2}$, $j=0,1,2,3,4; k=0,1$ will be treated independently. These cases have been substracted, but for each possibility $(j,k)$, the subloci where $x_4=0, x_5=0$ or both have been substracted and then treated separately.
  \item $(\chi -\chi_0-\chi_{45})\cdot (\chi(\cp 2)-1)\cdot\chi (\cp 0)$: this is the most general case when the two $\Dob$s coincide. One has to treat various loci separately: the case, when the two particles lie on the point $x_1=x_2=x_3=0$ (this is substracted by the term $-\chi_0$), and also the case when $x_4=x_5=0$ has been removed (the term $-\chi_{45}$) and will be treated separately.  The factor $\chi (\cp 2)$ arises from the blowup of a codimension $3$ locus. According to the analysis presented above, one does however need to substract one direction, for which there will be a constraint loss. In this case, two constraints on the tachyon are imposed, reducing the moduli space to $\cp 0$.
  \item $(\chi -\chi_0-\chi_{45})\cdot 1\cdot\chi (\cp 1)$: This is the case analogous to the previous, but associated to the blowup direction yielding a constraint loss.
 \item $\chi_{45}\cdot (\chi(\cp 2)-\chi(\cp 1))\cdot \chi (\cp 0)$:  When the two $\Dob$s coincide and $x_4=x_5=0$, a blowup is performed, but the previous analysis revealed two tangent directions associated to a constraint loss. That locus will be dealt with, next. In this case, two constraints are imposed on the tachyon, yielding a $\chi (\cp 0)$.
  \item $\chi_{45}\cdot \chi (\cp 1)\cdot \chi(C \mathbb{P}^1)$: This is the case, when the blowup is associated to a constraint loss, with the two  $\Dob$s coincident on a locus with $x_4=x_5=0$.
   \item $\chi_0\cdot \chi (\cp 2)\chi (\cp 1)$: When the two $\Dob$s lie on the point $x_1=x_2=x_3=0$, there is only one constraint on the tachyon, arising from the tangent directions after the blowup. Recall that there is no constraint from placing a particle at this point: this subtlety appeared already when considering the state $D4-\Dob$ on the decantic, previously.
 \item $(\chi -\chi_0)\cdot \chi_0\cdot \chi (\cp 1)$: This contribution arises, when one $\Dob$ is placed on the locus $x_1=x_2=x_3=0$ (this $\Dob$ will not impose a constraint on the tachyon), and the other one somewhere else.
 \item $\frac{1}{2}\cdot 9(\chi -\chi_0-\chi_4-\chi_5+\chi_{45})\cdot \chi (\cp 1)$: These are the cases, when the two $\Dob$s have identical coordinates $x_1,x_2,x_3$, but differ in at least one of the other coordinates, $x_4^{P_1}=e^{\frac{2\pi i j}{5}}x_4^{P_2}, x_5^{P_1}=e^{\frac{2\pi i k}{2}}x_5^{P_2}$. In these cases, there is only one constraint on the tachyon. The cases when $x_4=0$ or $x_5=0$ will be treated separately, though.
  \item $1\cdot \frac{1}{2}(\chi_4-\chi_{45})\cdot\chi (\cp 1)$: This is as the previous case, but additionally $x_4=0$.
  \item $4\cdot \frac{1}{2}(\chi_5-\chi_{45})\cdot\chi (\cp 1)$: Again, the conditions as previously, but with $x_5=0$.
 \ei
Collecting all the pieces linked to the value $0$ or $1$ for the tachyon index (up to a sign), one can state the correct index in the form
\be
   \Omega_{\textrm{exact}} =1\cdot (40'752+3'127)+2\cdot (-3'127)=37'625.
\ee
The \emph{Donaldson--Thomas partitions} $\mathcal{N}^{(1,2)}_{\textrm{DT}}(0,2)$ for the decantic thus read
\bea
   \mathcal{N}^{(g)}_{\textrm{DT}}(0,2)&=43'879&,\\
   \mathcal{N}^{(s)}_{\textrm{DT}}(0,2)&=-3'127&.
\eea
Again, note that the sum of the partitions yields $N_{\textrm{DT}}(0,2)$.

Clearly, the number $37'625$ is still very far off from the modular prediction $271'955$. The missing states cannot however be calculated at the moment. Namely, one finds a single flow for this charge system, so there is little hope of obtaining the correct index exclusively using the methods utilized in this paper. It is left as a problem for future research to enumerate the number of BPS states corresponding to this single flow, and (possibly) find and enumerate other split flow trees.

\section{Discussion}\label{sect:discussion}

In this work, we studied examples of BPS degeneracies of D--particle microstates, arising from a mixed ensemble of D--branes wrapped on algebraic one--modulus Calabi--Yau varieties. We have (presumably) found exact results, directly, and have explained and put to use a refined computational scheme, which allowed us also to find exact results for the other tractable cases. This indeed provides strong evidence for the split attractor flow tree conjecture, stating that (single and) split flows are an accurate classification of the BPS spectrum of type II string theory.

We will discuss whether one might be able to draw any conclusions about the supergravity spectrum associated to our D--particles, and in particular the topic of scaling solutions, in subsection \ref{scalingsolutions}. We will then discuss our most important results: the refined index to enumerate bound states and our computational scheme, in subsection \ref{subsect:meiosis}. We will support our discussion with a metaphoric interpretation for the need of a refined index for some polar states, along the lines of their interpretation as D--particle/black hole chromosomes given in section \ref{sect:background}.

\subsection{Absence of scaling solutions?}\label{scalingsolutions}
In this section, we will discuss some of our results (as well as results from \cite{Collinucci:2008ht}), and in particular whether they might shed any light on the mysterious nature of scaling solutions. We will start by reviewing some of their basic features for the reader. More details can be found e.g. in \cite{Denef:2007vg}.

\begin{figure}[h]
\setlength{\unitlength}{1cm}
\centering 
\begin{picture}(3.9,3.6)
\put(-1.25,0) {\includegraphics[width=0.35\textwidth,angle=0]{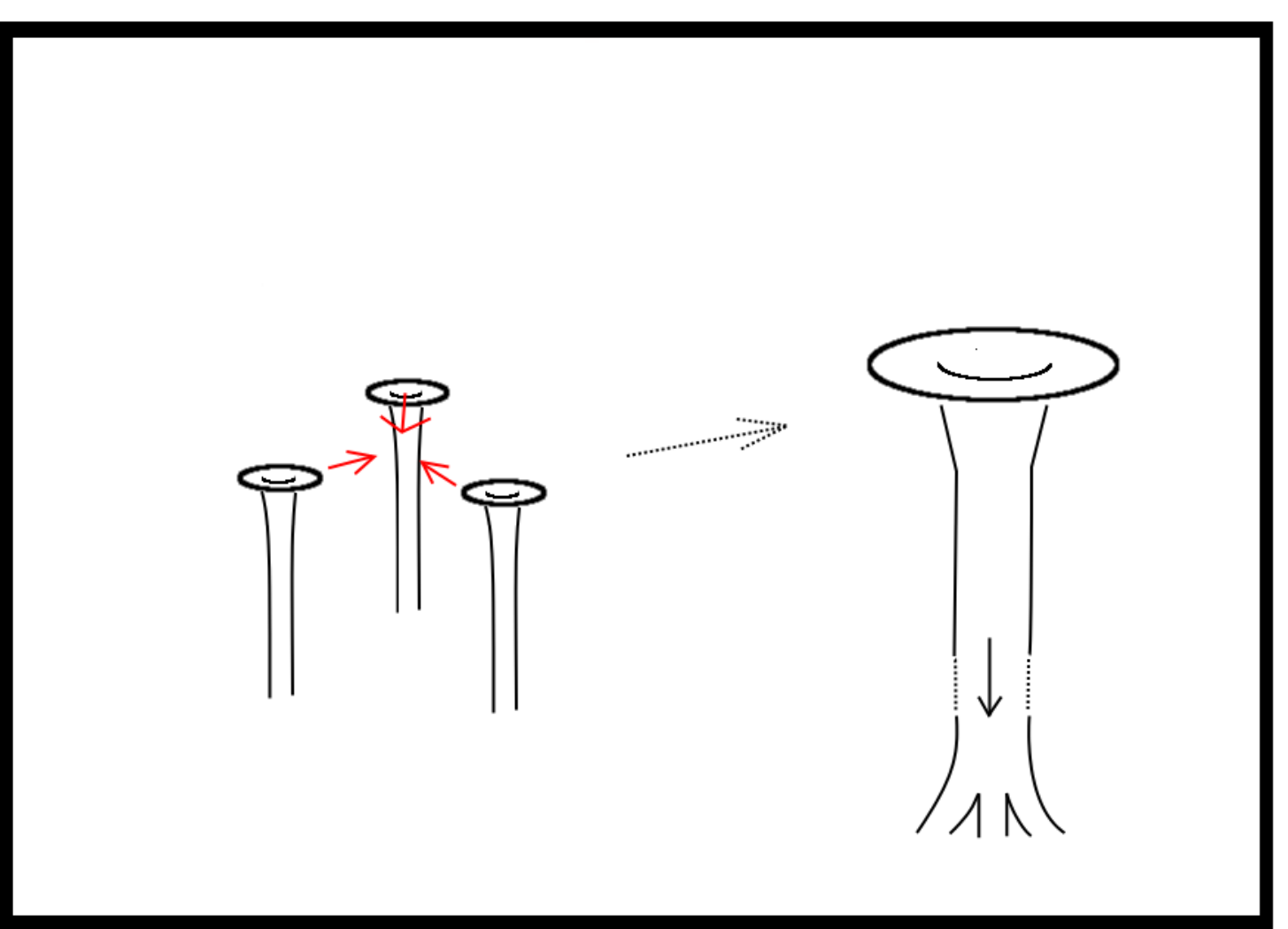}}
\end{picture}
\caption{\textbf{Scaling solution}: As the positions of several black holes come very close to each other, the throats belonging to the black holes melt together, eventually disappearing down the (infinitely long) throat of one big black hole. From the outside, it is indistinguishable from a single--centered black hole, for an observer placed somewhere down the throat it remains multi--centered.}
\label{scalingsolutionfigure}
\end{figure}

Scaling solutions for a given total charge $\Gamma$ are microstates that owe their name to the special feature that the distance between their centers is not fixed, but rather a `scaling' modulus. The appearance of scaling solutions can be understood easily using a concrete three--centered example. To make following equations transparent, the shorthand notation $\Gamma_{ij}=\langle \Gamma_i,\Gamma_j\rangle$ is introduced, as well as $h$ to label the constants appearing in harmonic functions, belonging to a multi--centered supergravity solution,
\be
    H(\vec{x})=\sum_{i=1}^N\frac{\Gamma_i}{|\vec{x}-\vec{x}_i|}+h,
\ee
where $h=-2\textrm{Im}(e^{-i\alpha}\Omega)_{\tau =0}$. One can find a scaling solution by treating  $|\vec{x}_i-\vec{x}_j|=\lambda\Gamma_{ij}$ as independent variables and sending $\lambda\rightarrow 0$: this explains why these solutions are called `scaling solutions'. The distances between the centers are not completely independent: in order for such a solution to exist, one must respect the triangle inequality,
\be\label{triangleinequality}
 \Gamma_{21}+\Gamma_{13}\geq\Gamma_{32} \qquad  + \textrm{cyclic permutations}.
 \ee
 In the limit $\lambda =0$ (or $\lambda$ infinitesimally small), the locations of the centers in spacetime become identical, and the black hole solution becomes indistinguishable from a single--centered black hole for a distant observer, whereas for an observer remaining at finite distance from the centers, the solution stays multi--centered. The interpretation is that the throats of the black holes have melted together and that the near observer has disappeared down the throat. This is illustrated in figure \ref{scalingsolutionfigure}.

We will now discuss some microstates of the D--particles investigated in this paper (as well as an example from \cite{Collinucci:2008ht}), for which one would expect scaling solutions (ignoring regimes of validity of the description for the moment) in the supergravity approximation. The condition for existence of scaling solutions for a D4--D2--D0 D--particle state, (\ref{triangleinequality}), consisting of a fluxed D6--brane, $\Gamma_1=e^{H}(1+\frac{c_2(X)}{24})$, a $\Dsb$--brane, $\Gamma_2=-(1+\frac{c_2(X)}{24})$, and $N$ $\Dob$'s, $\Gamma_3=(0,0,0,-N)$ reads
\bi
 \item $N\geq 2$ for the sextic: in particular, the charge labeled by $\Delta q=0,\Delta q_0=-2$ should allow for a scaling solution. Note that the $40'392$ BPS states we found should be the exact number of BPS quantum states (at least it matches with the prediction from modularity).
 \item $N\geq 2$ for the octic: again, the charge system $\Delta q=0,\Delta q_0=-2$ fulfills this requirement. Note that the $86'140$ BPS states should be exact.
    \item $N\geq \frac{3}{2}$ for the decantic: in this case, the charge system $\Delta q=0,\Delta q_0=-2$ should support scaling solutions. At this moment, we cannot address (at least predict and check) the number of BPS states conclusively.
     \item $N\geq \frac{5}{2}$ for the quintic, studied in \cite{Collinucci:2008ht}: this implies that the charge system $\Delta q=0,\Delta q_0=-3$ falls into this category. The number of $5'817'125$ BPS states found in \cite{Collinucci:2008ht} should be exact.
\ei

There are several possibilites: it might be, that some BPS states in the quantum spectrum might correspond to classical BPS states which are scaling solutions. Another is that scaling solutions exist, but are just not `valid' classical microstate descriptions of the D--particle, as they do not correspond to any quantum states. What is meant by `correspond'? The authors of \cite{deBoer:2008zn} studied the quantization of the phase space of smooth supergravity solutions. In well understood situations (type IIB compactifications and the Strominger--Vafa D1--D5--P black hole), it can be shown that the quantization of the phase space of smooth supergravity solutions (thus a restriction to the states carrying quantum numbers) is in one--to--one correspondence with the BPS microstates in the D--brane description. Microstates on the gravity side arise as wavefunctions, which localize on a unit volume of phase space. A classical solution can then be interpreted as the limit of this localized wavefunction. It can however be the case that there are quantum states in the spectrum, which do not localize on such a unit of phase space, and it is questionable if these states have a reasonable classical description. On the other hand, the authors also argue, that `nominally' classical solutions might occupy the same volume in quantized phase space (arise as limits of a state localized in the same volume), although they differ on a macroscopic scale in their gravitational description. This cannot be reasonable when taking into account Heisenberg's uncertainty principle, and thus `classical solutions' of this kind might not be `valid' geometries of the black hole. They might just be a peculiarity without physical significance, found in supergravity. This is exactly what the authors claim to be the case for scaling solutions. Thus, scaling solutions might not be valid classical limits of black hole microstates. Nevertheless, an index accounting for scaling solutions based on the quantization of phase space was constructed in \cite{deBoer:2008zn}, and this might help when trying to match classical and quantum BPS states.

For three of the four listed charge systems, the number of BPS states in the spectrum can be predicted by modularity, and the methods used in this paper allow for exact results to be derived, based on the prescriptions for refined calculations. One is led to believe that one has indeed exhausted the BPS spectrum of the quantum theory, in the sense that one has classified and enumerated all BPS microstates of the corresponding D--particles, carrying the desired charge. At the same time, this means that the split attractor flow conjecture worked perfectly. For none of the mentioned charge systems is there seemingly room for states, which would be described by scaling solutions, after taking a classical limit. This can be concluded from the fact, that an index (as suggested in \cite{deBoer:2008zn}) for scaling solutions seems to yield results that strongly deviate from any numbers contributing to the total number of BPS states one would expect. It seems reasonable to suppose that the exact numbers for BPS states of the D--particles investigated can be interpreted as an indication that the interpretation of \cite{deBoer:2008zn} on scaling solutions is indeed correct. It would be wrong however to make an overstatement, as these implications do remain quite speculative, and there are no properly legitimated methods to compute an `index for scaling solutions', taking into account the regime of validity needed in this case. One should at this point remain open to other ideas, one of which might be the wild proposal that `special states' could be quantum partners of scaling solutions. The authors believe this to be rather unlikely, but further investigation on this exciting topic remains for the future.

Scaling solutions might not be good classical limits of quantum microstates, or one might not understand them properly, yet. Maybe --- being very careful not to make any overstatement --- it is too much to hope for, that split attractor flow trees classify the full BPS spectrum. In any case, the split attractor flow tree conjecture is extremely accurate for a large number of BPS states in supergravity, and also classifies BPS bound states outside of the regime of supergravity: this claim was successfully put to various non--trivial tests in this work.

\subsection{The meiosis of D--particles/black holes}\label{subsect:meiosis}

Our main results involve a scheme for computing a refined index enumerating BPS bound states, and in connection with this we have calculated partitions of Donaldson--Thomas invariants, resulting from the distinction of constituent states as based on the perception of the tachyon fields gluing together a bound state. Basically, the fact that one needs to use a refined index to enumerate BPS bound states, results from the fact that the constituents are not fully independent, there is some sort of `interaction' between the constituents for which we would like to give a nice little metaphoric interpretation, using the idea to call polar states chromosomes of D--particles/black holes, as stated in the introduction of this paper.

The degeneracy of a polar state is naively believed to factorize into three factors, the degeneracy of the first `parent' constituent, the degeneracy of the second `parent' constituent, and the index accounting for the tachyonic degrees of freedom, gluing the state together. Although this analogy again has many shortcomings, one might --- when thinking of the suggested metaphor ---  count the number of possible outcomes of meiosis, which produces the four haploid gametes, necessary for the sexual reproduction of eukaryotic organisms. Naively, one might expect there to be only one outcome (with half of the gametes containing the father chromosome and the other half the mother chromosome). During the meiosis however, crossing--over (intrachromosomal recombination) takes places, referring to the process of exchange of pieces of the DNA on pair chromosomes (a sort of `interaction' between chromosomes). This means that the number of possibilities does not factorize. For our analogy, the condensation of two higher--dimensional branes is the parallel for the reductional division of two chromosomes leading to a (haploid) cell, a `gamete'. For a BPS bound state, constituent branes of course do not really perform a crossing--over, but the number of tachyonic strings (whose specific existence does depend on both of the branes) gluing the state together jumps. Still, this interpretation is meant to stress the point which the authors believe to be important: namely, that the constituents of a bound state are not independent, but in general \emph{do} perform some interaction. This can be taken into account when calculating indices by using the refined prescription, used in this paper.

\begin{figure}[h]
\setlength{\unitlength}{1cm}
\centering 
\begin{picture}(4.75,5)
\put(0,0) {\includegraphics[width=0.4\textwidth,angle=0]{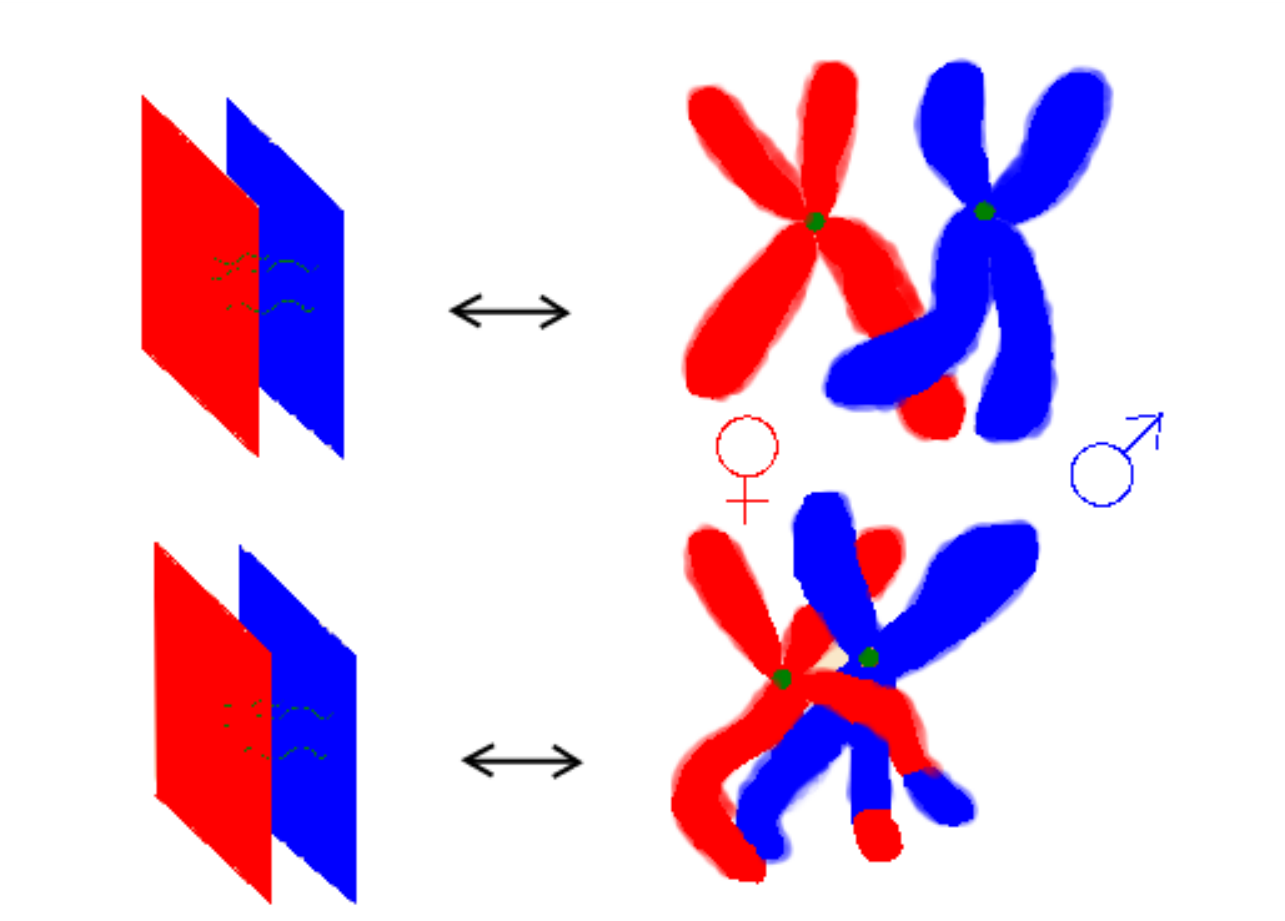}} 
\end{picture}
\caption{\textbf{Polar states, non--factorization and crossing--over:} \emph{1. Top}: the naive factorization of states. \emph{2. Bottom}: The number of gluing strings between two branes is not independent of the states in which the two branes reside. This ruins factorization of the total number of states for such a polar term. In meiosis, homologous pairs of chromosomes are lined up and can exchange pieces of their DNA (intrachromosomal recombination). This phenomenon is called crossing--over and is depicted for the lower pair of chromosomes. This means that the number of (possibilities to express the) genes, passed on to the next generation does not factorize between homologous chromosome pairs.}
\end{figure}

Finally we would like to state some implications and future directions of reseach suggested by our results:
\begin{enumerate}
   \item It would be extremely appealing to be able to find more certainty for the correctness of our new prediction of the elliptic genus for the decantic hypersurface in $\wpr{4}{11125}$. This would involve enumerating the remaining BPS states, amongst which those corresponding to a single flow.
 \item Probably, special loci as discussed in this paper will appear even more prominently, when working with Calabi--Yau varieties embedded in toric ambient spaces. As this feature gains more importance, it would be interesting to study tractable examples, and see how important this effect is. Apart from that, it would also be nice to predict elliptic genera for higher class divisors (also for one--modulus CY's), and also see what implications the refined BPS bound state index computations have for those examples. A goal is to develop the prescriptions for refined bound state indices to full generality, which includes constraint loss arising from curves and particles on all constituents of a bound state.
 \item Additionally, it is a question what the implications are for a factorization formula for BPS indices, and also of the OSV conjecture. The factorization of BPS states is only possible on the level of densities. Thus the question is whether one can also write down a closed formula in the spirit of OSV for a D--particle / black hole partition function, on the level of index densities.
\item Considering our interest in Donaldson--Thomas partitions, it would be interesting to study partition functions built from Donaldson--Thomas partitions. Given the fact that only three divergent asymptotic series (amongst which the Donaldson--Thomas partition function) approximating the topological string partition function are known, it would be interesting to study the convergence behavior of such a `partition function of Donaldson--Thomas partitions'.
 \item Using our metaphoric interpretation of chromosomes, one might call the quest for finding more elliptic genera for D--particles and eventually of course black holes, the black hole genome project. It would seem an interesting task to map mixed ensembles to IIB string theory, lift them to five dimensions, and in this way strengthen the link between the formalism of 4d polar states (the chromosomes of black holes), and the fuzzball program in 5d. A first step in the direction of mapping polar states to fuzzball microstates in 5d was taken in \cite{Raeymaekers:2008gk}, for BPS polar states in $\mathcal{N}=8$ (4d) supergravity theories.
\end{enumerate}

\acknowledgments{We would like to thank Andr\'es Collinucci for numerous helpful discussions. This work  is supported in part by the European Community's Human Potential Programme under contract MRTN-CT-2004-005104 �`Constituents, fundamental forces and symmetries of the universe', in part by the FWO--Vlaanderen, project G.0235.05, in part by the Federal Office for Scientific, Technical and Cultural Affairs through the Interuniversity Attraction Poles Programme Belgian Science Policy P6/11-P.}

\begin{appendix}

\section{Setup and conventions}\label{sect:setup}
As all Calabi--Yau 3--folds $X$ used in this paper are one--modulus examples, $h^{(1,1)}(X)=1$, we use $H$ to denote the basis element of $H^2(X,\mathbb{Z})$. We also use the symbol $H^{2*}(X,\mathbb{Z})$, or for short, $H^{2*}(X)$ to denote the even degree cohomology of $X$:
\be
  H^{2*}(X)=H^0(X)\oplus H^2(X)\oplus H^4(X)\oplus H^6(X).
\ee
For a type IIA string theory compactification, a vector bundle with fibers $H^{2*}(X)$ over the K\"ahler moduli space is defined, by virtue of the properties of \emph{special geometry}. These fibres are naturally endowed with a symplectic product (Dirac--Schwinger--Zwanziger), $\langle \cdot , \cdot \rangle :H^{2*}\times H^{2*}\rightarrow\mathbb{C}$, for $\Gamma_1,\Gamma_2\in H^{2*}(X)$:
\be
  \langle \Gamma_1, \Gamma_2\rangle =\int_X\Gamma_1\wedge\Gamma^{*}_2
\ee
where it is understood that only the wedge products of the right dimension remain as an integrand, and $\Gamma^{*}_2$ is defined under an operation inverting the sign of the two--form and the six--form components: $\Gamma=\Gamma^0+\Gamma^2 H+\Gamma^4H^2+\Gamma^6H^3\rightarrow\Gamma^{*}=\Gamma^0-\Gamma^2 H+\Gamma^4H^2-\Gamma^6H^3$.

The IIA complexified K\"ahler form will be written as\footnote{We will also refer to $t$ just as the \emph{K\"ahler modulus}.} $tH=(B+iJ)H$ using our basis, and allows us to form the holomorphic period vector
\be    \Omega_{\textsl{hol}}:=-\textrm{exp}(tH)=-1-tH-\frac{t^2}{2}H^2-\frac{t^3}{6}H^3.\nonumber
\ee
The K\"ahler potential of the K\"ahler moduli space reads $$K=-\textrm{ln}\left( i\langle\Omega_{\textrm{hol}},\overline{\Omega}_{\textrm{hol}}\rangle\right) =-\textrm{ln}\left( \frac{4}{3}\int_X(J\wedge J\wedge J)\right) ,$$
which also allows us to define the normalized period vector
\be\label{periodvector}
   \Omega :=\textrm{exp}(\frac{K}{2})\Omega_{\textrm{hol}}.
\ee

The general formula for the induced charges on a D--brane (due to the WZ term in the Born--Infeld action) wrapped on a (sub)--manifold $W$ reads
\begin{equation}
 S^{\rm Dbrane}_{W,C} = 2 \pi \int_W C \wedge e^{-B} \, {\rm Tr} \, e^F
 \sqrt{\frac{\widehat{A}(TW)}{\widehat{A}(NW)}},
\end{equation}
where $\widehat A$ is the A--roof characteristic class, $TW$ the tangent bundle of the brane, and $NW$ its normal bundle. From this formula, one obtains for a D6--brane carrying U(1)--flux with field--strength $F_1$, a $\Dtb$ of class $-\beta_1$ and $N_1$ $\Dob$'s the following polyform:
\begin{equation}
\Gamma_{D6} = e^{F_1}\,\Big(1-\beta_1-(\tfrac{1}{2}\,\chi(C_{\beta_1})+N_1)\,\omega \Big)\,\Big(1+\frac{c_2(X)}{24}\Big)\,,
\end{equation}
where $\beta \in H^4(X, \mathbb{Z})$, and $c_2(X)$ is the second Chern class of the tangent bundle of the CY threefold $X$. Similarly, a $\Dsb$ with flux $F_2$ will bind to a D2 of class $\beta_2$ and $N_2$ $\Dob$'s to give the following total charge vector:
\begin{equation}
\Gamma_{\Dsb} = -e^{F_2}\,\Big(1-\beta_2+(\tfrac{1}{2}\,\chi(C_{\beta_2})+N_2)\,\omega \Big)\,\Big(1+\frac{c_2(X)}{24}\Big)\,,
\end{equation}
The modification with respect to the general formula is the addition of D2 and D0 charge in the form of sheaves, which can be thought of as generalizations of bundles (U(1) fluxes). Notice that the D6 will bind with a $\Dtb$, the $\Dsb$ with a D2, but both will bind to $\Dob$'s.

For a D6--D4--D2--D0 brane system, we denote the charges as $(p_0,p,q,q_0)$, such that a polyform using our basis can be written as
\be
  \Gamma = p_0+pH+\frac{q}{\mathcal{H}}H^2+\frac{q_0}{\mathcal{H}}H^3,
\ee
where $\mathcal{H}:=\int_X H^3$. We will equivalently use a vector notation: $\Gamma \equiv(p_0,p,q,q_0)$.

The central charge of a D--brane system $\Gamma$ is defined using the period vector \eqref{periodvector},
\be
    Z=\langle\Gamma ,\Omega\rangle =\textrm{exp}(\frac{K}{2})\left( p_0\frac{\mathcal{H}}{6}t^3-p\frac{\mathcal{H}}{2}t^2+qt-q_0\right ) ,
\ee
The central charges of brane systems under investigation will be used for establishing the existence of single/split flows and thus of BPS states, as will be explained in subsection \ref{subsect:splitflows}.

\section{Modified elliptic genera and D--particles}\label{sect:ellGenera}

Modified elliptic genera arise as supersymmetric BPS indices of the $(0,4)$ Maldacena--Strominger--Witten CFT \cite{Maldacena:1997de}. A non--linear sigma model realization of this CFT has been derived from the dimensional reduction of the M5--brane worldvolume perspective. In this paper, we will work from the IIA perspective.
In the following, we use a basis $\Sigma_A \in H_4(X,\mathbb{Z})$ and define the intersection numbers as $6D_{ABC}$. The definition $D_{AB}\equiv D_{ABC}p^C$, its inverse $D^{AB}$ and $D\equiv D_{ABC}p^Ap^Bp^C$ will also come in handy. Using mixed ensembles of D4--D2--D0 branes with fixed magnetic D4--brane charge $p^A$ and variable electric charges ($q_A,q_0$), the central charges of the CFT read
\be
   c_L=P^3+c_2\cdot P,\quad c_R=P^3+\frac{1}{2}c_2\cdot P,
\ee
where $c_2$ is the second Chern class of the Calabi--Yau $X$ and $P$ is the Poincar\'e dual of the hypersurface $p$. The modified elliptic genus is defined as 
\be\label{modifiedellipticgenus}
   Z(q,\bar{q},y)=\textrm{Tr}_R\left( \frac{1}{2}F^2(-1)^Fq^{L_0-\frac{c_L}{24}}\bar{q}^{\bar{L}_0-\frac{c_R}{24}}e^{2\pi i y^Aq_A}\right) ,
\ee
and is constrained to be a \emph{weak Jacobi form} of weight $(-\frac{3}{2},\frac{1}{2})$. The flux on the D4--brane is an element of $H^2(P,\mathbb{Z})+\frac P 2$, where the $\frac P 2$ factor accounts for the Freed--Witten anomaly. Each flux can be decomposed as follows
\be\label{fluxdecomposition}
   F=\frac{P}{2}+f^{||}+f^{\perp}+\gamma ,
\ee
where $f^{||}\in L_X\equiv \imath^*H^2(X,\mathbb{Z})$ and $f^{\perp}\in L_X^{\perp}$, and $\gamma$ is given by gluing vectors in the flux lattice. The reader can consult \cite{Denef:2007vg} for more details. The modified elliptic genus can be decomposed as
\begin{equation}\label{ellipticgenusmodvector}
   Z(q,\bar{q},y) = \sum_{\gamma } Z_\gamma (q)\, \Theta_\gamma (q,\bar{q},y)\, ,
\end{equation}
where $\gamma$ runs through a finite set of gluing vectors. The $Z_{\gamma}(q)$ are meromorphic functions of the variable $q$ and the theta functions are so--called Siegel--Narain theta functions:
\be\label{gammathetafunctions}
   \Theta_{\gamma}(q ,\bar{q},y)=\sum_{q_A(\gamma)}(-1)^{p^A q_A}q^{\frac{1}{12}\left(\frac{p^Ap^B}{D}-D^{AB}\right) q_Aq_B} \bar{q}^{-\frac{1}{12}(p^Aq_A)^2}e^{2\pi i y^Aq_A}
\ee
where the summation over $q_A$ is understood to be $q_A=6D_{AB}(\frac{1}{2}p^B+k^B)+\gamma_A,\; k^A\in \mathbb{Z}^n$ and $n$ is the dimension of the flux lattice.

The dimensionality of the vector $Z_{\gamma}$ is thus given by the number of independent elements $\gamma$ of the discriminant group, the gluing vectors. Additionally, as discussed in \cite{Moore:1998pn,Moore:1998zu,Denef:2007vg,Manschot:2008zb}, it follows from modular invariance, that there is an identification
\be\label{gammaidentification}
    Z_{\gamma}=Z_{\delta},\qquad \textrm{for} \qquad \gamma =-\delta\;\textrm{mod}\; L_X.
\ee

The $Z_{\gamma}$'s can be given some physical interpretation: $Z_0$ corresponds to a sum over states with no added D2--charge, and increasing D0--charge as the powers of $q$ increase. Each coefficient in the $q$--expansion corresponds to the index of a state with fixed D0--charge. Similarly, the $Z_{\gamma}$'s correspond to states with added D2--charge\footnote{Note that by spectral flow, which transforms $q_A\rightarrow q_A-6D_{AB}k^B$ and $q_0\rightarrow q_0+k^Aq_A-3D_{AB}k^Ak^B$, most of the D2--charge can be absorped in $q_0$}. Schematically, the first few terms of $Z_{\gamma}$ will look as follows
\begin{equation}
Z_{\gamma}(q) = q^{-\alpha} (\#+ \# \, q+ \# \, q^2 + \ldots)\,,
\end{equation}
where $\hat{q}_0 \equiv q_0-\frac{1}{12}D^{AB}q_Aq_B$, and $\alpha$ is the highest possible value of $\hat{q}_0$ for a given $\gamma$. In this paper, only the $Z_{\gamma}(q)$ functions will be of concern, as they contain all the relevant information. A stringent mathematical property of weak Jacobi forms is the fact that they are entirely determined by their \emph{polar part}, i.e. terms with negative powers of $q$. These terms correspond to charge configurations that satisfy $\hat{q}_0=q_0-\frac{1}{12}D^{AB}q_Aq_B>0$. Such configurations will be referred to as \emph{polar states}. To construct elliptic genera, we used the method of generating modular representations as in the appendix of \cite{Gaiotto:2007cd}.

\section{The split attractor flow tree conjecture}\label{subsect:splitflows}
In the following, the reader will find a very brief presentation on the use and meaning of split flow trees. By the split attractor flow tree conjecture, \cite{Denef:2007vg}, single flows and split flow trees are believed to be an existence criterion and provide a complete classification\footnote{The statement that they classify all BPS states in string theory is sometimes referred to as the \emph{strong} form of the conjecture, while one can also formulate the weaker conjecture for BPS states in supergravity.} for BPS states in type II string theory. Single/split flows are graphical depictions of the flow of the K\"ahler moduli (or complex structure moduli in the mirror type IIB picture) belonging to a BPS solution of supergravity, but their meaning extends beyond the range of validity of the supergravity approximation to type II string theory, as suggested by the smooth interpolation between the supergravity picture and the quiver description of D--brane bound states found in \cite{Denef:2002ru}.\\
\\
In order to incorporate the use of split attractor flow trees also outside of the regime of validity of the supergravity approximation to string theory, the following definition of a split flow tree is given:\\
\\
A \textbf{split flow tree} belonging to the total charge $\Gamma$ and $n$ constituents $(\Gamma_1,...,\Gamma_{n})$ is a set of data
\begin{equation}
  (t_{\infty};t_{1, \textrm{\tiny{split}}},...,t_{n-1, \textrm{\tiny{split}}};t_{1 *},...,t_{n *}), \label{dataset}
\end{equation}
consisting of $2n$ points, a background value $t_{\infty}$, a set of $n-1$ split points $t_{j, \textrm{\tiny{split}}}$ ($j=(1,..,n-1)$), and n attractor points $t_{j *}$ ($j=1,...,n$), one for each center. Using this notation, a single flow, for example, is denoted as $(t_{\infty};t_{*})$. Note that this definition boils down to the most essential features, but the intuition, that the flow tree depicts the values of the scalars belonging to a supergravity solution in moduli space, remains valuable, when constructing split flow trees.\\
\\
\textbf{The construction of a split flow tree}\\
\\
A split flow tree is built as follows. One follows the \emph{incoming branch}\footnote{The term `incoming branch' refers to the part of the flow tree connecting the background point and the first split point.} of a flow tree from radial infinity towards a putative attractor point, until one hits a \emph{wall of marginal stability} for two \emph{non--local}\footnote{If one starts with local constituents, $\langle\Gamma_1,\Gamma_2\rangle =0$ (such as two branes of the same type), they are trivially mutually BPS, and one speaks of a threshold wall instead of a wall of marginal stability.} constituents $\langle\Gamma_1, \Gamma_2\rangle\neq 0$ such that $\Gamma=\Gamma_1+\Gamma_2$. This is defined as the hypersurface in moduli space where the phases of the central charges align, $\textrm{arg}(Z_1)=\textrm{arg}(Z_2)$. The modulus of a central charge measures the mass of the corresponding state, whereas the phase indicates which $\mathcal{N}=1$ supersymmetry of the original $\mathcal{N}=2$ supersymmetry is preserved by the state. If the phases of two central charges align, the two states are mutually BPS (i.e. preserve the same supersymmetry) and the binding energy of the BPS bound state vanishes,
\begin{equation}
  |Z_{1+2}|=|Z_1|+|Z_2|,
\end{equation}
or equivalently,
\begin{equation}
  \textrm{Re}(\bar{Z}_1Z_2)>0,\qquad \textrm{Im}(\bar{Z}_1Z_2)=0.
\end{equation}
If one reaches the wall from the side where $\langle\Gamma_1,\Gamma_2\rangle (\textrm{arg}(Z_1)-\textrm{arg}(Z_2))>0$, the decay of $\Gamma \rightarrow \Gamma_1+\Gamma_2$ is energetically favored. Microscopically, the non--zero value of $\langle\Gamma_1,\Gamma_2\rangle$ means that there are chiral strings stretched between the two constituent branes, which make the decay (or the recombination if one goes in reverse) possible. One then follows the flows of the constituents, which might decay again according to the same scheme, until every end branch flows towards an attractor point. This is illustrated by the following figure for a three--centered solution:
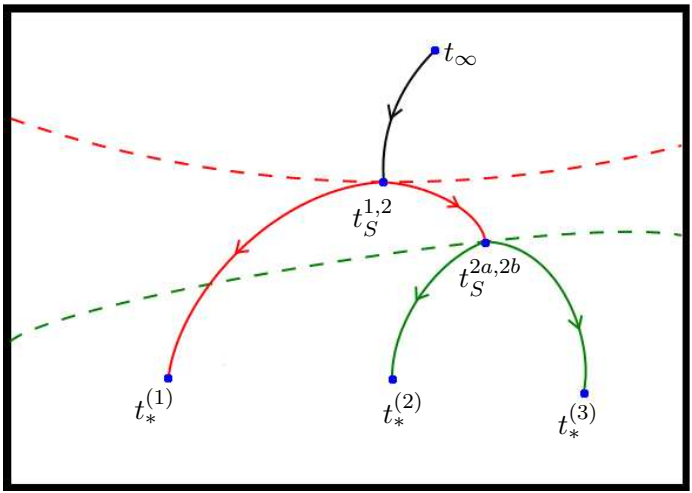
\begin{figure}[h]\label{splitflowBASICfigure}
\setlength{\unitlength}{1cm}
\centering 
\begin{picture}(7.5,6)
\put(-1,-0.2) {\includegraphics[width=0.6\textwidth,angle=0]{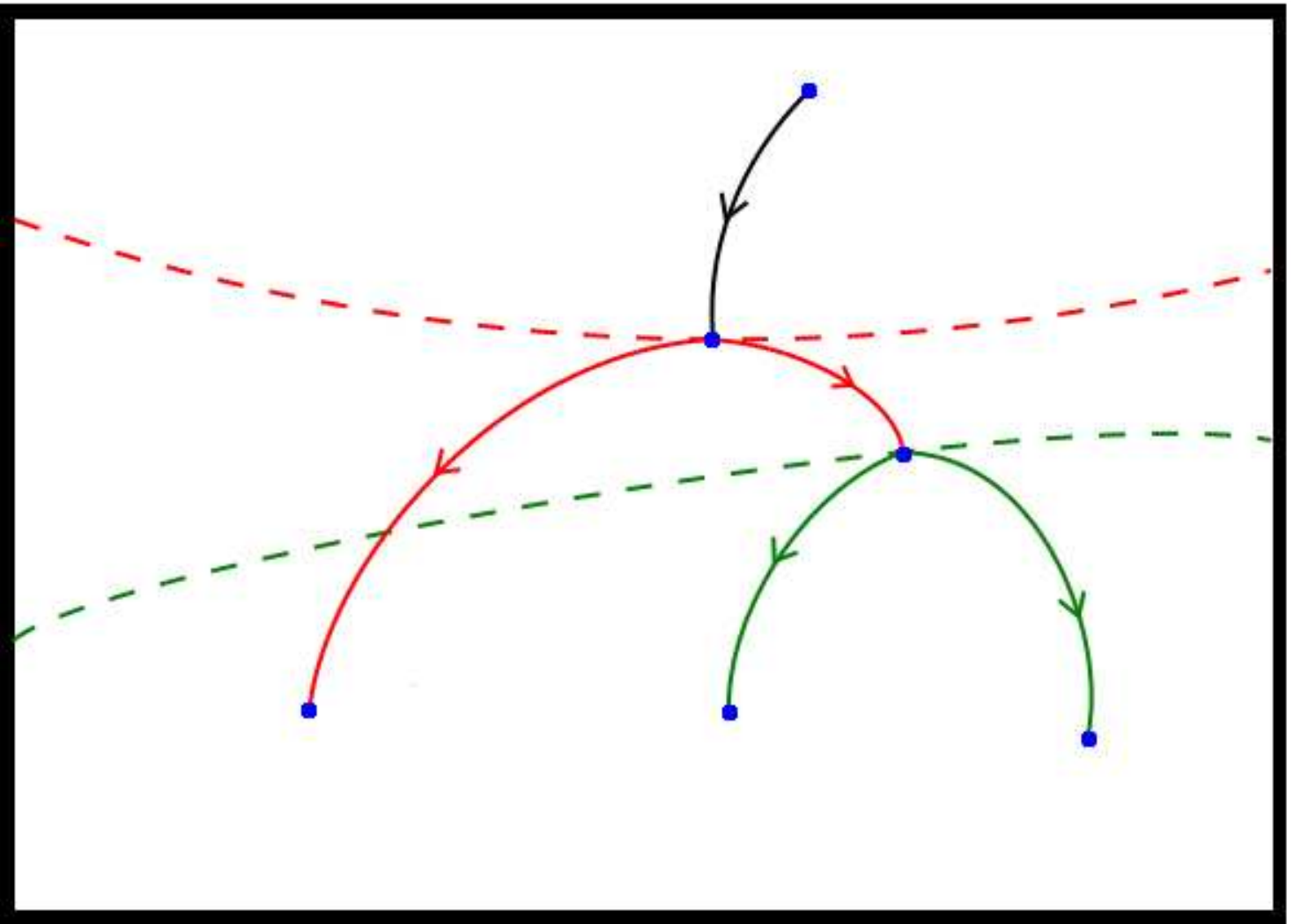}}
\put(4.8,5.5) {$t_{\infty}$}
\put(3.6,3.3) {$t_S^{1,2}$}
\put(5,2.5) {$t_S^{2a,2b}$}
\put(0.73,0.75) {$t_{*}^{(1)}$}
\put(4,0.7) {$t_{*}^{(2)}$}
\put(6.3,0.55) {$t_{*}^{(3)}$}
\end{picture}
\caption{\textbf{Split flow tree for a three--centered solution with charges $\Gamma =\Gamma_1+\Gamma_{2a}+\Gamma_{2b}$}: One starts at the background point, $t_{\infty}$, at the top, and follows the incoming branch (plotted in black) until one hits a wall of marginal stability (plotted in red) between $\Gamma_1$ and $\Gamma_2$. This is the first split point $t_S^{1,2}$, wherefrom single flows corresponding to the two centers continue (both branches are plotted in red). The first center reaches an attractor point, $t_{*}^{(1)}$, whereas the second branch again reaches a wall of marginal stability (plotted in green) at $t_S^{2a,2b}$. At this point, the charge $\Gamma_2$ splits into $\Gamma_{2a}$ and $\Gamma_{2b}$. Two branches, corresponding to these two charges, are plotted in green: they flow to their attractor points, $t_{*}^{(2)}$ and $t_{*}^{(3)}$.}
\end{figure}

For a given total charge $\Gamma$ one might in general find several different split flow trees, and maybe also a single flow, all contributing to the total index of BPS states with this total charge. In general, the BPS spectrum of states $\Omega(\Gamma)$ remains invariant under infinitesimal variations of the background moduli $t_{\infty}$, but it can jump when the moduli are driven through a wall of marginal stability. This is intuitively clear; a certain split might either not be possible anymore because one can only reach the appropriate wall of marginal stability from the unstable side, or, alternatively, a new type of split becomes possible, as one can now reach this wall from the stable side. In such a case one has taken the background into a different \emph{area}, and the fact that there are different basins of attraction in moduli space has led to the name \emph{area code} for the background. Hence, an index of BPS states should more precisely be denoted as $\Omega(\Gamma,t_{\infty})$.\\
\\
\textbf{Threshold walls}\\
\\
Apart from the walls of marginal stability, which separate regions in moduli space between which a BPS index can jump, there is a second type of wall that is of importance for our work: walls of \emph{threshold} stability. The distinction between these two kinds of walls is explained in more detail in \cite{deBoer:2008fk}. The threshold conditions are the same as for marginal stability
\begin{equation}
  \textrm{Re}(\bar{Z}_1Z_2)>0,\qquad \textrm{Im}(\bar{Z}_1Z_2)=0\,,
\end{equation}
however this time, the two charges are mutually local $\langle\Gamma_1,\Gamma_2\rangle =0$. Microscopically this means that there are no tachyonic strings between the two branes that can condense to merge $\Gamma_1$ with $\Gamma_2$. Most importantly, a BPS index cannot jump when crossing the threshold stability wall with the background modulus. What can happen is that the flow tree changes topology: one can imagine, for example, a three--centered solution, with one `satellite' center bound to either of the two others, which then changes sides and is bound to the other center after crossing the threshold wall with the background modulus.\\
\\
This allows the formulation of the split attractor flow tree conjecture (\cite{Denef:2007vg}):\\
\\
\textbf{The split attractor flow tree conjecture} (strong version)
\begin{enumerate}
   \item For a given background $t_{\infty}$ in K\"ahler/complex structure moduli space, the existence of a split attractor flow tree starting at the background $t_{\infty}$, with a given total charge $\Gamma$ and endpoints corresponding to $\Gamma_i$, is equivalent to the existence of a BPS bound state in string theory, with constituents $\Gamma_i$.
\item The number of split flow trees and hence the total number of states with a given charge $\Gamma$ in a fixed background is finite, at least when charge quantization is imposed.
\end{enumerate}
As will be discussed later on, our results offer support for a strong version of the conjecture.\\
\\
In this work, we use split flow trees to establish existence of BPS states in a regime where supergravity cannot be trusted. As the D--brane charges used are very low, the attractor equations will drive the horizon size to be very small with respect to the string scale. Thus, the supergravity approximation breaks down, and higher curvature corrections from perturbative $\alpha'$ effects become important. At the same time, the cycles on which the branes are wrapped also become very small, and worldsheet corrections to the central charges of the considered brane systems become dominant. Luckily, one can calculate exact central charges in the mirror picture, as the periods are exactly determined by classical geometry. The periods are determined as solutions to the Picard--Fuchs differential equations and cannot be written down and used analytically. Therefore, we used numerical techniques to tackle this problem. These techniques were developed in \cite{Denef:2001xn}, and are partially based on the techniques prepared in \cite{Greene:2000ci}. They were also used in the same way as in this work in \cite{Collinucci:2008ht}. The reader will find more explanations in appendix \ref{numericalmethods}.

\section{Methods to establish split flows outside the supergravity regime}\label{numericalmethods}
For configurations with low D--brane charges, the attractor flow equations drive the horizon size of solutions to very small sizes in string scale units. This automatically leads far outside the supergravity regime, requiring higher curvature corrections. However, as stated in the main text, the main tool of analysis, namely split attractor flow tree techniques, retains its meaningfulness as predicted by the strong split attractor flow conjecture. As the attractor equations will typically drive the cycles and the CY itself to stringy sizes, the central charges of brane systems receive important worldsheet instanton corrections as a consequence, and these need to be taken into account. In type IIA string theory this would be an impossible task. Luckily, mirror symmetry solves the problem as the central charges are exactly determined classically by the periods of the holomorphic three--form in type IIB string theory on the mirror CY manifold. The mass of a BPS saturated brane wrapping an even cycle corresponds exactly to the notion of quantum volume of that cycle from \cite{Greene:2000ci}. The general scheme is thus to identify an integral basis of three--cycles, calculate the periods of these cycles, find the explicit mirror map and in this way define the quantum volume of any even dimensional cycle of the mirror.

The periods of the mirror Calabi--Yau's are determined as solutions of the Picard--Fuchs equation, and are known as Meijer G--functions. Thus, it is not possible to write down analytic formulae for central charges, and certainly not possible to work with split attractor flow trees analytically. This hurdle is overcome in the following by means of numerical approximations with Mathematica\footnote{Special thanks to F. Denef for sharing his Mathematica code, \cite{Denef:2001xn}.}. The periods of the mirror CY are evaluated, and a lattice of points is created, from which the function can be approximated by interpolation. Split flow trees and single flows are then established numerically. More details on the applied technique can be found in \cite{Denef:2001xn}. The mirror symmetry induced monomial--divisor map is used to convert K\"ahler to complex structure modulus, and map $(D6,D4,D2,D0)$ brane systems $\Gamma_A$ into their $(D3,D3,D3,D3)$ brane mirrors $\Gamma_B$, $L:H^{2*}(X,\mathbb{Z})\rightarrow H^3(Y,\mathbb{Z})$, and then analyze the attractor flows of the exact central charges in complex structure moduli space (or more precisely in the n--fold cover w--plane).
In the IIB picture, BPS black holes are made of D3--branes wrapped along special Lagrangian three--cycles of the internal Calabi--Yau manifold, whereby these D3's (and their corresponding three--cycles) can split up into intersecting D3's, by moving in the complex structure moduli space of the CY across some `line of marginal stability'.

The mirror map $L$ can be found from a comparison of the IIA and IIB periods near the Large Complex Structure (LCS) point. It relates the even cycles of real dimension $2j$ on the CY $X_A$ to the three--cycles $\gamma_{i}$ of the mirror CY manifold $X_B$. The periods $\Pi_i = \int_{\gamma_i} \Omega$, of the holomorphic three--form on the $\gamma_{i}$ have leading logarithmic behavior $\textrm{log}^j(z)$ near $z=0$ (LCS point), using the coordinate $z=\psi^{-6}$, $\psi^{-8}$ and $\psi^{-10}$ on the mirror moduli space, for the sextic, the octic, and the decantic, respectively. The periods are solutions to the Picard--Fuchs equation,
\begin{equation}
  \Big{[}z\,\partial_z \prod_{i=1...q}(z\,\partial_z +\beta_i-1)-z\prod_{j=1...p}(z\,\partial_z+\alpha_j)\Big{]}u=0,
\end{equation}
where the $\alpha_j$ and the $\beta_i$ are model dependent constants, which read $\beta_i=1$, $i=1, 2, 3$, and $\alpha =(\frac{1}{6},\frac{1}{3},\frac{2}{3},\frac{5}{6})$ for the sextic, $(\frac{1}{8},\frac{3}{8},\frac{5}{8},\frac{7}{8})$ for the octic, and $(\frac{1}{10},\frac{3}{10},\frac{7}{10},\frac{9}{10})$ for the decantic hypersurface, used in this paper. The Meijer--functions (G--functions) $U_j(z)$ can be expressed as
\begin{equation}
  U_j(z)=\frac{1}{(2\pi i)^j} \oint \frac{\Gamma (-s)^{j+1}\prod_{i=1}^4\Gamma(s+\alpha_i)((-1)^{j+1}z)^s}{\Gamma (s+1)^{3-j}} ds.
\end{equation}
This particular basis of periods is related to three branching points (LCS point, conifold point and Gepner point) which are connected by appropriately chosen branch cuts.

Associated to these branch cuts are three types of monodromies, which can be expressed as matrices acting on the periods. The monodromy $T(0)$ around the LCS point ($z=0 ,\psi=\infty$) and the monodromy $T(\infty )$ around the Gepner point ($z=\infty ,\psi =0$) act on the period vector ${\bf U}(z)=(U_j(z))_{j=1...4}$ as follows,
\begin{eqnarray}
  {\bf U}(e^{2\pi i}z) &=&\, T(0)\, {\bf U}(z),\qquad |z| << 1 \nonumber\\
  {\bf U}(e^{2\pi i}z) &=&T(\infty )\, {\bf U}(z),\qquad  |z|>> 1 .
\end{eqnarray}
The third monodromy matrix of course follows directly from the other two, as a monodromy can always be seen either as `around one of the branching points' or, equivalently as a monodromy `around the two other branching points' in the appropriate directions. For the monodromy around the conifold point one has
\begin{eqnarray}
  T(1)&=&T(\infty )T(0)^{-1}\qquad \textrm{Im}(z)<0,\nonumber\\
  T(1)&=&T(0)^{-1}T(\infty )\qquad \textrm{Im}(z)>0 .
\end{eqnarray}
Using Mathematica, we evaluated the periods for our models on a lattice of points, the number of which can be adjusted to the degree of precision demanded. By interpolation this gives a numerical approximation of the periods.

\end{appendix}
\newpage

\bibliographystyle{JHEP}
\bibliography{ref}

\providecommand{\href}[2]{#2}\begingroup\raggedright\begin{thebibliography}{10}

\bibitem{Collinucci:2008ht}
A.~Collinucci and T.~Wyder, {\it {The elliptic genus from split flows and
  Donaldson-Thomas invariants}},  \href{http://xxx.lanl.gov/abs/0810.4301}{{\tt
  arXiv:0810.4301}}.

\bibitem{Gaiotto:2007cd}
D.~Gaiotto and X.~Yin, {\it {Examples of M5-brane elliptic genera}},  {\em
  JHEP} {\bf 11} (2007) 004,
  [\href{http://xxx.lanl.gov/abs/hep-th/0702012}{{\tt hep-th/0702012}}].

\bibitem{Denef:2007vg}
F.~Denef and G.~W. Moore, {\it {Split states, entropy enigmas, holes and
  halos}},  \href{http://xxx.lanl.gov/abs/hep-th/0702146}{{\tt
  hep-th/0702146}}.

\bibitem{Moore:1998pn}
G.~W. Moore, {\it {Arithmetic and attractors}},
  \href{http://xxx.lanl.gov/abs/hep-th/9807087}{{\tt hep-th/9807087}}.

\bibitem{Moore:1998zu}
G.~W. Moore, {\it {Attractors and arithmetic}},
  \href{http://xxx.lanl.gov/abs/hep-th/9807056}{{\tt hep-th/9807056}}.

\bibitem{Denef:2002ru}
F.~Denef, {\it {Quantum quivers and Hall/hole halos}},  {\em JHEP} {\bf 10}
  (2002) 023, [\href{http://xxx.lanl.gov/abs/hep-th/0206072}{{\tt
  hep-th/0206072}}].

\bibitem{Denef:2000ar}
F.~Denef, {\it {On the correspondence between D-branes and stationary
  supergravity solutions of type II Calabi-Yau compactifications}},
  \href{http://xxx.lanl.gov/abs/hep-th/0010222}{{\tt hep-th/0010222}}.

\bibitem{Denef:2000nb}
F.~Denef, {\it {Supergravity flows and D-brane stability}},  {\em JHEP} {\bf
  08} (2000) 050, [\href{http://xxx.lanl.gov/abs/hep-th/0005049}{{\tt
  hep-th/0005049}}].

\bibitem{Bates:2003vx}
B.~Bates and F.~Denef, {\it {Exact solutions for supersymmetric stationary
  black hole composites}},  \href{http://xxx.lanl.gov/abs/hep-th/0304094}{{\tt
  hep-th/0304094}}.

\bibitem{Denef:2001xn}
F.~Denef, B.~R. Greene, and M.~Raugas, {\it {Split attractor flows and the
  spectrum of BPS D-branes on the quintic}},  {\em JHEP} {\bf 05} (2001) 012,
  [\href{http://xxx.lanl.gov/abs/hep-th/0101135}{{\tt hep-th/0101135}}].

\bibitem{Ooguri:2004zv}
H.~Ooguri, A.~Strominger, and C.~Vafa, {\it {Black hole attractors and the
  topological string}},  {\em Phys. Rev.} {\bf D70} (2004) 106007,
  [\href{http://xxx.lanl.gov/abs/hep-th/0405146}{{\tt hep-th/0405146}}].

\bibitem{Pioline:2006ni}
B.~Pioline, {\it {Lectures on on black holes, topological strings and quantum
  attractors}},  {\em Class. Quant. Grav.} {\bf 23} (2006) S981,
  [\href{http://xxx.lanl.gov/abs/hep-th/0607227}{{\tt hep-th/0607227}}].

\bibitem{deBoer:2006vg}
J.~de~Boer, M.~C.~N. Cheng, R.~Dijkgraaf, J.~Manschot, and E.~Verlinde, {\it {A
  farey tail for attractor black holes}},  {\em JHEP} {\bf 11} (2006) 024,
  [\href{http://xxx.lanl.gov/abs/hep-th/0608059}{{\tt hep-th/0608059}}].

\bibitem{Manschot:2007ha}
J.~Manschot and G.~W. Moore, {\it {A Modern Farey Tail}},
  \href{http://xxx.lanl.gov/abs/0712.0573}{{\tt arXiv:0712.0573}}.

\bibitem{deBoer:2008fk}
J.~de~Boer, F.~Denef, S.~El-Showk, I.~Messamah, and D.~Van~den Bleeken, {\it
  {Black hole bound states in AdS$_3$ x S$^2$}},  {\em JHEP} {\bf 11} (2008)
  050, [\href{http://xxx.lanl.gov/abs/0802.2257}{{\tt arXiv:0802.2257}}].

\bibitem{deBoer:2009un}
J.~de~Boer, S.~El-Showk, I.~Messamah, and D.~Van~den Bleeken, {\it {A bound on
  the entropy of supergravity?}},
  \href{http://xxx.lanl.gov/abs/0906.0011}{{\tt arXiv:0906.0011}}.

\bibitem{deBoer:2008zn}
J.~de~Boer, S.~El-Showk, I.~Messamah, and D.~Van~den Bleeken, {\it {Quantizing
  N=2 Multicenter Solutions}},  {\em JHEP} {\bf 05} (2009) 002,
  [\href{http://xxx.lanl.gov/abs/0807.4556}{{\tt arXiv:0807.4556}}].

\bibitem{MNOP1}
A.~O. D.~Maulik, N.~Nekrasov and R.~Pandharipande, {\it Gromov-witten theory
  and donaldson-thomas theory},
  \href{http://xxx.lanl.gov/abs/math-AG/0312059}{{\tt math-AG/0312059}}.

\bibitem{MNOP2}
A.~O. D.~Maulik, N.~Nekrasov and R.~Pandharipande, {\it Gromov-witten theory
  and donaldson-thomas theory},
  \href{http://xxx.lanl.gov/abs/math-AG/0406092}{{\tt math-AG/0406092}}.

\bibitem{Gopakumar:1998ii}
R.~Gopakumar and C.~Vafa, {\it {M-theory and topological strings. I}},
  \href{http://xxx.lanl.gov/abs/hep-th/9809187}{{\tt hep-th/9809187}}.

\bibitem{Gopakumar:1998jq}
R.~Gopakumar and C.~Vafa, {\it {M-theory and topological strings. II}},
  \href{http://xxx.lanl.gov/abs/hep-th/9812127}{{\tt hep-th/9812127}}.

\bibitem{Dijkgraaf:2006um}
R.~Dijkgraaf, C.~Vafa, and E.~Verlinde, {\it {M-theory and a topological string
  duality}},  \href{http://xxx.lanl.gov/abs/hep-th/0602087}{{\tt
  hep-th/0602087}}.

\bibitem{Huang:2006hq}
M.-x. Huang, A.~Klemm, and S.~Quackenbush, {\it {Topological String Theory on
  Compact Calabi-Yau: Modularity and Boundary Conditions}},  {\em Lect. Notes
  Phys.} {\bf 757} (2009) 45--102,
  [\href{http://xxx.lanl.gov/abs/hep-th/0612125}{{\tt hep-th/0612125}}].

\bibitem{Klemm:2004km}
A.~Klemm, M.~Kreuzer, E.~Riegler, and E.~Scheidegger, {\it {Topological string
  amplitudes, complete intersection Calabi-Yau spaces and threshold
  corrections}},  {\em JHEP} {\bf 05} (2005) 023,
  [\href{http://xxx.lanl.gov/abs/hep-th/0410018}{{\tt hep-th/0410018}}].

\bibitem{Collinucci:2008pf}
A.~Collinucci, F.~Denef, and M.~Esole, {\it {D-brane Deconstructions in IIB
  Orientifolds}},  {\em JHEP} {\bf 02} (2009) 005,
  [\href{http://xxx.lanl.gov/abs/0805.1573}{{\tt arXiv:0805.1573}}].

\bibitem{Vonk:2005yv}
M.~Vonk, {\it {A mini-course on topological strings}},
  \href{http://xxx.lanl.gov/abs/hep-th/0504147}{{\tt hep-th/0504147}}.

\bibitem{Witten:1982im}
E.~Witten, {\it {Supersymmetry and Morse theory}},  {\em J. Diff. Geom.} {\bf
  17} (1982) 661--692.

\bibitem{Candelas:1989hd}
P.~Candelas, M.~Lynker, and R.~Schimmrigk, {\it {Calabi-Yau Manifolds in
  Weighted P(4)}},  {\em Nucl. Phys.} {\bf B341} (1990) 383--402.

\bibitem{Raeymaekers:2008gk}
J.~Raeymaekers, W.~Van~Herck, B.~Vercnocke, and T.~Wyder, {\it {5D fuzzball
  geometries and 4D polar states}},  {\em JHEP} {\bf 10} (2008) 039,
  [\href{http://xxx.lanl.gov/abs/0805.3506}{{\tt arXiv:0805.3506}}].

\bibitem{Maldacena:1997de}
J.~M. Maldacena, A.~Strominger, and E.~Witten, {\it {Black hole entropy in
  M-theory}},  {\em JHEP} {\bf 12} (1997) 002,
  [\href{http://xxx.lanl.gov/abs/hep-th/9711053}{{\tt hep-th/9711053}}].

\bibitem{Manschot:2008zb}
J.~Manschot, {\it {On the space of elliptic genera}},  {\em Commun. Num. Theor.
  Phys.} {\bf 2} (2008) 803--833,
  [\href{http://xxx.lanl.gov/abs/0805.4333}{{\tt arXiv:0805.4333}}].

\bibitem{Greene:2000ci}
B.~R. Greene and C.~I. Lazaroiu, {\it {Collapsing D-branes in Calabi-Yau moduli
  space. I}},  {\em Nucl. Phys.} {\bf B604} (2001) 181--255,
  [\href{http://xxx.lanl.gov/abs/hep-th/0001025}{{\tt hep-th/0001025}}].

\end{thebibliography}\endgroup

\end{document}